\documentclass[pdflatex,sn-mathphys-num]{sn-jnl}

\usepackage{amsmath,amssymb,amsthm}
\usepackage{enumitem}
\usepackage{booktabs}
\usepackage{tikz}
\usetikzlibrary{arrows.meta,decorations.pathreplacing}
\setlist[enumerate]{label=(\arabic*)}

\theoremstyle{thmstyleone}
\newtheorem{theorem}{Theorem}[section]
\newtheorem{proposition}[theorem]{Proposition}
\newtheorem{lemma}[theorem]{Lemma}
\newtheorem{corollary}[theorem]{Corollary}
\theoremstyle{thmstyletwo}
\newtheorem{remark}[theorem]{Remark}
\newtheorem{problem}[theorem]{Problem}
\theoremstyle{thmstylethree}
\newtheorem{definition}[theorem]{Definition}

\newcommand{\F}{\mathbb F}
\newcommand{\PG}{\mathrm{PG}}
\newcommand{\PGL}{\mathrm{PGL}}
\newcommand{\PGaL}{\mathrm{P\Gamma L}}
\newcommand{\GL}{\mathrm{GL}}
\newcommand{\Sym}{\mathrm{Sym}}
\newcommand{\cC}{\mathcal C}
\newcommand{\cD}{\mathcal D}
\newcommand{\Wh}{\widehat W}

\DeclareMathOperator{\Aut}{Aut}

\hypersetup{bookmarksdepth=2}

\definecolor{orcidgreen}{HTML}{A6CE39}
\gdef\orcidlogo{\tikz[baseline=-0.6ex]{\fill[orcidgreen] (0,0) circle (0.85ex);
  \node[white,font=\sffamily\bfseries\fontsize{4.6}{4.6}\selectfont,inner sep=0pt] at (0,-0.02ex) {iD};}}

\begin{document}

\title[A second family of NMDS codes holding 4-designs]{A second infinite family of near-MDS codes holding
$4$-designs}

\author[1,3]{\fnm{Jae-Hyun} \sur{Baek}\,\orcid{https://orcid.org/0009-0006-2765-636X}}\email{jhbaek@sogang.ac.kr}
\author*[2,3]{\fnm{Jon-Lark} \sur{Kim}\,\orcid{https://orcid.org/0000-0002-0517-9359}}\email{jlkim@sogang.ac.kr}
\affil[1]{\orgdiv{Department of Artificial Intelligence}, \orgname{Sogang University}, \orgaddress{\city{Seoul},
  \country{South Korea}}}
\affil*[2]{\orgdiv{Department of Mathematics}, \orgname{Sogang University}, \orgaddress{\city{Seoul},
  \country{South Korea}}}
\affil[3]{\orgdiv{Institute for Mathematical and Data Sciences}, \orgname{Sogang University},
  \orgaddress{\city{Seoul}, \country{South Korea}}}

\abstract{\unboldmath Tang and Ding gave the first infinite family of linear codes holding $4$-designs. For $q=2^h$ with
$h\ge5$ odd and $\beta$ of order $q+1$, their code is the cyclic code of length $q+1$ over $\F_q$ with zeros
$\beta^{\pm1},\beta^{\pm2},\beta^{\pm3}$. It is near maximum distance separable (near-MDS), and its minimum-weight
codewords support $4$-$(q+1,6,(q-8)/2)$ designs. With $\gamma=\beta+\beta^{-1}$, the cyclic code with zeros
$\beta^{\pm(s-1)},\beta^{\pm s},\beta^{\pm(s+1)}$, for a power $s$ of $2$, has generator polynomial
$(x^2+\gamma^sx+1)(x^4+\gamma^{s+1}x^3+(\gamma^{2s}+\gamma^2)x^2+\gamma^{s+1}x+1)$, and $s=2$ gives their code. We show
that $s=2^{(h-1)/2}$ gives a second infinite family with the same parameters. The proof uses the evaluation code of the
binary forms $A(t)+t^{2s}B(t)$ with $\deg A,\deg B\le2$ on the projective line, whose dual is equivalent to this
cyclic code. It attaches to each form a sextic, whose zeros are the first coordinates of the intersection points of
two curves on $\mathbb P^1\times\mathbb P^1$, and uses an involution of these zeros defined by $t\mapsto t^{2s}$. Both
designs have full automorphism group $\PGaL(2,q)$, and they are not isomorphic. They have no block in common, although
they cover the same $5$-sets. We also determine all weight shells and weight distributions of the new codes. Finally,
every $[q+1,6]$ code invariant under a monomial action of $\PGL(2,q)$ inducing its natural action on the projective
line is equivalent to the dual of one of these cyclic codes. For $h\le100$ only $s=2$ and $s=2^{(h-1)/2}$ give
near-MDS codes, and for even $h$ none does. The theorems are formally verified in the Lean proof assistant, with five
cited theorems and a computer certificate taken as axioms.}



\maketitle

\section{Introduction}

Let $\cC$ be a linear code of length $n$. For each weight $w$, the supports of the codewords of weight $w$ form a set
$\mathcal B_w(\cC)$ of $w$-subsets of the coordinates. When it is a $t$-$(n,w,\lambda)$ design, we say that $\cC$
holds this design~\cite{DT22}. The classical sources of such designs are the Assmus--Mattson theorem~\cite{AM69} and
$t$-homogeneous automorphism groups. Designs of strength $4$ held by linear codes remained sporadic for a long time.

Tang and Ding~\cite{TD21} found the first infinite family. Let $q=2^h$ with $h\ge5$ odd (Tang and Ding write $q=2^m$), and let $\beta\in\F_{q^2}$
have order $q+1$. Since $\beta^q=\beta^{-1}$, the minimal polynomial of $\beta^{\pm j}$ over $\F_q$ is
$x^2+(\beta^j+\beta^{-j})x+1$. Tang and Ding considered the BCH code of length $q+1$ over $\F_q$ with defining set
$\pm\{1,2,3\}$, the cyclic code generated by the product of these polynomials for $j=1,2,3$. This code and its dual
are near-MDS, and the supports of its codewords of minimum weight $6$ form a $4$-$(q+1,6,(q-8)/2)$ design. Yan and Zhou~\cite{YZ22} found further
$4$-designs in the same codes. Ding, Tang and Tonchev~\cite{DTT21} and Liu, Ding, Mesnager, Tang and
Tonchev~\cite{LDMTT22} studied other codes of length $q+1$ related to the action of $\PGL(2,q)$ on the projective
line; these hold $3$-designs. In a recent study of near-MDS codes supporting designs~\cite{ZZWL25}, Tang and
Ding's construction is the only infinite family of near-MDS codes holding $4$-designs that is mentioned, and further
such families are asked for; we know of no other (see also~\cite{DT20,HW23,XWLC26}). Bierbrauer~\cite{Bie95}
constructed simple $4$-$(2^h+1,6,10)$ designs for odd $h$; the designs of the present paper have the same number of
points and the same block size, but $\lambda=(q-8)/2$, and they are held by linear codes.

In this paper we give a second family with the same parameters, obtained by changing the exponents in the generator
polynomial. Put $\gamma=\beta+\beta^{-1}\in\F_q$ and, for a power $s$ of $2$,
\begin{equation}\label{eq:Gs}
G_s(x)=(x^2+\gamma^sx+1)\bigl(x^4+\gamma^{s+1}x^3+(\gamma^{2s}+\gamma^2)x^2+\gamma^{s+1}x+1\bigr).
\end{equation}
This is the product of the minimal polynomials of $\beta^{\pm(s-1)}$, $\beta^{\pm s}$ and $\beta^{\pm(s+1)}$. Indeed,
with $D_k=\beta^k+\beta^{-k}$, we have $D_s=\gamma^s$ because $s$ is a power of $2$, and
$D_{s-1}+D_{s+1}=D_sD_1=\gamma^{s+1}$ and $D_{s-1}D_{s+1}=D_{2s}+D_2=\gamma^{2s}+\gamma^2$. For $2\le s\le q/4$ the six
roots are distinct $(q+1)$-th roots of unity, so $G_s$ divides $x^{q+1}-1$ and generates a cyclic code $\mathcal K_s$
of length $q+1$ and dimension $q-5$ over $\F_q$. For $s=2$ the roots are $\beta^{\pm1},\beta^{\pm2},\beta^{\pm3}$,
and $\mathcal K_2$ is the code of Tang and Ding. We show that $s=2^{(h-1)/2}$, the square root of $q/2$, gives a
second family, and that for $h\le100$ no other $s$ gives a near-MDS code (Theorem~\ref{thm:unique}).
Section~\ref{sec:q32} writes out $G_2$ and $G_4$ for $q=32$.

In our proofs these cyclic codes are realised, up to monomial equivalence, as the duals of evaluation codes of binary
forms on the projective line. Let $\sigma=2^{(h+1)/2}$, so that the new value of $s$ is $\sigma/2$. Let $W_\sigma$ be the $6$-dimensional space of polynomials
\[
P(t)=A(t)+t^\sigma B(t),\qquad \deg A,\deg B\le2,
\]
regarded as binary forms of degree $\sigma+2$, and let $\cC_\sigma\subseteq\F_q^{q+1}$ be the code formed by their
values at the $q+1$ points of $\PG(1,q)$. With $t^4$ in place of $t^\sigma$, the same construction gives a code
$\cC_4$. Up to monomial equivalence, $\cC_\sigma^\perp$ is $\mathcal K_{\sigma/2}$ and $\cC_4^\perp$ is
$\mathcal K_2$, the code of Tang and Ding (Proposition~\ref{prop:cyclic}).

\begin{theorem}\label{thm:main}
Let $h\ge5$ be odd.
\begin{enumerate}[nosep]
\item $\cC_\sigma$ is a $[q+1,6,q-5]$ code and $\cC_\sigma^\perp$ is a $[q+1,q-5,6]$ code, so both are near-MDS.
\item The supports of the codewords of weight $6$ in $\cC_\sigma^\perp$ form a $4$-$(q+1,6,(q-8)/2)$ design
  $\cD_\sigma$, which is invariant under $\PGaL(2,q)$.
\item Up to monomial equivalence, $\cC_\sigma^\perp$ is the cyclic code of length $q+1$ over $\F_q$ with defining
  set $\pm\{1,\sigma-1,\sigma+1\}$.
\end{enumerate}
\end{theorem}

With respect to $\beta^{s}$, $s=\sigma/2$, the defining set $\pm\{1,\sigma-1,\sigma+1\}$ of part~3 is the defining
set $\pm\{s-1,s,s+1\}$ with respect to $\beta$, since $s\sigma=q\equiv-1\pmod{q+1}$. So part~3 says that
$\cC_\sigma^\perp$ is monomially equivalent to $\mathcal K_{\sigma/2}$, the cyclic code generated by
$G_{\sigma/2}$ in~\eqref{eq:Gs}. Thus $\mathcal K_{\sigma/2}$ is a near-MDS $[q+1,q-5,6]$ code whose codewords of
weight $6$ support a design isomorphic to $\cD_\sigma$, just as $\mathcal K_2$ holds the design of Tang and Ding.

We also determine every weight shell of $\cC_\sigma$ and $\cC_\sigma^\perp$ (Theorem~\ref{thm:shells}) and their
weight distributions (Corollary~\ref{cor:weights}). Besides $\cD_\sigma$, the code $\cC_\sigma^\perp$ holds a
$4$-design with blocks of size $7$, and $\cC_\sigma$ holds the complements of $\cD_\sigma$ and of a $4$-$(q+1,5,5)$
design formed by the $5$-sets in no block of $\cD_\sigma$. All other nonempty shells are complete.

Write $\cD_4$ for the design of Tang and Ding. Our second result shows that the two families are genuinely different.

\begin{theorem}\label{thm:noniso}
For every odd $h\ge5$, the full automorphism group of both $\cD_\sigma$ and $\cD_4$ is $\PGaL(2,q)$, and the two
designs are not isomorphic.
\end{theorem}

Consequently no semilinear map carries $\cC_\sigma^\perp$ onto the code of Tang and Ding (Corollary~\ref{cor:codes}).
To our knowledge, Theorem~\ref{thm:noniso} also determines the automorphism group of their design for the first
time. In the notation of~\cite{TD21}, $\cC_{(q,n,\delta,1)}$ is the narrow-sense BCH code over $\F_q$ of length $n$
and designed distance $\delta$, and their code is $\cC_{(q,q+1,4,1)}$. More generally, $\cC_{(q,n,\delta,b)}$ has
the zeros $\beta^b,\dots,\beta^{b+\delta-2}$ and their conjugates, so that $\mathcal K_s=\cC_{(q,q+1,4,s-1)}$; codes
$\cC_{(q,q+1,3,b)}$ with $b\neq1$ are studied in~\cite{XWLC26}. Liu et al.~\cite{LDMTT22} show that $\PGL(2,q)$
acts by monomial automorphisms on the codes $\cC_{(q,q+1,\delta,1)}$ when $q$ is a power of $\delta$, which for
$\delta=4$ requires $q$ to be an even power of $2$, and leave their full automorphism groups open. Wu and Zhou give
such an action on the dual of the code of Tang and Ding~\cite{WZ26a} and, more generally, on the codes
$\cC_{(q,q+1,\delta,1)}$ for every power $\delta<q$ of the characteristic~\cite{WZ26b}, and show that these codes have
$3$-transitive automorphism groups. Tang and Ding call
the determination of the automorphism groups of their codes an interesting problem~\cite{TD21}.

Tang and Ding also remark that a proof of the design property through automorphism groups looks
impossible~\cite{TD21}. Indeed $\PGaL(2,q)$ does not force a $4$-design. Every $4$-set is equivalent to some
$\{0,1,\infty,c\}$. Each double transposition of a $4$-set preserves the cross-ratio and is therefore induced by an
element of $\PGL(2,q)$. So $\{0,1,\infty,c\}$ and $\{0,1,\infty,c'\}$ are equivalent only if $c'$ is the image of $c$
under one of the $6h$ elements of $\PGaL(2,q)$ that preserve $\{0,1,\infty\}$, and the group has at least
$(q-2)/(6h)$ orbits on $4$-sets. It is $4$-homogeneous only for $q=32$~\cite{Kan72}.

Nevertheless, the two designs are closely related. They have no block in common, so their union is a $4$-$(q+1,6,q-8)$ design
(Proposition~\ref{prop:differ}). Through a $4$-set $\{0,1,\infty,c\}$ they leave out the same five points, so they
cover the same $5$-sets (Theorem~\ref{thm:fivesets}). Hence the $4$-$(q+1,5,5)$ design of
Theorem~\ref{thm:shells}, formed by the $5$-sets in no block, is the same for both designs, and by Yan and Zhou~\cite{YZ22} it is one of Alltop's
$4$-designs~\cite{Alt69}.

The two families also differ in how the design property arises. Six points of $\F_q$ form a block of $\cD_4$ iff their
third elementary symmetric function $e_3$ vanishes, and they form a block of $\cD_\sigma$ iff a Schur function of degree
$3(\sigma-3)$ vanishes (Proposition~\ref{prop:blocks}). Tang and Ding work with the first condition in the group
$U_{q+1}$ of $(q+1)$-th roots of unity. They complete a $4$-set $\{u_1,\dots,u_4\}$ by
$u_6=e_3(u_1,\dots,u_5)/e_2(u_1,\dots,u_5)$, a M\"obius function of $u_5$, and exclude the four given points and
five further ones with trace criteria for quadratic equations. They write $\sigma_{k,\ell}$ for the elementary
symmetric polynomial $e_\ell$ in $k$ variables, so these are their conditions $\sigma_{6,3}=0$ and
$u_6=\sigma_{5,3}/\sigma_{5,2}$. In this paper $\sigma$ always denotes the exponent $2^{(h+1)/2}$ and the field
automorphism $t\mapsto t^\sigma$. For $\cD_\sigma$ the condition has degree
$3(\sigma-3)\approx3\sqrt{2q}$, and we do not solve it for the sixth point. Instead, the zeros of a member of
$W_\sigma$ are zeros of an auxiliary form of degree $6$, which describes the intersection of two curves on
$\mathbb P^1\times\mathbb P^1$, and an involution of its zeros, defined by the field automorphism $t\mapsto t^\sigma$,
makes the number of blocks through a $4$-set constant. In the proof of Theorem~\ref{thm:main} the representation as a
cyclic code is needed only for part~3; in Section~\ref{sec:compare} it identifies $\cD_4$ with the design of Tang and Ding.

Our third result shows that, at least for $h\le100$, the two families are the only ones with this symmetry. Let
$\PGL(2,q)$ act on $\F_q^{q+1}$ by monomial maps that permute the coordinates as it permutes the points of
$\PG(1,q)$. By Lemma~\ref{lem:Wbasic} it acts in this way on $\cC_\sigma$ and on $\cC_4$. For a power $\tau=2^j$ let
$\cC_\tau$ be the evaluation code of the forms $A(t)+t^\tau B(t)$ with $\deg A,\deg B\le2$.

\begin{theorem}\label{thm:unique}
Let $q=2^h$ with $h\ge5$, odd or even.
\begin{enumerate}[nosep]
\item Every $[q+1,6]$ code over $\F_q$ that is invariant under such an action of $\PGL(2,q)$ is semilinearly
  equivalent to $\cC_{2^j}$ for some $2\le j\le h-1$.
\item If $h\le100$, then $\cC_{2^j}$ is near-MDS if and only if $h$ is odd and $j\in\{2,(h+1)/2\}$, that is,
  $2^j\in\{4,\sigma\}$.
\end{enumerate}
\end{theorem}

So for $h\le100$ the codes $\cC_4$ and $\cC_\sigma$ are, up to semilinear equivalence, the only near-MDS codes of
length $q+1$ and dimension $6$ with this symmetry, and for even $h$ there are none. Since $\cC_{2s}^\perp$ is
monomially equivalent to $\mathcal K_s$ for $2\le s\le q/4$ (Section~\ref{sec:nmdstwists}), part~2 says that among
the cyclic codes $\mathcal K_s$ of~\eqref{eq:Gs} only $\mathcal K_2$ and $\mathcal K_{\sigma/2}$ are near-MDS when
$h\le100$. The code $\cC_{q/2}$ is semilinearly
equivalent to the doubly extended Reed--Solomon code of dimension $6$. Part~2 rests on linearised polynomials. A
polynomial $x^{2^j}+ax^2+bx$ with $8$ roots in $\F_q$ yields a member of the form space with $9$ zeros. For
$2^j\in\{4,\sigma,q/2\}$ no such polynomial exists, because $x$, $x^2$ and $x^{2^j}$ then span a Gabidulin code. This is the precise sense in which the two families correspond to
the generators $t\mapsto t^2$ and $t\mapsto t^\sigma$ of $\mathrm{Gal}(\F_q/\F_2)$. For every other $j$ we find such a
polynomial whenever $h\le100$, by a computer search with an independently checked certificate. For every $h$ we find
one when $j<(h-2)/4$ or $j>(3h+6)/4$, by a theorem of Bartoli and Zhou on Moore exponent sets~\cite{BZ20}
(Theorem~\ref{thm:nmdstwists}).

Section~\ref{sec:setting} fixes notation and describes the group $\PGaL(2,q)$. Section~\ref{sec:proof} proves Theorems~\ref{thm:main}
and~\ref{thm:shells}, and Section~\ref{sec:compare} proves Theorems~\ref{thm:noniso} and~\ref{thm:fivesets}.
Section~\ref{sec:unique} proves Theorem~\ref{thm:unique}. Section~\ref{sec:examples} gives examples: the twists
for $h\le13$, both codes for $q=32$, the incidence matrices of the blocks through a $4$-set, and an elementary invariant that
separates the two designs for $q\le512$. Section~\ref{sec:remarks}
collects computational evidence, the formal verification in Lean of the numbered results
(Section~\ref{sec:lean}), and open problems.

\section{Preliminaries}\label{sec:setting}

\subsection{Codes and designs}

We recall the notions from design theory and coding theory that are used below; see~\cite{DT22} for background. All
designs in this paper are simple, that is, they have no repeated blocks.

\begin{definition}\label{def:design}
Let $X$ be a set of $n$ points and $\mathcal B$ a set of $k$-subsets of $X$, called \emph{blocks}.
\begin{enumerate}[nosep]
\item The pair $\cD=(X,\mathcal B)$ is a \emph{$t$-$(n,k,\lambda)$ design} if every $t$-subset of $X$ lies in exactly
  $\lambda$ blocks.
\item The design is \emph{complete} if $\mathcal B$ consists of all $k$-subsets of $X$.
\item An \emph{isomorphism} between two designs is a bijection between their point sets that maps blocks onto blocks.
  The isomorphisms of $\cD$ onto itself form its \emph{automorphism group} $\Aut(\cD)$.
\end{enumerate}
\end{definition}

A complete design is a $t$-design for every $t\le k$, so only designs that are not complete are of interest. The
\emph{complement} of a design $(X,\mathcal B)$ is $(X,\{X\setminus B:B\in\mathcal B\})$; it is a $t$-design whenever
$(X,\mathcal B)$ is one. Linear codes give designs through the supports of their codewords.

\begin{definition}\label{def:codes}
Let $\cC$ be a linear code of length $n$ over $\F_q$, whose coordinates we identify with a set $X$ of $n$ points.
\begin{enumerate}[nosep]
\item For a weight $w$, the \emph{weight shell} $\mathcal B_w(\cC)$ is the set of supports of the codewords of
  weight $w$.
\item The code $\cC$ \emph{holds} a design on $X$ if the set of blocks of the design is $\mathcal B_w(\cC)$ for some
  $w$.
\end{enumerate}
\end{definition}

By the Singleton bound, an $[n,k,d]$ code satisfies $d\le n-k+1$, with equality for MDS codes. The codes in this paper
miss this bound by one, and so do their duals.

\begin{definition}\label{def:nmds}
Let $\cC$ be a linear $[n,k,d]$ code.
\begin{enumerate}[nosep]
\item $\cC$ is \emph{almost MDS} if $d=n-k$.
\item $\cC$ is \emph{near-MDS} (NMDS) if both $\cC$ and its dual $\cC^\perp$ are almost MDS.
\end{enumerate}
\end{definition}

If $\cC$ is an NMDS $[n,k,n-k]$ code, then $\cC^\perp$ is an $[n,n-k,k]$ code, so the minimum-weight shells of $\cC$ and
$\cC^\perp$ consist of $(n-k)$-sets and $k$-sets. For the codes of Theorem~\ref{thm:main}, the minimum-weight
shell of $\cC_\sigma^\perp$ holds the design $\cD_\sigma$, and that of $\cC_\sigma$ holds its complement. Finally, we compare codes by the following equivalences.

\begin{definition}\label{def:semilinear}
Consider codes of length $n$ over $\F_q$.
\begin{enumerate}[nosep]
\item A \emph{semilinear map} $\F_q^n\to\F_q^n$ is a permutation of the coordinates combined with nonzero rescalings
  of the coordinates and a field automorphism applied to every entry. It is \emph{monomial} if the field automorphism
  is the identity.
\item Two codes are \emph{semilinearly} (respectively \emph{monomially}) \emph{equivalent} if a semilinear
  (respectively monomial) map carries one onto the other.
\end{enumerate}
\end{definition}

A semilinear map preserves weights and carries supports to supports, so it carries every weight shell of a code onto
the corresponding weight shell of its image. Equivalent codes therefore hold isomorphic designs, and this is how
Corollary~\ref{cor:codes} follows from Theorem~\ref{thm:noniso}.

\subsection{The exponent \texorpdfstring{$\sigma$}{sigma}}

The construction uses one particular automorphism of $\F_q$, which exists because $h$ is odd.

\begin{lemma}\label{lem:sigma}
Let $h=2e+1\ge5$, $q=2^h$, $s=2^e$ and $\sigma=2s=2^{e+1}$. Then $\sigma^2=2q$, $\sigma^2\equiv2\pmod{q-1}$ and
$s\sigma\equiv1\pmod{q-1}$. Hence $t\mapsto t^\sigma$ and $t\mapsto t^s$ are mutually inverse automorphisms of
$\F_q$, and $(t^\sigma)^\sigma=t^2$ for $t\in\F_q$.
\end{lemma}

\begin{proof}
We have $\sigma^2=2^{2e+2}=2q=2(q-1)+2$ and $s\sigma=2^{2e+1}=q=(q-1)+1$.
\end{proof}

The automorphism $t\mapsto t^\sigma$ is the one that defines the Suzuki--Tits ovoid. The notation $h,e,q,s,\sigma$ of
Lemma~\ref{lem:sigma} is fixed from now on.

\subsection{Binary forms}

The codes of Theorem~\ref{thm:main} are evaluation codes on the projective line. We work with polynomials of bounded
degree, viewed as binary forms, so that they have values and zeros at every point of $\PG(1,q)$, including $\infty$.

\begin{definition}\label{def:forms}
Points of $\PG(1,q)$ are written $t\in\F_q\cup\{\infty\}$.
\begin{enumerate}[nosep]
\item A \emph{form of degree $d$} is a polynomial $P$ of degree at most $d$. Its value at $\infty$ is its coefficient
  of $t^d$.
\item For a point $t$, $m_t(P)$ is the multiplicity of $t$ as a root of $P$; at infinity, $m_\infty(P)=d-\deg P$.
\item The form $P$ is \emph{tangent} at $t$ if $m_t(P)\ge2$.
\end{enumerate}
\end{definition}

For finite $t$, tangency means $P(t)=P'(t)=0$, where $P'$ is the formal derivative. At $t=\infty$ it means that the
two top coefficients vanish. Equivalently, we identify a form $P$ of degree $d$ with the binary form $v^dP(u/v)$ and
write points as $w=[u:v]$, with $t=u/v$; values and multiplicities are those of this binary form. The codes of this paper come from the following space of forms.

\begin{definition}\label{def:W}
Let $\sigma=2^{e+1}$ be as in Lemma~\ref{lem:sigma}.
\begin{enumerate}[nosep]
\item $W_\sigma$ is the space of forms $P=A+t^\sigma B$ of degree $\sigma+2$, where $A=a_0+a_1t+a_2t^2$ and
  $B=b_0+b_1t+b_2t^2$ have coefficients in $\F_q$.
\item $\cC_\sigma=\{(P(t))_{t\in\PG(1,q)}:P\in W_\sigma\}\subseteq\F_q^{q+1}$ is the evaluation code of $W_\sigma$.
\item $\Wh_\sigma$ is the auxiliary space of forms $a_0+a_1t+t^\sigma(b_0+b_1t)$ of degree $\sigma+1$.
\end{enumerate}
\end{definition}

The value of $P\in W_\sigma$ at $\infty$ is $b_2$. The auxiliary space $\Wh_\sigma$ arises when a linear factor is
removed from a member of $W_\sigma$; Section~\ref{sec:proof} uses it to bound the zeros of members of $W_\sigma$
(Lemma~\ref{lem:zeros}(1)).

We fix the basis $t^{a+b\sigma}$, $a\in\{0,1,2\}$, $b\in\{0,1\}$, of $W_\sigma$. The \emph{column} of a point is the
vector of values of these six forms there: $e(t)=(1,t,t^2)\otimes(1,t^\sigma)$ for finite $t$, and at $\infty$ the
vector with entry $1$ at $t^{\sigma+2}$ and $0$ elsewhere. These are the columns of a generator matrix of
$\cC_\sigma$. A set of points is \emph{independent}, \emph{dependent} or \emph{of rank $r$} according to its columns.
Equivalently, a set $S$ is independent iff it imposes $|S|$ independent linear conditions on $W_\sigma$.

\begin{lemma}\label{lem:Wbasic}
The space $W_\sigma$ has the following two properties.
\begin{enumerate}[nosep]
\item Every $P=A+t^\sigma B\in W_\sigma$ satisfies $P'=a_1+t^\sigma b_1$ and $P''=0$.
\item The space $W_\sigma$ is stable under $\GL_2(q)$ acting on forms by linear substitution, and under applying a field
automorphism to the coefficients. Consequently $\PGaL(2,q)$ acts on $\cC_\sigma$ by semilinear maps, which are
monomial for $\PGL(2,q)$. These maps carry supports to supports, and linear substitutions preserve the
multiplicities of zeros.
\end{enumerate}
\end{lemma}

\begin{proof}
For part 1, since $2=0$ in $\F_q$ and $\sigma$ is even, the derivatives of $t^2$ and $t^\sigma$ vanish. Hence
$P'=a_1+t^\sigma b_1$, and the derivative of this form is $0$.

For part 2, let $V$ be the natural module of $\GL_2$ and $V^{(\sigma)}$ its Frobenius twist. The multiplication map
$\Sym^2V\otimes V^{(\sigma)}\to\Sym^{\sigma+2}V$, $f\otimes\ell\mapsto f\,\ell^\sigma$, is $\GL_2(q)$-equivariant,
because raising a linear form to the power $\sigma$ commutes with linear substitution in characteristic $2$. Its
image is spanned by the monomials $t^{a+b\sigma}$ with $a\in\{0,1,2\}$ and $b\in\{0,1\}$, so it is $W_\sigma$. Since
$W_\sigma$ is spanned by monomials, it is also stable under field automorphisms applied to the coefficients. For
$g\in\PGL(2,q)$, the values of $P\circ g$ are those of $P$, permuted by $g$ and rescaled by nonzero factors that
come from the choice of homogeneous coordinates, which is a monomial map. A field automorphism applied to the
coefficients and to the points gives the remaining semilinear maps. Linear substitution preserves multiplicities.
\end{proof}

The proof identifies $W_\sigma$ with $\Sym^2V\otimes V^{(\sigma)}$. With $4$ in place of $\sigma$ it gives the same
statements for the space $W_4=\Sym^2V\otimes V^{(4)}$ of forms $A+t^4B$ of degree $6$, which appears in
Section~\ref{sec:compare}. In particular $t\mapsto1/t$ preserves the exponent set $\{0,1,2,\sigma,\sigma+1,\sigma+2\}$ of $W_\sigma$. Since
$\PGL(2,q)$ is transitive on $\PG(1,q)$, in each argument below we may assume that a given point is $\infty$, or that a
given set of fewer than $q+1$ points avoids $\infty$. The rescalings in Lemma~\ref{lem:Wbasic}(2) cannot be avoided:
by~\cite[Theorem~11]{DTT21}, the only linear codes of length $q+1$ over $\F_q$ that are invariant under $\PGL(2,q)$
acting by coordinate permutations are the zero code, the whole space, the repetition code and its dual.

\subsection{The group \texorpdfstring{$\PGaL(2,q)$}{PGammaL(2,q)}}\label{sec:group}

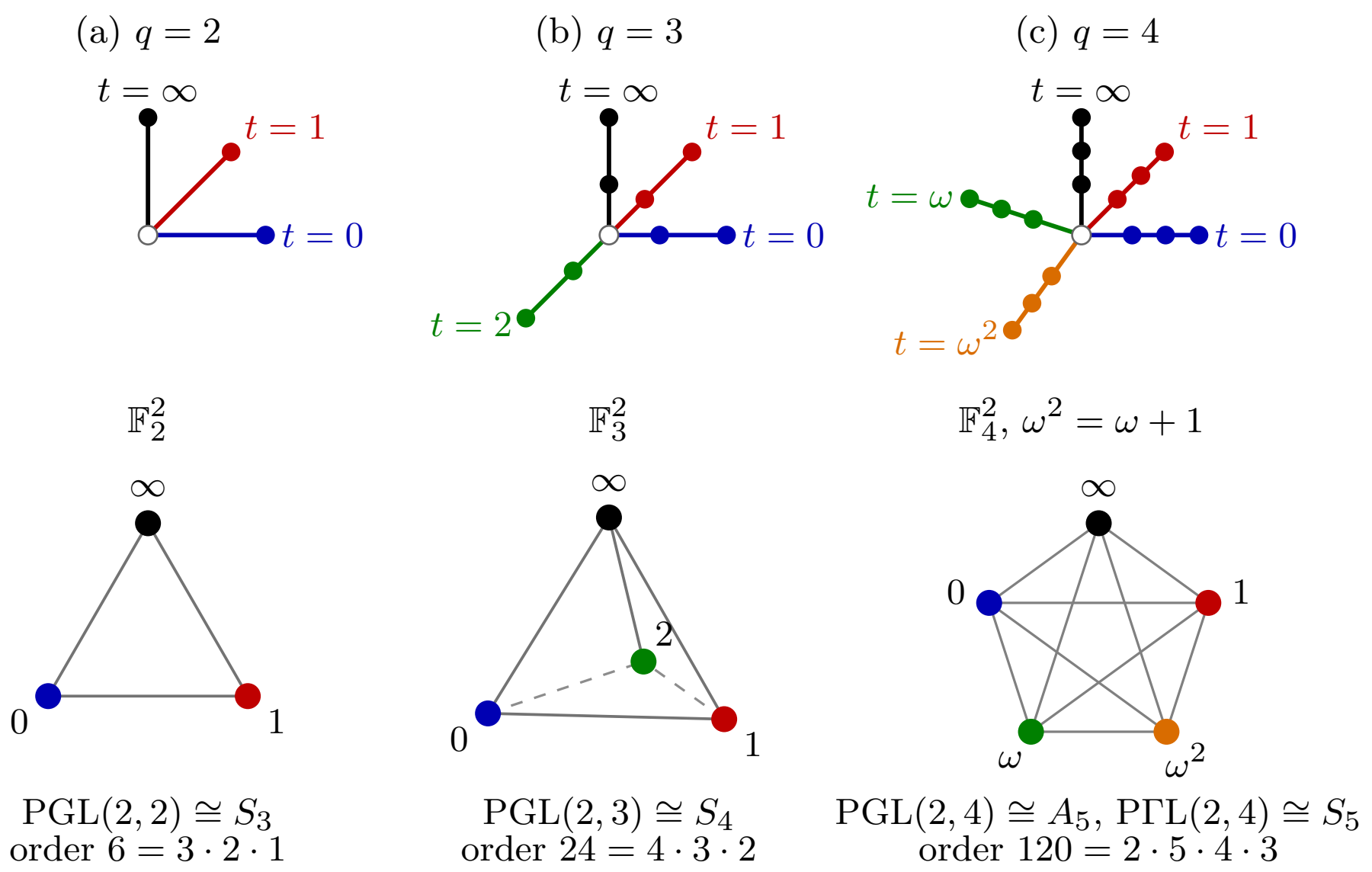
\begin{figure}[t]
\centering
\begin{tikzpicture}[x=1cm,y=1cm,
  gp/.style={circle,inner sep=0pt,minimum size=4.6pt},
  og/.style={circle,draw=black!60,fill=white,inner sep=0pt,minimum size=4.6pt,line width=0.5pt},
  vx/.style={circle,inner sep=0pt,minimum size=6.4pt},
  lab/.style={font=\scriptsize}]
\draw[blue!70!black,line width=1.1pt] (1.35,-0.35) -- (2.370,-0.350);
\node[gp,fill=blue!70!black] at (2.370,-0.350) {};
\node[lab,text=blue!70!black,anchor=west,inner sep=1.5pt] at (2.450,-0.350) {$t=0$};
\draw[red!75!black,line width=1.1pt] (1.35,-0.35) -- (2.071,0.371);
\node[gp,fill=red!75!black] at (2.071,0.371) {};
\node[lab,text=red!75!black,anchor=south west,inner sep=1.5pt] at (2.128,0.428) {$t=1$};
\draw[black,line width=1.1pt] (1.35,-0.35) -- (1.350,0.670);
\node[gp,fill=black] at (1.350,0.670) {};
\node[lab,text=black,anchor=south,inner sep=1.5pt] at (1.350,0.750) {$t=\infty$};
\node[og] at (1.35,-0.35) {};
\node[lab] at (1.35,-1.95) {$\F_2^2$};
\draw[blue!70!black,line width=1.1pt] (5.35,-0.35) -- (6.370,-0.350);
\node[gp,fill=blue!70!black] at (5.790,-0.350) {};
\node[gp,fill=blue!70!black] at (6.370,-0.350) {};
\node[lab,text=blue!70!black,anchor=west,inner sep=1.5pt] at (6.450,-0.350) {$t=0$};
\draw[red!75!black,line width=1.1pt] (5.35,-0.35) -- (6.071,0.371);
\node[gp,fill=red!75!black] at (5.661,-0.039) {};
\node[gp,fill=red!75!black] at (6.071,0.371) {};
\node[lab,text=red!75!black,anchor=south west,inner sep=1.5pt] at (6.128,0.428) {$t=1$};
\draw[black,line width=1.1pt] (5.35,-0.35) -- (5.350,0.670);
\node[gp,fill=black] at (5.350,0.090) {};
\node[gp,fill=black] at (5.350,0.670) {};
\node[lab,text=black,anchor=south,inner sep=1.5pt] at (5.350,0.750) {$t=\infty$};
\draw[green!50!black,line width=1.1pt] (5.35,-0.35) -- (4.629,-1.071);
\node[gp,fill=green!50!black] at (5.039,-0.661) {};
\node[gp,fill=green!50!black] at (4.629,-1.071) {};
\node[lab,text=green!50!black,anchor=east,inner sep=1.5pt] at (4.572,-1.128) {$t=2$};
\node[og] at (5.35,-0.35) {};
\node[lab] at (5.35,-1.95) {$\F_3^2$};
\draw[blue!70!black,line width=1.1pt] (9.45,-0.35) -- (10.470,-0.350);
\node[gp,fill=blue!70!black] at (9.890,-0.350) {};
\node[gp,fill=blue!70!black] at (10.180,-0.350) {};
\node[gp,fill=blue!70!black] at (10.470,-0.350) {};
\node[lab,text=blue!70!black,anchor=west,inner sep=1.5pt] at (10.550,-0.350) {$t=0$};
\draw[red!75!black,line width=1.1pt] (9.45,-0.35) -- (10.171,0.371);
\node[gp,fill=red!75!black] at (9.761,-0.039) {};
\node[gp,fill=red!75!black] at (9.966,0.166) {};
\node[gp,fill=red!75!black] at (10.171,0.371) {};
\node[lab,text=red!75!black,anchor=south west,inner sep=1.5pt] at (10.228,0.428) {$t=1$};
\draw[black,line width=1.1pt] (9.45,-0.35) -- (9.450,0.670);
\node[gp,fill=black] at (9.450,0.090) {};
\node[gp,fill=black] at (9.450,0.380) {};
\node[gp,fill=black] at (9.450,0.670) {};
\node[lab,text=black,anchor=south,inner sep=1.5pt] at (9.450,0.750) {$t=\infty$};
\draw[green!50!black,line width=1.1pt] (9.45,-0.35) -- (8.480,-0.035);
\node[gp,fill=green!50!black] at (9.032,-0.214) {};
\node[gp,fill=green!50!black] at (8.756,-0.124) {};
\node[gp,fill=green!50!black] at (8.480,-0.035) {};
\node[lab,text=green!50!black,anchor=east,inner sep=1.5pt] at (8.404,-0.010) {$t=\omega$};
\draw[orange!85!black,line width=1.1pt] (9.45,-0.35) -- (8.850,-1.175);
\node[gp,fill=orange!85!black] at (9.191,-0.706) {};
\node[gp,fill=orange!85!black] at (9.021,-0.941) {};
\node[gp,fill=orange!85!black] at (8.850,-1.175) {};
\node[lab,text=orange!85!black,anchor=east,inner sep=1.5pt] at (8.803,-1.240) {$t=\omega^2$};
\node[og] at (9.45,-0.35) {};
\node[lab] at (9.45,-1.95) {$\F_4^2$, $\omega^2=\omega+1$};
\draw[black!55,line width=0.7pt] (1.350,-2.850) -- (0.484,-4.350) -- (2.216,-4.350) -- cycle;
\node[vx,fill=black] at (1.350,-2.850) {};
\node[vx,fill=blue!70!black] at (0.484,-4.350) {};
\node[vx,fill=red!75!black] at (2.216,-4.350) {};
\node[lab] at (1.350,-2.550) {$\infty$};
\node[lab] at (0.234,-4.570) {$0$};
\node[lab] at (2.466,-4.570) {$1$};
\node[lab,align=center] at (1.35,-5.5) {$\PGL(2,2)\cong S_3$\\ order $6=3\cdot2\cdot1$};
\draw[black!55,line width=0.7pt] (5.350,-2.800) -- (4.300,-4.500);
\draw[black!55,line width=0.7pt] (5.350,-2.800) -- (6.350,-4.550);
\draw[black!55,line width=0.7pt] (4.300,-4.500) -- (6.350,-4.550);
\draw[black!55,line width=0.7pt] (5.350,-2.800) -- (5.650,-4.050);
\draw[black!45,dashed,line width=0.6pt] (4.300,-4.500) -- (5.650,-4.050);
\draw[black!45,dashed,line width=0.6pt] (6.350,-4.550) -- (5.650,-4.050);
\node[vx,fill=black] at (5.350,-2.800) {};
\node[vx,fill=blue!70!black] at (4.300,-4.500) {};
\node[vx,fill=red!75!black] at (6.350,-4.550) {};
\node[vx,fill=green!50!black] at (5.650,-4.050) {};
\node[lab] at (5.350,-2.500) {$\infty$};
\node[lab] at (4.050,-4.720) {$0$};
\node[lab] at (6.600,-4.770) {$1$};
\node[lab] at (5.830,-3.810) {$2$};
\node[lab,align=center] at (5.35,-5.5) {$\PGL(2,3)\cong S_4$\\ order $24=4\cdot3\cdot2$};
\draw[black!50,line width=0.6pt] (9.600,-2.850) -- (10.551,-3.541);
\draw[black!50,line width=0.6pt] (9.600,-2.850) -- (10.188,-4.659);
\draw[black!50,line width=0.6pt] (9.600,-2.850) -- (9.012,-4.659);
\draw[black!50,line width=0.6pt] (9.600,-2.850) -- (8.649,-3.541);
\draw[black!50,line width=0.6pt] (10.551,-3.541) -- (10.188,-4.659);
\draw[black!50,line width=0.6pt] (10.551,-3.541) -- (9.012,-4.659);
\draw[black!50,line width=0.6pt] (10.551,-3.541) -- (8.649,-3.541);
\draw[black!50,line width=0.6pt] (10.188,-4.659) -- (9.012,-4.659);
\draw[black!50,line width=0.6pt] (10.188,-4.659) -- (8.649,-3.541);
\draw[black!50,line width=0.6pt] (9.012,-4.659) -- (8.649,-3.541);
\node[vx,fill=black] at (9.600,-2.850) {};
\node[lab] at (9.600,-2.550) {$\infty$};
\node[vx,fill=red!75!black] at (10.551,-3.541) {};
\node[lab] at (10.836,-3.448) {$1$};
\node[vx,fill=orange!85!black] at (10.188,-4.659) {};
\node[lab] at (10.364,-4.902) {$\omega^2$};
\node[vx,fill=green!50!black] at (9.012,-4.659) {};
\node[lab] at (8.836,-4.902) {$\omega$};
\node[vx,fill=blue!70!black] at (8.649,-3.541) {};
\node[lab] at (8.364,-3.448) {$0$};
\node[lab,align=center] at (9.6,-5.5) {$\PGL(2,4)\cong A_5$, $\PGaL(2,4)\cong S_5$\\ order $120=2\cdot5\cdot4\cdot3$};
\node[font=\small] at (1.35,1.4) {(a) $q=2$};
\node[font=\small] at (5.35,1.4) {(b) $q=3$};
\node[font=\small] at (9.5,1.4) {(c) $q=4$};
\end{tikzpicture}
\caption{The projective line $\PG(1,q)$ and the group $\PGaL(2,q)$ for $q=2,3,4$. Top: the $q+1$ lines through the
origin of $\F_q^2$, which are the points $t\in\F_q\cup\{\infty\}$ of $\PG(1,q)$; each is drawn as a ray from the origin
carrying its $q-1$ nonzero points $(x,tx)$, $x\in\F_q^\times$, or $(0,x)$ for $t=\infty$. Bottom:
the $q+1$ points as the vertices of a regular simplex, whose symmetry group $S_{q+1}$ is $\PGaL(2,q)$ for these $q$.}
\label{fig:group}
\end{figure}

The invertible semilinear maps of $\F_q^2$ act on $\PG(1,q)$, the scalar maps act trivially, and the quotient is
$\PGaL(2,q)$. In the coordinate $t$ its elements are the maps
\[
t\mapsto\frac{a\,t^{\phi}+b}{c\,t^{\phi}+d},\qquad a,b,c,d\in\F_q,\quad ad+bc\neq0,\quad
\phi\in\{t\mapsto t^{2^k}:0\le k<h\},
\]
where $\infty$ goes to $a/c$ and the zero of $c\,t^\phi+d$ goes to $\infty$. The maps with $\phi=\mathrm{id}$ form
$\PGL(2,q)$, of order $q(q^2-1)$, and $\PGaL(2,q)$, the extension of $\PGL(2,q)$ by the field automorphisms, has order
$hq(q^2-1)$. For a primitive element $\alpha$ of $\F_q$, the group $\PGL(2,q)$ is generated by $t\mapsto t+1$,
$t\mapsto\alpha t$ and $t\mapsto1/t$, and $t\mapsto t^2$ together with these generates $\PGaL(2,q)$. The group
$\PGL(2,q)$ is sharply $3$-transitive. For distinct $x_1,x_2,x_3\in\F_q$ the only element mapping them to
$0,1,\infty$ is
\[
t\mapsto\frac{(t+x_1)(x_2+x_3)}{(t+x_3)(x_2+x_1)},
\]
and it maps a fourth point $x_4$ to the cross-ratio of $x_1,\dots,x_4$. This is why a $4$-set may always be assumed to
be $\{0,1,\infty,c\}$.

Figure~\ref{fig:group} shows the smallest cases. The points of $\PG(1,q)$ are the $q+1$ lines through the origin of
$\F_q^2$, the point $t$ being the line $\{(x,tx)\}$ and $\infty$ the line $x=0$. For $q\le4$ the group $\PGaL(2,q)$
acts on these points as the full symmetric group: $\PGL(2,2)\cong S_3$, $\PGL(2,3)\cong S_4$, $\PGL(2,4)\cong A_5$ and
$\PGaL(2,4)\cong S_5$. Drawn as the vertices of a regular simplex, the projective line then has $\PGaL(2,q)$ as the
full symmetry group of the simplex. For larger $q$ the group, of order $hq(q^2-1)$, is a small subgroup of the
symmetric group of order $(q+1)!$, and in general it has many orbits on $4$-sets (Section~1). For example, let
$q=32$ and let $\alpha$ be a root of $x^5+x^2+1$. Then $\PGaL(2,32)$ has order $163{,}680$, and the stabiliser of the
block $\{\alpha,\alpha^3,\alpha^{11},\alpha^{19},\alpha^{23},\alpha^{26}\}$ of $\cD_8$ has order $5$. So the orbit
of this block has $32{,}736$ elements, which is the number $b$ of blocks (Corollary~\ref{cor:weights}): for $q=32$
the design $\cD_8$ is a single orbit of $\PGaL(2,32)$ (see also Section~\ref{sec:invariant}).

\section{\texorpdfstring{Proof of Theorem~\ref{thm:main}}{Proof of Theorem 1.1}}\label{sec:proof}

We begin with an outline of the proof of parts 1 and 2. Lemma~\ref{lem:zeros}(2) and Proposition~\ref{prop:zerosets}(1) show that a nonzero
member of $W_\sigma$ has at most six zeros and that any five points impose independent conditions on $W_\sigma$.
Together they identify the blocks of $\cD_\sigma$ with the zero sets of the members of $W_\sigma$ that have exactly six
zeros (Proposition~\ref{prop:zerosets}). Fix a $4$-set $T$. The members vanishing on $T$ form a pencil $\Lambda_T$, and
every point outside $T$ lies on exactly one member, which has five or six zeros. Counting the $q-3$ points outside $T$
gives $q-3=2N_6+N_5$, where $N_k$ is the number of members of $\Lambda_T$ with exactly $k$ zeros. Since $N_6$ is the
number of blocks through $T$, the design property amounts to $N_5=5$ for every $T$.

The key input is Corollary~\ref{cor:never5}: if $A$ and $B$ are coprime, the multiplicities of the zeros of
$P=A+t^\sigma B$, each capped at two, never sum to five. We prove it in Section~\ref{sec:curves}. To each member $P$ we attach a form $G$ of
degree $6$ whose zeros include those of $P$, with multiplicities capped at two. An involution $\Phi$ of the zeros
of $G$, built from the automorphism $t\mapsto t^\sigma$, fixes exactly the zeros of $P$. So the zeros of $G$ that are not
zeros of $P$ come in pairs, whereas a capped sum of five would leave exactly one.
Consequently a member with exactly five zeros is tangent at exactly one of them.
The tangency is either at a point of $T$, which happens for exactly four members, or at the fifth zero, which
happens for exactly one member (Proposition~\ref{prop:N5}). Hence $N_5=5$
and $N_6=(q-8)/2$.

\subsection{A sextic and an involution}\label{sec:curves}

Let $P=A+t^\sigma B\in W_\sigma$, where $A$ and $B$ are coprime as binary quadratic forms: they have no common
factor of positive degree, and they do not both vanish at $\infty$. Equivalently, they have no common zero on the
projective line over $\overline\F_q$. Then $R=A/B$ is a well-defined map $\PG(1,q)\to\PG(1,q)$, with $R(w)=\infty$
when $B(w)=0$. For a quadratic form $Q=c_0+c_1z+c_2z^2$ we write
$Q^{(\sigma)}=c_0^\sigma+c_1^\sigma z+c_2^\sigma z^2$ for the form whose coefficients are raised to the
power $\sigma$, and $Q^{(\sigma)}(Y_1,Y_2)=c_0^\sigma Y_2^2+c_1^\sigma Y_1Y_2+c_2^\sigma Y_1^2$ for its
homogeneous version. A superscript $\sigma$ without parentheses always denotes a power; thus
$Q^{(\sigma)}(z^\sigma)=Q(z)^\sigma$. The \emph{sextic} of $P$ is the form of degree $6$
\[
G(w)=w^2B^{(\sigma)}\bigl(A(w),B(w)\bigr)+A^{(\sigma)}\bigl(A(w),B(w)\bigr).
\]
In homogeneous coordinates $w=[u:v]$ it reads $G(u,v)=u^2B^{(\sigma)}(A,B)+v^2A^{(\sigma)}(A,B)$, where $A=A(u,v)$ and
$B=B(u,v)$, and $P(u,v)=v^\sigma A(u,v)+u^\sigma B(u,v)$.

The sextic has the following geometric interpretation, which motivates the argument but is not used in the proofs. On
$\mathbb P^1\times\mathbb P^1$, with points $(w,z)$, consider the curves $C_1:zB(w)+A(w)=0$ of class $(2,1)$, the graph
of $R$, and $C_2:w^2B^{(\sigma)}(z)+A^{(\sigma)}(z)=0$ of class $(2,2)$. Where $B(w)\neq0$ we have
$G(w)=B(w)^2\bigl(w^2B^{(\sigma)}(R(w))+A^{(\sigma)}(R(w))\bigr)$, so the zeros of $G$ are the first coordinates of
the points of $C_1\cap C_2$, and the degree $6$ of $G$ is the intersection number $(2,1)\cdot(2,2)$ given by
B\'ezout's theorem. A zero $t$ of $P$ gives the point $(t,t^\sigma)$ of $C_1\cap C_2$: it lies on $C_2$ because
raising $A(t)+t^\sigma B(t)=0$ to the power $\sigma$ gives $A^{(\sigma)}(t^\sigma)+t^2B^{(\sigma)}(t^\sigma)=0$, as
$t^{\sigma^2}=t^2$. The map $\Phi$ of the next lemma is the involution $(w,z)\mapsto(z^s,w^\sigma)$ of $C_1\cap C_2$,
read on the first coordinate.

\begin{lemma}\label{lem:curves}
Let $P=A+t^\sigma B\in W_\sigma$ with $A,B$ coprime, and let $G$ be its sextic.
\begin{enumerate}[nosep]
\item $G\neq0$.
\item If $t_0$ is a zero of $P$, then $m_{t_0}(G)=\min(m_{t_0}(P),2)$.
\item Put $\Phi(w)=R(w)^s$ for $w\in\PG(1,q)$. If $G(w)=0$, then $G(\Phi(w))=0$ and $\Phi(\Phi(w))=w$, and
  $\Phi(w)=w$ if and only if $P(w)=0$.
\end{enumerate}
\end{lemma}

\begin{proof}
Put $L_j=b_j^sw+a_j^s$ for $j=0,1,2$, so that $L_j^2=b_j^\sigma w^2+a_j^\sigma$ because $\sigma=2s$. Collecting the
terms of $G$ by $B^2$, $AB$ and $A^2$, where $A=A(w)$ and $B=B(w)$, gives
\begin{equation}\label{eq:sextic}
G=L_0^2B^2+L_1^2AB+L_2^2A^2=(L_0B+L_2A)^2+L_1^2AB .
\end{equation}

For part 1, suppose that $G=0$. In characteristic $2$ the derivative of a square vanishes, and $A'=a_1$, $B'=b_1$, so
$0=G'=L_1^2(a_1B+b_1A)$. If $a_1B+b_1A=0$ with $(a_1,b_1)\neq(0,0)$, then $A$ and $B$ are proportional or one of
them is $0$, which is impossible for coprime forms. So in both cases $a_1=b_1=0$, and $L_1=0$. Then
$G=(L_0B+L_2A)^2$, so $L_0B=L_2A$, and coprimality gives $A\mid L_0$ and $B\mid L_2$, where $\deg L_j\le1$. Since $A$
and $B$ do not both vanish at $\infty$, one of them has degree $2$. If $\deg A=2$, then $L_0=0$, hence $L_2=0$ and
$a_2=0$, a contradiction. If $\deg B=2$, then $L_2=0$ and $b_2=0$, again a contradiction.

For part 2, note first that $t\mapsto1/t$ replaces $(A,B)$ by $(t^2B(1/t),t^2A(1/t))$ and $G$ by $w^6G(1/w)$, so
we may take $t_0$ finite. We rewrite~\eqref{eq:sextic} in terms of $P=P(w)$. Since
$L_0^2+w^\sigma L_1^2+w^{2\sigma}L_2^2=A^\sigma+w^2B^\sigma$, substituting $A=P+w^\sigma B$ gives
$G=B^2(A^\sigma+w^2B^\sigma)+PBL_1^2+P^2L_2^2$. Finally $A^\sigma=P^\sigma+w^{2q}B^\sigma$, because $\sigma^2=2q$, so
\[
G=B^2P^\sigma+B^{\sigma+2}(w^q+w)^2+PB\,L_1^2+P^2L_2^2 .
\]
Let $m=m_{t_0}(P)$. Then $B(t_0)\neq0$, since $A$ and $B$ have no common zero. The second term has order exactly $2$ at
$t_0$, because $w^q+w=\prod_{a\in\F_q}(w+a)$ has a simple root at $t_0$. The first and last terms have order at
least $\sigma m\ge4$ and $2m$. For the third term, $L_1(t_0)^2=a_1^\sigma+b_1^\sigma t_0^{\sigma^2}=P'(t_0)^\sigma$ by
Lemma~\ref{lem:Wbasic}(1). If $m=1$, then $P'(t_0)\neq0$, so the third term has order exactly $1$ and the others
order at least $2$, and $m_{t_0}(G)=1$. If $m\ge2$, then $P'(t_0)=0$, so the third term has order at least $m+1\ge3$,
the last has order at least $4$, and $m_{t_0}(G)=2$.

For part 3, let $w=[u:v]$ be a zero of $G$ and put $z=R(w)=[z_1:z_2]$ with $z_1=A(u,v)$ and $z_2=B(u,v)$, so that
$u^2B^{(\sigma)}(z_1,z_2)+v^2A^{(\sigma)}(z_1,z_2)=0$. Applying the automorphism $x\mapsto x^s$, the inverse of
$x\mapsto x^\sigma$, gives $u^\sigma B(z_1^s,z_2^s)+v^\sigma A(z_1^s,z_2^s)=0$. So
$\bigl(A(z_1^s,z_2^s),B(z_1^s,z_2^s)\bigr)=\mu(u^\sigma,v^\sigma)$ for some $\mu\neq0$, that is,
$R(\Phi(w))=w^\sigma$, and $\Phi(\Phi(w))=(w^\sigma)^s=w$. Moreover
\[
G(\Phi(w))=\mu^2\bigl(z_1^\sigma B^{(\sigma)}(u^\sigma,v^\sigma)+z_2^\sigma A^{(\sigma)}(u^\sigma,v^\sigma)\bigr)
=\mu^2\bigl(z_1B(u,v)+z_2A(u,v)\bigr)^\sigma=0 .
\]
Finally $\Phi(w)=w$ if and only if $R(w)=w^\sigma$, that is, $v^\sigma A(u,v)+u^\sigma B(u,v)=0$, which says
$P(w)=0$.
\end{proof}

\begin{corollary}\label{cor:never5}
If $A,B$ are coprime, then $\sum_{t:\,P(t)=0}\min(m_t(P),2)\ne5$.
\end{corollary}

\begin{proof}
By Lemma~\ref{lem:curves}(1), $G$ is a nonzero form of degree $6$, so it is a product of six linear forms over
$\overline\F_q$. By Lemma~\ref{lem:curves}(2), the zeros of $P$ account for exactly $\sum\min(m_t(P),2)$ of these
factors, all defined over $\F_q$. Suppose this sum is $5$. Then the remaining factor is defined over $\F_q$ as well,
so its zero $w_1$ lies in $\PG(1,q)$, and $w_1$ is not a zero of $P$, because part 2 gives the exact multiplicity of
$G$ at each zero of $P$. So $w_1$ is the only zero of $G$ outside the zeros of $P$. By
Lemma~\ref{lem:curves}(3), $\Phi(w_1)$ is a zero of $G$ and $\Phi(w_1)\neq w_1$. It is not a zero of $P$ either,
since otherwise it would be fixed by $\Phi$, and then $w_1=\Phi(\Phi(w_1))=\Phi(w_1)$. This is a second zero of $G$
outside the zeros of $P$, a contradiction.
\end{proof}

\subsection{Zero bounds}

\begin{lemma}\label{lem:zeros}
Counted with multiplicities capped at two, the zeros of nonzero forms are bounded as follows.
\begin{enumerate}[nosep]
\item For nonzero $Q\in\Wh_\sigma$, $\sum_t\min(m_t(Q),2)\le3$.
\item For nonzero $P\in W_\sigma$, $\sum_t\min(m_t(P),2)\le6$, so $P$ has at most six zeros. If $A$ and $B$ are not
coprime, the sum is at most $4$.
\end{enumerate}
\end{lemma}

\begin{proof}
For part 1, the space $\Wh_\sigma$, spanned by $t^{a+b\sigma}$ with $a,b\in\{0,1\}$, is stable under linear
substitutions by the proof of Lemma~\ref{lem:Wbasic}(2), and these preserve multiplicities. If $Q$ has at most one
zero, the sum is at most $2$. Otherwise we move two zeros of $Q$ to $0$ and $\infty$. Then the constant term and the
coefficient of $t^{\sigma+1}$ vanish, so $Q=t\,(a_1+b_0t^{\sigma-1})$. If $b_0=0$, then $Q=a_1t$ has a simple zero at
$0$ and a zero of multiplicity $\sigma$ at $\infty$. If $a_1=0$, then $Q=b_0t^\sigma$ has a zero of multiplicity
$\sigma$ at $0$ and a simple zero at $\infty$. In both cases the sum is $3$. If $a_1b_0\neq0$, then $0$
and $\infty$ are simple zeros, and the remaining zeros solve $t^{\sigma-1}=a_1/b_0$. Since
$\gcd(2^{e+1}-1,2^{2e+1}-1)=2^{\gcd(e+1,2e+1)}-1=1$, the map $t\mapsto t^{\sigma-1}$ is a bijection of
$\F_q^\times$, so there is exactly one such zero. It is simple, because the derivative $b_0t^{\sigma-2}$ of
$a_1+b_0t^{\sigma-1}$ does not vanish on $\F_q^\times$. Again the sum is $3$.

For part 2, if $A$ and $B$ are coprime, Lemma~\ref{lem:curves}(1) and~(2) show that the zeros of $P$, with
multiplicities capped at two, are zeros of the nonzero sextic $G$ of degree $6$, so the sum is at most $6$.
Otherwise $A$ and $B$, as binary quadratic forms, have a common factor $H$ of positive degree, which is defined
over $\F_q$ because it is stable under the Galois group. If $\deg H=1$, then $P=H\cdot\widetilde P$ with
$\widetilde P\in\Wh_\sigma$. If $\deg H=2$, then $P=H\cdot(a+bt^\sigma)$, where
$a+bt^\sigma$ is either a constant or $b(t+(a/b)^s)^\sigma$; this case includes $A\equiv0$ and $B\equiv0$. In both
cases the inequality $\min(x+y,2)\le\min(x,2)+\min(y,2)$ and part 1 bound the sum by $1+3$ or $2+2$.
\end{proof}

\begin{proposition}\label{prop:zerosets}
The following hold for the space $W_\sigma$ and the code $\cC_\sigma$.
\begin{enumerate}[nosep]
\item Any five points of $\PG(1,q)$ impose independent conditions on $W_\sigma$.
\item For a $4$-set $T$, the members of $W_\sigma$ vanishing on $T$ form a $2$-dimensional space $\Lambda_T$, and each
  point $u\notin T$ lies on exactly one member $P_u$ of $\Lambda_T$, up to scalars.
\item The supports of the weight-$6$ codewords of $\cC_\sigma^\perp$ are exactly the dependent $6$-sets of columns of
  $\cC_\sigma$.
\item A $6$-set is dependent if and only if it is the zero set of a member of $W_\sigma$. Hence the blocks of $\cD_\sigma$ are the
  zero sets of the members of $W_\sigma$ with exactly six zeros, and the blocks through a $4$-set $T$ correspond
  one-to-one to the members of $\Lambda_T$ with six zeros.
\end{enumerate}
\end{proposition}

\begin{proof}
For part 1, take the points finite, $t_1,\dots,t_5$, and suppose that $\sum_ic_ie(t_i)=0$ with $(c_i)\neq0$. With
$v_i=(1,t_i,t_i^2)$, this says that both $(c_i)$ and $(c_it_i^\sigma)$ are linear dependencies of $v_1,\dots,v_5$.
These dependencies form the $2$-dimensional space of the vectors $(\ell(t_i)/\delta_i)$ with $\deg\ell\le1$ and
$\delta_i=\prod_{j\ne i}(t_i-t_j)$, because $\sum_if(t_i)/\delta_i=0$ whenever $\deg f\le3$ (a fourth divided
difference). Hence $c_i=\ell(t_i)/\delta_i$ and $c_it_i^\sigma=\ell^*(t_i)/\delta_i$ with $\ell\neq0$, and
$\ell^*+t^\sigma\ell$ is a nonzero element of $\Wh_\sigma$ vanishing at $t_1,\dots,t_5$. This contradicts
Lemma~\ref{lem:zeros}(1).

By part 1, the four points of $T$ impose independent conditions on the $6$-dimensional space $W_\sigma$,
so $\dim\Lambda_T=2$, and a further point $u\notin T$ imposes one more independent condition, which leaves a single
member up to scalars. This proves part 2. For part 3, a dependent $6$-set of columns has rank exactly $5$, again by part 1. Its dependency is therefore unique up to scalars, and it has full support because any five of
the columns are independent. For part 4, exactly one member $P$ of $W_\sigma$ vanishes on a dependent $6$-set $S$, up
to scalars, and by Lemma~\ref{lem:zeros}(2) the zero set of $P$ is $S$. Conversely, six common zeros of a
nonzero member make the six columns dependent.
\end{proof}

\subsection{Blocks through a \texorpdfstring{$4$}{4}-set}

Fix a $4$-set $T=\{t_1,\dots,t_4\}$. Let $N_k$ be the number of members of $\Lambda_T$ (counted projectively) with
exactly $k$ zeros. By Proposition~\ref{prop:zerosets}, each of the $q-3$ points outside $T$ lies on exactly one member,
and that member has five or six zeros. So
\begin{equation}\label{eq:count}
q-3=2N_6+N_5 ,
\end{equation}
and $N_6$ is the number of blocks through $T$. Call $u\notin T$ \emph{unpaired} if $P_u$ has only five zeros. Put
\[
U_a=\{u: P_u\text{ is tangent at some }t_i\},\qquad U_b=\{u: P_u\text{ is tangent at }u\}.
\]
Both sets are defined intrinsically, so we may change coordinates freely when computing them.

\begin{proposition}\label{prop:N5}
Let $T$ be a $4$-set. A member of $\Lambda_T$ with exactly five zeros is tangent at exactly one of them, and no nonzero
member of $\Lambda_T$ is tangent at two points of $T$. Moreover $|U_a|=4$ and $|U_b|=1$, so $N_5=5$.
\end{proposition}

\begin{proof}
The sets $U_a$ and $U_b$, and the members of $\Lambda_T$ tangent at a given point, do not depend on coordinates, so
each of the four steps below may use its own normalisation.

We first show that a member $P_u$ with exactly five zeros is tangent at exactly one of them. Every member $P_u$ has at least five zeros, so $A$ and $B$ are coprime by Lemma~\ref{lem:zeros}(2). With exactly five zeros,
$\sum\min(m,2)\in\{5,6\}$, and it is not $5$ by Corollary~\ref{cor:never5}. Two tangencies would give a sum of at
least $7$, contradicting Lemma~\ref{lem:zeros}(2). Conversely, if $P_u$ is tangent at some point, then the zeros in
$T\cup\{u\}$ already contribute $6$ to the capped sum, so $P_u$ has no further zero by Lemma~\ref{lem:zeros}(2), and
$u$ is unpaired. Consequently the unpaired points form the disjoint union of $U_a$ and $U_b$, and $N_5=|U_a|+|U_b|$.

Second, no nonzero member $P$ of $\Lambda_T$ is tangent at two points $t_i,t_j$ of $T$. Take both points finite. From $P'(t_i)=P'(t_j)=0$ and Lemma~\ref{lem:Wbasic}(1) we get $b_1(t_i^\sigma+t_j^\sigma)=0$, so $a_1=b_1=0$. Then $P=H^2$
with $H\in\mathrm{span}\{1,t,t^s,t^{s+1}\}$, and $H$ vanishes on $T$.

Put $\widetilde H(y)=H(y^\sigma)$. Using $y^q=y$, this is an element of $\Wh_\sigma$ in the variable $y$, as a
function on $\PG(1,q)$. The top coefficients of $H$ and $\widetilde H$ agree, and $t\mapsto t^s$ is a bijection of
$\PG(1,q)$, so $\widetilde H$ has four zeros, contrary to Lemma~\ref{lem:zeros}(1).

Third, we show that $|U_a|=4$. For this step we take $T$ finite. On $\Lambda_T$ the linear functional $\varphi_i(P)=P'(t_i)$ detects tangency at $t_i$. It is not
identically zero: otherwise, for $j\ne i$, a nonzero member of $\ker\varphi_j$ would be tangent at both $t_i$ and
$t_j$, contrary to the second step. Hence $\ker\varphi_i$ is spanned by a single member $Q_i$.

We claim that the quadratic forms $A$ and $B$ of $Q_i$ are coprime. If they have a common factor of degree $2$, then
$Q_i$ has at most three zeros, as in the proof of Lemma~\ref{lem:zeros}(2), which is impossible since it vanishes on
$T$. If they have a common linear factor $L$, with zero $w_0$, then $Q_i=L\widetilde Q$ with $\widetilde
Q\in\Wh_\sigma$, and each possible location of $w_0$ contradicts Lemma~\ref{lem:zeros}(1). If $w_0\notin T$, then
$\widetilde Q$ vanishes on the four points of $T$. If $w_0=t_i$, then $Q_i'(t_i)=\widetilde Q(t_i)$, so $\widetilde Q$
again vanishes on $T$. If $w_0=t_j$ with $j\ne i$, then $\widetilde Q$ vanishes on $T\setminus\{t_j\}$ and
$\widetilde Q'(t_i)=0$, so $\sum\min(m,2)\ge4$ for $\widetilde Q$.

Next, $Q_i$ is tangent at $t_i$ and, by the second step, has simple zeros at the other three points of $T$, so
$T$ contributes exactly $5$ to $\sum_t\min(m_t(Q_i),2)$. By Corollary~\ref{cor:never5} the sum is not $5$, so $Q_i$
has a further zero $u_i\notin T$, and by Lemma~\ref{lem:zeros}(2) this zero is simple and unique. Hence $Q_i=P_{u_i}$ and
$u_i\in U_a$. If $u_i=u_j$ for some $j\ne i$, then $Q_i=Q_j$ would be tangent at $t_i$ and at $t_j$, contrary to
the second step. Conversely, every $u\in U_a$ satisfies $P_u=Q_i$ for some $i$, so $u=u_i$. Thus
$U_a=\{u_1,\dots,u_4\}$ has four elements.

Finally, we show that $|U_b|=1$. For this step we assume instead that $\infty\in T$. Then every $u\notin T$ is finite, and every member
of $\Lambda_T$ has $b_2=0$. Let $M_0,M_1$ be a basis of $\Lambda_T$. For $u\notin T$, the member
$P_u=M_1(u)M_0+M_0(u)M_1$ vanishes at $u$, and it is nonzero by Proposition~\ref{prop:zerosets}(1). Its derivative at $u$ is
$P_u'(u)=\mathcal W(u)$, where $\mathcal W=M_0'M_1+M_0M_1'$ is the Wronskian of the pencil. Hence
$U_b=\{u\notin T:\mathcal W(u)=0\}$.

The Wronskian is a square. Indeed $\mathcal W'=M_0''M_1+M_0M_1''=0$ by Lemma~\ref{lem:Wbasic}(1), so $\mathcal W\in\F_q[t^2]$, and its exponents
lie in $\{0,1,2\}+\sigma\{0,1,2\}$. Hence $\mathcal W=H^2$ with $H$ in the span of
$1,t,t^s,t^{s+1},t^\sigma,t^{\sigma+1}$, which we read as a form of degree $\sigma+1$. Put
$\widetilde H(y)=H(y^\sigma)$. For $t=y^\sigma$ with $y\in\F_q$ we have $t^s=y$ and $t^\sigma=y^2$, so $\widetilde H$
lies in $W_\sigma$ in the variable $y$, as a form of degree $\sigma+2$. The coefficient of $t^{\sigma+1}$ in $H$
equals the coefficient of $y^{\sigma+2}$ in $\widetilde H$, so the zeros of $H$ and $\widetilde H$ correspond under
$t\mapsto t^s$, including the point $\infty$.

The form $H$ vanishes on $T$. It vanishes at the finite points of $T$, which are zeros of $\mathcal W$. It also
vanishes at $\infty$: since $b_2=0$ we have $\deg\mathcal W\le2\sigma+1$, and since $\mathcal W$ is a polynomial in
$t^2$, $\deg\mathcal W\le2\sigma$, so $\deg H\le\sigma$. Moreover $H\ne0$, for if $\mathcal W\equiv0$, then the member
$Q_1$ of the third step would be tangent at $t_1$ and at $u_1$, contrary to the first step.

By Lemma~\ref{lem:zeros}(2), $\widetilde H$ has at most six zeros, and four of them come from $T$, so $|U_b|\le2$. On the
other hand, \eqref{eq:count} together with $N_5=|U_a|+|U_b|$ and $|U_a|=4$ gives $q-3=2N_6+4+|U_b|$. Since $q$
is even, $|U_b|$ is odd, and therefore $|U_b|=1$.
\end{proof}

\begin{proof}[Proof of Theorem~\ref{thm:main}, parts 1 and 2]
By Proposition~\ref{prop:N5}, $N_5=5$, and by~\eqref{eq:count} every $4$-set lies in
$N_6=(q-8)/2$ blocks. This proves part 2; the design is $\PGaL(2,q)$-invariant by Lemma~\ref{lem:Wbasic}(2). For part
1, evaluation is injective on $W_\sigma$, since a nonzero form has at most six zeros (Lemma~\ref{lem:zeros}(2)), so
$\dim\cC_\sigma=6$. Lemma~\ref{lem:zeros}(2) also gives $d(\cC_\sigma)\ge q-5$, and Proposition~\ref{prop:zerosets}(1) gives
$d(\cC_\sigma^\perp)\ge6$. Both bounds are attained because blocks exist for $q\ge32$.
\end{proof}

\begin{remark}
The relations $\sigma=2s$ and $s\sigma\equiv1\pmod{q-1}$ of Lemma~\ref{lem:sigma}, which amount to
$\sigma^2\equiv2\pmod{q-1}$, are used in all three parts of Lemma~\ref{lem:curves}: they give the
decomposition~\eqref{eq:sextic} in part~1, the factor $(w^q+w)^2$ in part~2, and the involution in part~3. The bound
$q\ge32$ ensures that $\lambda>0$.
\end{remark}

\subsection{The other weight shells}

\begin{theorem}\label{thm:shells}
The weight shells of $\cC_\sigma$ and $\cC_\sigma^\perp$ are as follows.
\begin{enumerate}[nosep]
\item The codewords of weight $7$ in $\cC_\sigma^\perp$ hold a $4$-$(q+1,7,\lambda_7)$ design, where
  \[
  \lambda_7=\binom{q-3}3-\frac76(q-8)(q-5)=\frac{(q-5)(q^2-14q+68)}6 .
  \]
\item The $5$-sets that lie in no block of $\cD_\sigma$ form a $4$-$(q+1,5,5)$ design, and the codewords of weight
  $q-4$ in $\cC_\sigma$ hold its complement.
\item The codewords of weight $q-5$ in $\cC_\sigma$ hold the complement of $\cD_\sigma$.
\item Every other nonempty weight shell of $\cC_\sigma$ and of $\cC_\sigma^\perp$ is complete.
\end{enumerate}
\end{theorem}

\begin{proof}
We first treat $\cC_\sigma$. The codeword $(P(t))_t$ of a member $P$ has weight $q+1-|Z(P)|$, where $Z(P)$ is the zero set of
$P$, and $|Z(P)|\le6$ by Lemma~\ref{lem:zeros}(2). Zero sets of size six are the blocks of $\cD_\sigma$
(Proposition~\ref{prop:zerosets}), so the supports of weight $q-5$ form the complement of $\cD_\sigma$. A $5$-set $Z$
is the exact zero set of some member iff it lies in no block: by Proposition~\ref{prop:zerosets}(1), $Z$ determines a unique member
$P$ up to scalars, and $Z$ lies in a block iff $P$ has a sixth zero. For a $4$-set $T$, the exact zero sets of size
five containing $T$ are the sets $T\cup\{u\}$ with $u$ one of the $N_5=5$ unpaired points. So these sets form a
$4$-$(q+1,5,5)$ design, and the supports of weight $q-4$ form its complement. Every $4$-set is an exact zero set,
since $N_4=q+1-N_5-N_6=q/2>0$. Let $I$ be a $z$-set with $1\le z\le3$. By Proposition~\ref{prop:zerosets}(1), the members of
$W_\sigma$ vanishing on $I$ form a projective space with $(q^{6-z}-1)/(q-1)$ points. Those with a further zero vanish
on some $I\cup\{u\}$ with $u\notin I$, and for each $u$ these form a projective space with $(q^{5-z}-1)/(q-1)$ points.
So there are at most $(q+1-z)(q^{5-z}-1)/(q-1)$ of them, and this number is
smaller, so $I$ is an exact zero set. Finally, the shell of weight $q+1$ is either empty or consists of the full point
set. The shells of weights $1,\dots,q-6$ are empty. So every nonempty shell of $\cC_\sigma$ other than those of
weights $q-5$ and $q-4$ is complete.

We turn to $\cC_\sigma^\perp$, whose weight-$6$ shell is $\cD_\sigma$. Any seven points have rank $6$, since otherwise
a nonzero member would have seven zeros, contrary to Lemma~\ref{lem:zeros}(2). So seven points have a unique dependency up
to scalars, and its coefficient at a point $x$ vanishes iff the other six points are dependent, that is, form a
block. Hence the supports of weight $7$ are the $7$-sets that contain no block. A $7$-set contains at most one block:
two blocks in it would share five points, which are independent, and both blocks would lie in the span of those five,
so the $7$-set would have rank $5$. Now fix a $4$-set $T$ and count the $7$-sets $X\supseteq T$ that contain a block
$B$. Either $T\subseteq B$ and $X=B\cup\{x\}$, with $q-5$ choices of $x$, or $|B\cap T|=3$ and $X=B\cup T$. With
$\lambda_4=(q-8)/2$ and $\lambda_3=\lambda_4(q-2)/3$, the number of supports of weight $7$ through $T$ is
\[
\binom{q-3}3-\lambda_4(q-5)-4(\lambda_3-\lambda_4)=\binom{q-3}3-\frac76(q-8)(q-5),
\]
which does not depend on $T$.

Let $8\le w\le q$. A $w$-set has rank $6$, so its dependencies form a space of dimension $w-6\ge2$. Every point of the
set has a nonzero coefficient in some dependency, because the other $w-1$ points still have rank $6$. Hence each coordinate functional on the space of
dependencies has a hyperplane as its kernel. At most $q$ hyperplanes cannot cover a space of dimension at least $2$
over $\F_q$, so a full-support dependency exists and the shell is complete. The shells of weights $1,\dots,5$ are
empty, and the shell of weight $q+1$ is trivial.
\end{proof}

The same statements hold for Tang and Ding's code. There they are due to Yan and Zhou~\cite{YZ22}, and their value
of $\lambda_7$ agrees with Theorem~\ref{thm:shells}(1).

The weight distributions follow from the same facts. Write $A_w$ and $A_w^\perp$ for the numbers of codewords of
weight $w$ in $\cC_\sigma$ and $\cC_\sigma^\perp$.

\begin{corollary}\label{cor:weights}
Let $n=q+1$, and let $b=\frac{q-8}{30}\binom{q+1}4$ be the number of blocks of $\cD_\sigma$.
\begin{enumerate}[nosep]
\item For $0\le j\le6$,
  \[
  A_{n-j}=\sum_{z=j}^{6}(-1)^{z-j}\binom zjM_z,
  \]
  where $M_z=\binom nz\bigl(q^{6-z}-1\bigr)$ for $z\le5$ and $M_6=(q-1)b$. Moreover $A_w=0$ for $1\le w\le n-7$.
\item For every $w\ge0$,
  \[
  A_w^\perp=\binom nw\sum_{r=0}^{w}(-1)^{w-r}\binom wrq^{\max(r-6,0)}+(-1)^w(q-1)\,b\binom{n-6}{w-6},
  \]
  where the last term is $0$ for $w<6$.
\end{enumerate}
\end{corollary}

\begin{proof}
For part 1, let $Z$ be a set of $z\le6$ points. The codewords of $\cC_\sigma$ that vanish on $Z$ are the values of
the members of $W_\sigma$ that vanish on $Z$. By Proposition~\ref{prop:zerosets}(1) there are $q^{6-z}$ of them for
$z\le5$, and for $z=6$ there are $q$ of them if $Z$ is a block and one otherwise, by
Proposition~\ref{prop:zerosets}(4). A nonzero codeword with exactly $j$ zeros vanishes on $\binom jz$ sets of size
$z$, and $j\le6$ by Lemma~\ref{lem:zeros}(2). Counting the pairs $(c,Z)$ with $c\ne0$ vanishing on $Z$ therefore gives
$\sum_{j=z}^{6}\binom jzA_{n-j}=M_z$ for $0\le z\le6$, and binomial inversion gives the formula. The weights
$1,\dots,n-7$ do not occur because $d(\cC_\sigma)=q-5$.

For part 2, the codewords of $\cC_\sigma^\perp$ with support in a set $I$ of points are the linear dependencies among
the columns of $\cC_\sigma$ at the points of $I$, so they form a space of dimension $|I|-\operatorname{rk}I$. Here
$\operatorname{rk}I=\min(|I|,6)$, except that $\operatorname{rk}I=5$ when $I$ is a block: sets of at most five points
are independent by Proposition~\ref{prop:zerosets}(1), and sets of at least seven points have rank $6$, as shown in
the proof of Theorem~\ref{thm:shells}. By inclusion--exclusion, the number of codewords with support exactly $S$ is
$\sum_{I\subseteq S}(-1)^{|S|-|I|}q^{|I|-\operatorname{rk}I}$. Summing over the sets $S$ of size $w$, using
$\binom nr\binom{n-r}{w-r}=\binom nw\binom wr$, and noting that each block lies in $\binom{n-6}{w-6}$ sets of size $w$,
gives part 2. Finally, counting the pairs $(T,B)$ of a $4$-set $T$ and a block $B\supseteq T$ gives
$15b=\frac{q-8}2\binom{q+1}4$.
\end{proof}

Expanding part 1 gives
\[
\begin{aligned}
A_{q-5}&=\tfrac1{720}\,q(q-8)(q-2)(q-1)^2(q+1),\\
A_{q-4}&=\tfrac1{24}\,q(q-2)(q-1)^2(q+1),\\
A_{q-3}&=\tfrac1{48}\,q^2(q-2)(q-1)^2(q+1),\\
A_{q-2}&=\tfrac1{36}\,q(q-1)^2(q+1)(2q^2+7q+20),\\
A_{q-1}&=\tfrac1{48}\,q(q-1)(q+1)(9q^3+13q^2-6q+80),\\
A_q&=\tfrac1{120}\,(q-1)(q+1)(44q^4+21q^3+49q^2-114q+120),\\
A_{q+1}&=\tfrac1{144}\,q(q-1)^2(q+1)(53q^2-26q+32).
\end{aligned}
\]
The proof of Corollary~\ref{cor:weights} uses only that $\cC_\sigma$ is a near-MDS $[q+1,6]$ code and the number $b$
of blocks. The code $\cC_4$, whose dual is the code of Tang and Ding, is also near-MDS, and $\cD_4$ has the same
number of blocks~\cite{TD21}. So $\cC_4$ and $\cC_4^\perp$ have the same weight distributions as $\cC_\sigma$ and
$\cC_\sigma^\perp$.

\subsection{The cyclic representation}\label{sec:cyclic}

\begin{proof}[Proof of Theorem~\ref{thm:main}, part 3]
Let $U_{q+1}\subseteq\F_{q^2}^\times$ be the group of $(q+1)$-th roots of unity. Fix $\xi\in\F_{q^2}\setminus\F_q$
and put $X=t+\xi$ and $Y=t+\xi^q$. For $t\in\F_q$ we have $(X/Y)^q=Y/X$, so $u=X/Y$ lies in $U_{q+1}$, and
$t\mapsto u$, $\infty\mapsto1$, is a bijection $\PG(1,q)\to U_{q+1}$. Over $\F_{q^2}$ the linear forms $X$ and $Y$ form
a basis, so $W_\sigma=\Sym^2V\otimes V^{(\sigma)}$ is spanned by the forms $X^aY^{2-a}X^\sigma$ and
$X^aY^{2-a}Y^\sigma$ with $a=0,1,2$. After division by $(XY)^{s+1}$ these become the powers $u^j$ with
\[
j\in\{-1,0,1\}+\{\pm s\}=\pm\{s-1,s,s+1\}=:J .
\]
For $t\in\F_q$ the factor $XY=N_{\F_{q^2}/\F_q}(t+\xi)$ lies in $\F_q^\times$. So, after rescaling the
coordinates by elements of $\F_q^\times$, the extension of $\cC_\sigma$ to $\F_{q^2}$ is spanned by the vectors
$(u^j)_{u\in U_{q+1}}$ with $j\in J$, and $\cC_\sigma^\perp$ is monomially equivalent to
$\{c\in\F_q^{q+1}:\sum_uc_uu^j=0\text{ for }j\in J\}$. Since $q\equiv-1\pmod{q+1}$, we have $qJ=J$, so this is the
cyclic code over $\F_q$ with defining set $J$: the intersection with $\F_q^{q+1}$ of the cyclic code over $\F_{q^2}$
with defining set $J$, which is stable under $x\mapsto x^q$. Finally, the automorphism $u\mapsto u^\sigma$ of
$\F_{q^2}$ permutes $U_{q+1}$, and since $s\sigma=q\equiv-1\pmod{q+1}$ it carries $J$ to $\pm\{1,\sigma-1,\sigma+1\}$.
\end{proof}

Up to the final substitution, the computation does not use $s=\sigma/2$. With $2s$ in place of $\sigma$ it gives the
following statement, which for $s=2$ identifies $\cC_4^\perp$ with the code of Tang and Ding.

\begin{proposition}\label{prop:cyclic}
Let $s$ be a power of $2$ with $2\le s\le q/4$.
\begin{enumerate}[nosep]
\item Up to monomial equivalence, $\cC_{2s}^\perp$ is the cyclic code of length $q+1$ over $\F_q$ with defining set
  $\pm\{s-1,s,s+1\}$, that is, $\mathcal K_s$.
\item In particular, $\cC_4^\perp$ is equivalent to $\mathcal K_2$, the BCH code $\cC_{(q,q+1,4,1)}$ of Tang and
  Ding, and a bijection between $\PG(1,q)$ and the coordinates of $\mathcal K_2$ carries the dependent $6$-sets of
  $\cC_4$ onto the supports of the codewords of weight $6$ in $\mathcal K_2$.
\end{enumerate}
\end{proposition}

\begin{proof}
For part 1, the forms $X^aY^{2-a}X^{2s}$ and $X^aY^{2-a}Y^{2s}$, $a=0,1,2$, span $\Sym^2V\otimes V^{(2s)}$ over
$\F_{q^2}$, and after division by $(XY)^{s+1}$ they become the powers $u^j$ with $j\in\pm\{s-1,s,s+1\}$; the rest is
as in the proof of part 3 of Theorem~\ref{thm:main}, without the final substitution. For part 2, a monomial map
carries the support of a codeword to the support of its image. The code of Tang and Ding has minimum distance
$6$~\cite{TD21}, so by part 1 any five columns of $\cC_4$ are independent, and the argument of
Proposition~\ref{prop:zerosets}(3) shows that the supports of the codewords of weight $6$ in $\cC_4^\perp$ are the
dependent $6$-sets of $\cC_4$.
\end{proof}

\section{Comparison with the design of Tang and Ding}\label{sec:compare}

\subsection{Block conditions}

For a partition $\lambda$ with at most $n$ parts, the bialternant formula~\cite{Mac95} gives
$\det[x_i^{\lambda_j+n-j}]=\prod_{i<j}(x_i-x_j)\,s_\lambda(x_1,\dots,x_n)$, where $s_\lambda$ is the Schur polynomial.
For $\tau\in\{4,\sigma\}$ let $\cC_\tau$ be the evaluation code on $\PG(1,q)$ of the forms $A+t^\tau B$ of degree
$\tau+2$ with $\deg A,\deg B\le2$, which span $\Sym^2V\otimes V^{(\tau)}$, and let $\cD_\tau$ be the set of dependent
$6$-sets of $\cC_\tau$. For $\tau=\sigma$ these are the code and the design of Theorem~\ref{thm:main}. For $\tau=4$,
Proposition~\ref{prop:cyclic}(2) shows that $\cD_4$ is the design of Tang and Ding: the block set
$\mathcal B_{\sigma_{6,3},q+1}$ of~\cite{TD21}, carried to $\PG(1,q)$ by a bijection of the coordinates. As in the introduction, $e_\ell$ denotes the elementary
symmetric polynomial of degree $\ell$, written $\sigma_{k,\ell}$ in~\cite{TD21}.

\begin{proposition}\label{prop:blocks}
Let $\tau\in\{4,\sigma\}$ and $k=\tau-3$.
\begin{enumerate}[nosep]
\item Six distinct points $t_1,\dots,t_6\in\F_q$ form a block of $\cD_\tau$ if and only if $s_{(k,k,k)}(t_1,\dots,t_6)=0$.
\item For distinct $x_1,\dots,x_4\in\F_q^\times$, the set $\{0,\infty,x_1,\dots,x_4\}$ is a block of $\cD_\tau$ if and only if
  $s_{(k,k)}(x_1,\dots,x_4)=0$.
\item Let $J=\pm\{s-1,s,s+1\}$ for $\tau=\sigma$, with $s=2^e$ as in Lemma~\ref{lem:sigma}, and $J=\pm\{1,2,3\}$
  for $\tau=4$. Six distinct points $u_1,\dots,u_6\in U_{q+1}$ give linearly dependent vectors $(u_i^j)_{j\in J}$
  if and only if $s_{(k,k,k)}(u_1,\dots,u_6)=0$. By Section~\ref{sec:cyclic}, these vectors are the columns of the
  cyclic representation of $\cC_\tau$ (for $\tau=\sigma$ before the substitution $u\mapsto u^\sigma$), so part 1 also
  holds for six points of $U_{q+1}$ in this representation.
\end{enumerate}
For $\tau=4$ the two conditions are $e_3=0$, the block condition $\sigma_{6,3}=0$ of~\cite{TD21}, and $e_2=0$.
For $\tau=\sigma$ the polynomials $s_{(k,k,k)}$ and $s_{(k,k)}$ are homogeneous of degrees $3(\sigma-3)$ and
$2(\sigma-3)$.
\end{proposition}

\begin{proof}
Six points form a block iff some nonzero member of $\Sym^2V\otimes V^{(\tau)}$ vanishes on all of them. For part 1,
the column at a finite point $t$ is $(t^a)_{a\in E}$ with $E=\{0,1,2,\tau,\tau+1,\tau+2\}$, so the condition is
$\det[t_i^a]=0$. Writing the elements of $E$ in decreasing order as $\lambda_j+6-j$ gives $\lambda=(k,k,k,0,0,0)$, and
the Vandermonde factor is nonzero. For part 2, a member vanishing at $0$ and $\infty$ has $a_0=b_2=0$, so it equals
$t\,(a_1+a_2t+b_0t^{\tau-1}+b_1t^\tau)$. The condition is then that the matrix $[x_i^{\mu}]$ with
$\mu\in\{0,1,\tau-1,\tau\}$ is singular, and here $\lambda=(k,k,0,0)$. For part 3, since $u_i^{q+1}=1$,
multiplying the vector of $u_i$ by $u_i^{s+1}$, or by $u_i^3$ when $\tau=4$, turns the exponents in $J$ into $E$, and
the computation of part 1 applies. The proof of part 3 of Theorem~\ref{thm:main} in Section~\ref{sec:cyclic} shows
that the columns of $\cC_\tau$ are, up to scalars in $\F_q^\times$ and one invertible $\F_{q^2}$-linear change of
coordinates, these vectors, and the rank of vectors over $\F_q$ does not change over $\F_{q^2}$.
\end{proof}

\subsection{Automorphism groups}

\begin{lemma}\label{lem:aut}
Let $q=2^h$ with $h\ge5$ odd. Let $\cD$ be a design on $\PG(1,q)$ with at least one block and with block size
$k$, $2\le k\le q-1$, invariant under $\PGaL(2,q)$ and not complete. Then $\Aut(\cD)=\PGaL(2,q)$. Moreover $\PGaL(2,q)$ is self-normalising in the
symmetric group on $\PG(1,q)$.
\end{lemma}

\begin{proof}
The group $G=\Aut(\cD)$ contains $\PGL(2,q)$, which is sharply $3$-transitive, so $G$ is $3$-transitive. By
Burnside's theorem~\cite{DM96}, $G$ is affine or almost simple. It is not affine: its degree $2^h+1$ would then be a
prime power, and since $3\mid2^h+1$, a power $3^m$. Then $3^m\equiv1\pmod4$, so $m=2j$ is even, and
$(3^j-1)(3^j+1)=2^h$ forces $3^j-1=2$, that is, $h=3$. So $G$ is almost simple. By the list of $2$-transitive almost
simple groups~\cite{Cam81}, which rests on the classification of finite simple groups, the
socles of $3$-transitive almost simple groups are $A_n$, $\mathrm{PSL}(2,r)$ in degree $r+1$, and Mathieu groups of
degree at most $24$. In degree $q+1\ge33$, $G$ therefore has socle $A_{q+1}$ or $\mathrm{PSL}(2,q)$. If the socle is
$A_{q+1}$, then $G$ is transitive on $k$-sets and $\cD$ is complete, which is excluded. So the socle is
$\mathrm{PSL}(2,q)$, and $G$ lies in its normaliser. This normaliser is $\PGaL(2,q)$, since
$\Aut(\mathrm{PSL}(2,q))=\PGaL(2,q)$ and the centraliser of a $2$-transitive group is trivial. Finally, the normaliser
of $\PGaL(2,q)$ normalises its socle $\mathrm{PSL}(2,q)$, so it is $\PGaL(2,q)$ as well.
\end{proof}

\subsection{The block sets are disjoint}

\begin{proposition}\label{prop:differ}
Let $h\ge5$ be odd.
\begin{enumerate}[nosep]
\item No $6$-set is a block of both $\cD_\sigma$ and $\cD_4$. In particular $\cD_\sigma\neq\cD_4$.
\item $\cD_\sigma\cup\cD_4$ is a $4$-$(q+1,6,q-8)$ design. A $5$-set lies in no block of it or in exactly two, one
  of $\cD_\sigma$ and one of $\cD_4$.
\end{enumerate}
\end{proposition}

\begin{proof}
For part 1, suppose that $B$ is a block of both designs. Both are invariant under $\PGL(2,q)$ by
Lemma~\ref{lem:Wbasic}(2), which holds for $W_4$ as well, and $\PGL(2,q)$ is
$3$-transitive, so we may assume that $B=\{0,1,\infty,c,u,v\}$ with $c,u,v\in\F_q$. Put
\[
D=1+c+u,\qquad N=c+u+cu,\qquad \varepsilon=1+cu,\qquad \eta=c^2+u,\qquad \zeta=c+u^2 .
\]
By Proposition~\ref{prop:blocks}(2) with $\tau=4$, $B\in\cD_4$ means that $e_2(1,c,u,v)=N+vD$ vanishes. If $D=0$,
then $u=1+c$ and $N=1+c+c^2$, which is nonzero because $X^2+X+1$ has no root in $\F_q$ for odd $h$. So $D\neq0$ and
$v=N/D$. Then $N=vD$, $\varepsilon=N+D=(v+1)D$, $\eta=N+cD=(v+c)D$ and $\zeta=N+uD=(v+u)D$, and since
$v\notin\{0,1,c,u\}$,
\begin{equation}\label{eq:K}
K=u(1+u)(c+u)\,DN\varepsilon\eta\zeta^2\neq0 .
\end{equation}

By Proposition~\ref{prop:blocks}(2) with $\tau=\sigma$, $B\in\cD_\sigma$ means that the $4\times4$ matrix with rows
$(x,x^2,x^\sigma,x^{\sigma+1})$ for $x=1,c,u,v$ is singular. Put $C=c^\sigma$ and $U=u^\sigma$, so that
$D^\sigma=1+C+U$ and $N^\sigma=C+U+CU$. Multiplying the row of $v=N/D$ by $D^2D^\sigma\neq0$ and expanding the
determinant shows that $B\in\cD_\sigma$ iff $F(c,u,C,U)=0$, where
\[
F(c,u,C,U)=(C\eta^2+c^2\varepsilon^2)\,U^2+\zeta^2(C+c)^2\,U+u^2C\,(C\varepsilon^2+\eta^2).
\]
Applying the automorphism $x\mapsto x^\sigma$ to $F(c,u,C,U)=0$, and using $c^{\sigma^2}=c^{2q}=c^2$ and
$u^{\sigma^2}=u^2$, gives $F(C,U,c^2,u^2)=0$. In $F(C,U,c^2,u^2)$ every variable occurs only to even powers, and
indeed
\[
F(C,U,c^2,u^2)=\bigl(r_2U^2+r_1U+r_0\bigr)^2,\qquad r_2=C\eta+c\varepsilon,\quad r_1=\zeta(C^2+c),\quad
r_0=uC(C\varepsilon+\eta).
\]
So $U$ is a common root of $f=a_2X^2+a_1X+a_0$, with $a_2,a_1,a_0$ the coefficients of $F(c,u,C,U)$ above, and of
$g=r_2X^2+r_1X+r_0$. The Sylvester matrix $S$ with rows $(a_2,a_1,a_0,0)$, $(0,a_2,a_1,a_0)$, $(r_2,r_1,r_0,0)$ and
$(0,r_2,r_1,r_0)$ maps the nonzero vector $(U^3,U^2,U,1)$ to $(Uf(U),f(U),Ug(U),g(U))=0$, so $\det S=0$, also when
leading coefficients vanish. Expanding,
\[
\det S=K\cdot C(C+1)(C+c)^2(C+c^2)(C^2+C+1).
\]
By~\eqref{eq:K}, $K\neq0$. But $C=c^\sigma\notin\{0,1\}$ because $c\notin\{0,1\}$. Next, $C\neq c$, since
$c^{2^{e+1}}=c$ would put $c$ in $\F_{2^{\gcd(e+1,2e+1)}}=\F_2$. Further $C\neq c^2$, since $c^\sigma=c^2$ gives
$c^s=c$ and $c\in\F_{2^{\gcd(e,2e+1)}}=\F_2$. Finally $C^2+C+1\neq0$. This contradiction proves part 1. The three
polynomial identities used here (the expansion giving $F$, the square, and $\det S$) were checked by computer algebra
over $\F_2$, and they are part of the formal proof.

For part 2, let $T$ be a $4$-set. A $5$-set $S$ lies in at most one block of each design. Indeed, the points of $S$
are independent for both codes (Proposition~\ref{prop:zerosets}(1), and~\cite{TD21} for $\cC_4$), so two blocks
containing $S$ would lie in the span of $S$, and their union, of seven points, would have rank $5$. Then a nonzero
member would vanish on seven points, which is impossible because both $\cC_\sigma$ and $\cC_4$ have minimum distance
$q-5$ (Theorem~\ref{thm:main} and~\cite{TD21}). So in each design the
blocks through $T$ pair up the points outside $T$ that are not unpaired. By Proposition~\ref{prop:N5} there are five
unpaired points for $\cD_\sigma$, and by Theorem~\ref{thm:fivesets}(2) below the unpaired points for $\cD_4$ are the
same. Hence each design has $(q-8)/2$ blocks through $T$, and by part 1 the union has $q-8$. Finally, a $5$-set lies
in a block of $\cD_\sigma$ iff it lies in a block of $\cD_4$, by Theorem~\ref{thm:fivesets}(2), and by part 1 these
two blocks are different.
\end{proof}

The proof of part 1 uses neither design property. It uses only the block conditions of
Proposition~\ref{prop:blocks}(2) and, through $c^{\sigma^2}=c^2$, the relation $\sigma^2\equiv2\pmod{q-1}$.

\begin{proof}[Proof of Theorem~\ref{thm:noniso}]
By Proposition~\ref{prop:differ}, $\cD_\sigma\neq\cD_4$. Both are $\PGaL(2,q)$-invariant, have blocks
(Theorem~\ref{thm:main} and~\cite{TD21}) and are not complete, so both have automorphism group $\PGaL(2,q)$ by Lemma~\ref{lem:aut}. Suppose $\pi$ is an isomorphism
$\cD_\sigma\to\cD_4$. Then $\pi$ conjugates $\Aut(\cD_\sigma)$ onto $\Aut(\cD_4)$, so it normalises $\PGaL(2,q)$ and
lies in it. But then $\pi(\cD_\sigma)=\cD_\sigma\neq\cD_4$, a contradiction.
\end{proof}

\begin{corollary}\label{cor:codes}
Let $h\ge5$ be odd and $\tau\in\{4,\sigma\}$.
\begin{enumerate}[nosep]
\item No semilinear map carries $\cC_\sigma^\perp$ onto $\cC_4^\perp$, or $\cC_\sigma$ onto $\cC_4$.
\item The coordinate permutations induced by the semilinear automorphisms of $\cC_\tau^\perp$ form the group
  $\PGaL(2,q)$.
\end{enumerate}
\end{corollary}

\begin{proof}
A semilinear map preserves weights and carries supports to supports, so it induces an isomorphism between the
weight-$6$ designs. A map from $\cC_\sigma^\perp$ onto $\cC_4^\perp$ would therefore give $\cD_\sigma\cong\cD_4$,
contradicting Theorem~\ref{thm:noniso}, and a map from $\cC_\sigma$ onto $\cC_4$ gives one between the duals, with the
rescalings inverted. This proves part 1. For part 2, the induced permutations lie in $\Aut(\cD_\tau)=\PGaL(2,q)$.
Conversely $\PGaL(2,q)$ acts on $\cC_\tau$, and hence on $\cC_\tau^\perp$, by Lemma~\ref{lem:Wbasic}(2), which holds
for $W_4$ as well.
\end{proof}

\subsection{The covered \texorpdfstring{$5$}{5}-sets}

A $5$-set lies in at most one block of $\cD_\sigma$: the members of $W_\sigma$ vanishing on it form a
$1$-dimensional space (Proposition~\ref{prop:zerosets}(1)), and a block containing it is the zero set of such a member
(Proposition~\ref{prop:zerosets}(4)). For a $4$-set $T$ and $\tau\in\{4,\sigma\}$, call $u\notin T$
\emph{unpaired for $\cD_\tau$} if no block of $\cD_\tau$ contains $T\cup\{u\}$. For $\tau=\sigma$ these are the
unpaired points of Section~\ref{sec:proof}, by Proposition~\ref{prop:zerosets}(4). Every $c\in\F_q$ has a unique
square root $\sqrt c=c^{q/2}$.

\begin{theorem}\label{thm:fivesets}
Let $h\ge5$ be odd and $\tau\in\{4,\sigma\}$.
\begin{enumerate}[nosep]
\item For $c\in\F_q\setminus\{0,1\}$, the unpaired points of $T=\{0,1,\infty,c\}$ for $\cD_\tau$ are $1+c$,
  $c/(1+c)$, $1/c$, $c^2$ and $\sqrt c$.
\item A $5$-set lies in a block of $\cD_\sigma$ if and only if it lies in a block of $\cD_4$.
\end{enumerate}
\end{theorem}

\begin{proof}
The five points of part 1 lie outside $T$ and are distinct: each coincidence forces $c\in\{0,1\}$ or $c^2+c+1=0$,
and $c^2+c+1$ has no root in $\F_q$ because $h$ is odd.

Let $\tau=4$. For $u\in\F_q\setminus\{0,1,c\}$ and $v\in\F_q\setminus\{0,1,c,u\}$, the set $T\cup\{u,v\}$ is a
block iff $e_2(1,c,u,v)=0$ by Proposition~\ref{prop:blocks}(2), that is, iff $(1+c+u)v=c+u+cu$. For $u=1+c$ the left
side vanishes and the right side is $1+c+c^2\neq0$, so $u$ is unpaired. Otherwise the only candidate is
$v=(c+u+cu)/(1+c+u)$, and $u$ is unpaired iff $v\in\{0,1,c,u\}$. These four cases are $u=c/(1+c)$, $u=1/c$,
$u=c^2$ and $u^2=c$.

Let $\tau=\sigma$. The involutions $g_1(t)=c/t$, $g_2(t)=c(t+1)/(t+c)$ and $g_3(t)=(t+c)/(t+1)$ in $\PGL(2,q)$ map
$T$ onto itself and carry $1+c$ to $c/(1+c)$, $c^2$ and $1/c$. They preserve $\cD_\sigma$ by
Lemma~\ref{lem:Wbasic}(2), so they permute the unpaired points of $T$. We show that $1+c$ is unpaired. Put
$a_2=c+c^\sigma$, $b_0=c+c^2$ and
\[
Q=(a_2+b_0)t+a_2t^2+b_0t^\sigma\in W_\sigma .
\]
Then $Q\neq0$, and $Q$ vanishes at $\infty$ and is tangent there, since its coefficients of $t^{\sigma+1}$ and
$t^{\sigma+2}$ vanish. For every $u$ we have $Q(u)=a_2(u+u^2)+b_0(u+u^\sigma)$, so $Q(0)=Q(1)=0$ and
$Q(c)=a_2b_0+b_0a_2=0$. At $u=1+c$ we have $u+u^2=c+c^2$ and $u+u^\sigma=c+c^\sigma$, so $Q(1+c)=Q(c)=0$. Hence the
zeros of $Q$ on $T\cup\{1+c\}$, with multiplicities capped at two, already add up to $6$, and $Q$ has no further zero
by Lemma~\ref{lem:zeros}(2). By Proposition~\ref{prop:zerosets}, a block through $T\cup\{1+c\}$ would be the zero
set of a member vanishing on these five points, and every such member is a multiple of $Q$. So $1+c$ is unpaired,
and so are $c/(1+c)$, $c^2$ and $1/c$. By Proposition~\ref{prop:N5} there is exactly one further unpaired point $x$.
The involution $g_1$ permutes the unpaired points and the four points above, so it fixes $x$. Then $c/x=x$, and
$x=\sqrt c$.

For part 2, every $5$-set is the image of a set $T\cup\{u\}$ as in part 1 under some element of $\PGL(2,q)$, since
$\PGL(2,q)$ is $3$-transitive. Both $\cD_\sigma$ and $\cD_4$ are invariant under $\PGL(2,q)$ by
Lemma~\ref{lem:Wbasic}(2), which holds for $W_4$ as well. By part 1, $T\cup\{u\}$ lies
in a block of $\cD_\sigma$ iff it lies in a block of $\cD_4$.
\end{proof}

By part 2, a $5$-set lies in no block of $\cD_\sigma$ iff it lies in no block of $\cD_4$. So the $4$-$(q+1,5,5)$
design of Theorem~\ref{thm:shells}(2) is also formed by the $5$-sets in no block of $\cD_4$, which Yan and
Zhou~\cite{YZ22} identify with one of Alltop's $4$-designs~\cite{Alt69}.

\section{\texorpdfstring{Uniqueness: proof of Theorem~\ref{thm:unique}}{Uniqueness: proof of Theorem 1.3}}\label{sec:unique}

In this section $h\ge5$ may be odd or even, $q=2^h$. For $1\le j\le h-1$ and $\tau=2^j$ let $W_\tau$ be the
space of forms $A+t^\tau B$ of degree $\tau+2$ with $\deg A,\deg B\le2$, and let $\cC_\tau$ be its evaluation code.
For $\tau=\sigma$ and $\tau=4$ these are the spaces of Sections~\ref{sec:proof} and~\ref{sec:compare}, and
Lemma~\ref{lem:Wbasic}(2) holds for every $W_\tau$ with the same proof. Section~\ref{sec:twists} shows that every
code in Theorem~\ref{thm:unique}(1) is some $\cC_\tau$. Section~\ref{sec:nmdstwists} decides which $\cC_\tau$ are
near-MDS, and Section~\ref{sec:certificate} describes the computation used there. Wang, Wu and Hu~\cite{WWH26}
obtain $3$-designs from $\GL_2(\F_q)$-invariant spaces of binary forms of degree $k$, with blocks of size $k$, and
also pass to the unit circle $U_{q+1}$ by the Cayley transform.

\subsection{Invariant codes are twists}\label{sec:twists}

Write $v_\infty=(1,0)$ and $v_t=(t,1)$ for $t\in\F_q$. For $d\in\mathbb Z/(q-1)$ let $M_d$ be the space of maps
$f\colon\F_q^2\setminus\{0\}\to\F_q$ with $f(cv)=c^df(v)$ for $c\in\F_q^\times$. The group $\GL(2,q)$ acts on $M_d$ by
$(gf)(v)=f(g^{-1}v)$, and evaluation at the vectors $v_P$ identifies $M_d$ with $\F_q^{q+1}$. A form $P$ of degree $d$
is the element $(u,v)\mapsto v^dP(u/v)$ of $M_d$, in accordance with Definition~\ref{def:forms}.

\begin{lemma}\label{lem:induced}
Let $\rho$ be a homomorphism from $\PGL(2,q)$ to the group of monomial maps of $\F_q^{q+1}$ such that $\rho(g)$
permutes the coordinates as $g$ permutes $\PG(1,q)$. Then for some $d\in\mathbb Z/(q-1)$, after rescaling the
coordinates, the $\rho$-invariant subspaces of $\F_q^{q+1}$ are exactly the evaluations of the
$\GL(2,q)$-submodules of $M_d$.
\end{lemma}

\begin{proof}
Let $B$ be the stabiliser of $\infty$ in $\PGL(2,q)$, the group of maps $t\mapsto at+b$. It acts on the coordinate
vector $e_\infty$ through a character $\chi\colon B\to\F_q^\times$. The translations are commutators in $B$, so
$\chi(t\mapsto at+b)=a^k$ for some $k$. For every point $P$ fix $g_P\in\PGL(2,q)$ with $g_P(\infty)=P$, and rescale the
coordinates so that $e_P=\rho(g_P)e_\infty$. Then $\rho(g)e_P=\chi(g_{gP}^{-1}gg_P)\,e_{gP}$ for all $g$ and $P$, so in
these coordinates $\rho$ is determined by $\chi$. Now put $m=-k$ and $d=2m$. The rule $(g\cdot f)(v)=\det(g)^m
f(g^{-1}v)$ defines an action of $\GL(2,q)$ on $M_d$ in which the centre acts trivially. After evaluation it is a
monomial action of $\PGL(2,q)$ that permutes the coordinates as $\PGL(2,q)$ permutes the points. The map
$t\mapsto at+b$, represented by $\left(\begin{smallmatrix}a&b\\0&1\end{smallmatrix}\right)$, multiplies the coordinate
at $\infty$ by $a^m\cdot a^{-d}=a^k$. By the first part, after a rescaling of the coordinates this action coincides
with $\rho$. The factor $\det(g)^m$ does not change which subspaces are invariant.
\end{proof}

From now on fix a representative $1\le d\le q-2$ of a class $d\not\equiv0$. For $0\le a\le q-2$ let $m_a\in M_d$ be given
by $m_a(v_t)=t^a$ for $t\in\F_q$, where $0^0=1$, and $m_a(v_\infty)=1$ if $a=d$ and $0$ otherwise. For $a\le d$ this is
the form $t^a$ of degree $d$. Let $\delta_P$ be the indicator function of the point $P$. For integers $a,b\ge0$ write
$b\preceq a$ if every binary digit of $b$ is at most the corresponding digit of $a$. By Lucas's theorem,
$\binom ab$ is odd exactly when $b\preceq a$.

\begin{lemma}\label{lem:weights}
Let $W$ be a $\GL(2,q)$-submodule of $M_d$ with $W\ne M_d$ that contains $m_a$ for some $a\notin\{0,d\}$, and let
$S=\{a:m_a\in W\}$. Then $W$ is spanned by the $m_a$ with $a\in S$. Moreover $S$ contains $0$ and $d$, and it is
closed under $a\mapsto b$ for $b\preceq a$ and under $a\mapsto(d-a)\bmod(q-1)$.
\end{lemma}

\begin{proof}
The diagonal matrix $\mathrm{diag}(\alpha,\beta)$ multiplies $m_a$ by $\alpha^{-a}\beta^{a-d}$, $\delta_0$ by $\beta^{-d}$,
and $\delta_\infty$ by $\alpha^{-d}$. These characters are distinct for distinct $a\in\mathbb Z/(q-1)$. So the common
eigenspaces of the diagonal matrices are the lines $\langle m_a\rangle$ for $a\notin\{0,d\}$, together with
$\langle m_0,\delta_0\rangle$ and $\langle m_d,\delta_\infty\rangle$. These matrices form a group of odd order, so $W$ is
the sum of its intersections with these eigenspaces. If $W$ contained $\delta_0$ or $\delta_\infty$, it would contain
every $\delta_P$, since $\PGL(2,q)$ is transitive, and so $W=M_d$. Hence the first claim follows once we know that
$m_0,m_d\in W$.

Let $u_\lambda=\left(\begin{smallmatrix}1&\lambda\\0&1\end{smallmatrix}\right)$. Then $(u_\lambda f)(v_t)=f(v_{t+\lambda})$
and $(u_\lambda f)(v_\infty)=f(v_\infty)$. For $a\in S\setminus\{0,d\}$ this gives
\[
u_\lambda m_a=\sum_{b\preceq a}\lambda^{a-b}m_b+\varepsilon\,\lambda^{a-d}\delta_\infty,
\]
where $\varepsilon=1$ if $d\preceq a$ and $\varepsilon=0$ otherwise. The exponents $a-b$ are distinct and smaller than
$q$, so the functions $\lambda\mapsto\lambda^{a-b}$ on $\F_q$ are linearly independent. Hence $W$ contains $m_b$ for
every $b\preceq a$ with $b\ne d$, and it contains $m_d+\delta_\infty$ if $d\preceq a$. In particular $m_0\in W$. Let
$w=\left(\begin{smallmatrix}0&1\\1&0\end{smallmatrix}\right)$. Then $(wf)(v_t)=t^df(v_{1/t})$ for $t\ne0$, while $w$
exchanges the values at $0$ and $\infty$. So $wm_0=m_d$, and $wm_a=m_{(d-a)\bmod(q-1)}$ for $a\notin\{0,d\}$. Thus
$m_d\in W$ and $S$ is closed under $a\mapsto d-a$. If $d\preceq a$ for some $a\in S\setminus\{0,d\}$, then also
$\delta_\infty\in W$, which is excluded. So $S$ is closed under $\preceq$ at every $a\ne d$. At $a=d$ it is closed as
well, because $u_\lambda m_d=\sum_{b\preceq d}\lambda^{d-b}m_b$, since $m_d$ is the form $t^d$.
\end{proof}

Multiplication by $2$ modulo $q-1$ permutes the $h$ binary digits of the integers in $[0,q-2]$ cyclically. It
therefore preserves $\preceq$ and the number of nonzero digits, the \emph{weight}. Put $T_j=\{0,1,2,2^j,2^j+1,2^j+2\}$,
the exponent set of $W_{2^j}$.

\begin{lemma}\label{lem:closed}
Let $h\ge5$ and $1\le d\le q-2$. Let $S\subseteq\{0,\dots,q-2\}$ have six elements, contain $0$ and $d$, and be
closed under $a\mapsto b$ for $b\preceq a$ and under $a\mapsto(d-a)\bmod(q-1)$. Then there are $k\ge0$ and
$2\le j\le h-1$ with $S=\{2^ka\bmod(q-1):a\in T_j\}$ and $d\equiv2^k(2^j+2)\pmod{q-1}$.
\end{lemma}

\begin{proof}
Multiplying $S$ and $d$ by a power of $2$ preserves the hypotheses, so we may rotate the digits. The elements $b\preceq
d$ lie in $S$, and there are $2^{\mathrm{wt}(d)}$ of them, so $\mathrm{wt}(d)\le2$. For the same reason every element of
$S$ has weight at most $2$.

Suppose first that $\mathrm{wt}(d)=1$; after rotation $d=1$. For $2\le a\le q-2$ we have $1-a\equiv2^h-a$. If $a=2^w$,
then $2^h-a$ has weight $h-w$. If $a=2^w+2^v$ with $w>v$, then $2^h-a$ has weight $h-v-1$. So every power $2^w\in S$ other than $1$ has $w\ge h-2$. An
element $2^w+2^v\in S$ with $w>v$ has $v\ge h-3\ge2$, so $2^w$ and $2^v$ are powers in $S$ other than $1$, and
$\{w,v\}=\{h-1,h-2\}$. Thus
$S\subseteq\{0,1,2^{h-2},2^{h-1},2^{h-1}+2^{h-2}\}$, which has only five elements.

So $\mathrm{wt}(d)=2$, and after rotation $d=1+2^u$ with $1\le u\le h-1$. Then $S$ contains $D=\{0,1,2^u,d\}$. The map
$c(a)=(d-a)\bmod(q-1)$ is an involution of $S$ that preserves $D$, so it permutes the two elements of $R=S\setminus D$.

Suppose $c$ fixes some $a\in R$. Then $2a\equiv d$, so $a=2^{h-1}+2^{u-1}$ and $2^{h-1},2^{u-1}\in S$. At most one of
these two powers lies in $D$: $2^{u-1}=1$ only for $u=1$, and $2^{h-1}=2^u$ only for $u=h-1$. So the other element $p$
of $R$ is a power of $2$. As $c$ permutes $R$ and fixes $a$, it also fixes $p$, so $2p\equiv d$, which is impossible
for a power of $2$. Hence $R=\{a,b\}$ with $a+b\equiv d$ and $a\ne b$.

If $a$ and $b$ both had weight $2$, their digits would lie among those of the only powers $1$ and $2^u$ in $S$,
and both would equal $d$. If both were powers of $2$, then $a+b$ and $d$ would be integers below $q-1$ with $a+b\equiv d\pmod{q-1}$, so
$a+b=d$ and $\{a,b\}=\{1,2^u\}$. So we may assume that $b=2^z$ with $z\notin\{0,u\}$ and that $a$ has weight $2$. The
digits of $a$ then lie in $\{0,u,z\}$ and include $z$.
\begin{itemize}
\item If $a=2^z+2^u$, then $a+b\equiv d$ gives $2^{z+1}\equiv1$, so $z=h-1$ and $u\le h-2$. Then
  $S=\{0,1,2^u,1+2^u,2^{h-1},2^{h-1}+2^u\}=\{2^{h-1}a\bmod(q-1):a\in T_{u+1}\}$.
\item If $a=2^z+1$, then $2^{z+1}\equiv2^u$, so $z=u-1$ and $u\ge2$. Then
  $S=\{0,1,2^{u-1},1+2^{u-1},2^u,1+2^u\}=\{2^{u-1}a\bmod(q-1):a\in T_{h-u+1}\}$.\qedhere
\end{itemize}
\end{proof}

\begin{proposition}\label{prop:twists}
Let $h\ge5$ and let $W$ be a $6$-dimensional $\GL(2,q)$-submodule of some $M_d$. Then a field automorphism, applied to
every coordinate, carries the evaluation code of $W$ onto $\cC_{2^j}$ for some $2\le j\le h-1$.
\end{proposition}

\begin{proof}
For $d\equiv0$ the action of $\PGL(2,q)$ on $M_0$ permutes the coordinates. By~\cite[Theorem~11]{DTT21} it has no
invariant subspace of dimension $6$. So let $d\not\equiv0$.
If $W$ contained no $m_a$ with $a\notin\{0,d\}$, it would lie in $\langle m_0,\delta_0\rangle+\langle
m_d,\delta_\infty\rangle$ by the proof of Lemma~\ref{lem:weights}, which has dimension $4$. Also $W\ne M_d$. So
Lemma~\ref{lem:weights} applies, and $W$ is spanned by the $m_a$ with $a$ in a $6$-set $S$. By Lemma~\ref{lem:closed},
$S=\{2^ka\bmod(q-1):a\in T_j\}$ and $d\equiv2^k(2^j+2)$. For $k=0$, the $m_a$ with $a\in T_j$ are the forms $t^a$ of
degree $2^j+2$ that span $W_{2^j}$. Raising the values to the power $2^k$ maps $m_a\in M_{2^j+2}$ to
$m_{2^ka\bmod(q-1)}\in M_d$. So the evaluation code of $W$ is obtained from $\cC_{2^j}$ by applying $c\mapsto c^{2^k}$ to
every coordinate, and its inverse carries it back.
\end{proof}

For $j=h-1$ we have $2T_{h-1}\equiv\{0,1,2,3,4,5\}\pmod{q-1}$, because $2^h\equiv1$. So $\cC_{q/2}$ is semilinearly
equivalent to the evaluation code of all forms of degree $5$, the doubly extended Reed--Solomon code of dimension $6$.

\subsection{Which twists are near-MDS}\label{sec:nmdstwists}

\begin{lemma}\label{lem:edgetwists}
\begin{enumerate}[nosep]
\item $\cC_2$ has dimension $5$.
\item $\cC_{q/2}$ is MDS, so it is not near-MDS.
\item If $g=\gcd(j,h)\ge2$, any $5$ points of the subline $\PG(1,2^g)$ are dependent for $\cC_{2^j}$, so $\cC_{2^j}$ is
  not near-MDS.
\end{enumerate}
\end{lemma}

\begin{proof}
(1) The six monomials of $W_2$ are $1,t,t^2,t^2,t^3,t^4$. (2) A nonzero form of degree $5$ has at most $5$ zeros on
$\PG(1,q)$; now use the remark after Proposition~\ref{prop:twists}. (3) For $u,v\in\F_{2^g}$ we have $u^{2^j}=u$ and
$v^{2^j}=v$. So on $\PG(1,2^g)$ every member of $W_{2^j}$, viewed as a binary form, agrees with a binary form of
degree $3$. These span a space of dimension $4$, so the columns of any $5$ of the $2^g+1\ge5$ points of $\PG(1,2^g)$ are
dependent.
\end{proof}

The remaining twists are handled through linearised polynomials. A polynomial $L(x)=\sum_ic_ix^{2^i}$ with
$c_i\in\F_q$ defines an $\F_2$-linear map of $\F_q$, whose kernel is the set of its roots in $\F_q$.

\begin{lemma}\label{lem:ninezeros}
Let $\tau=2^j$. If some nonzero $L(x)=cx^\tau+ax^2+bx$ with $a,b,c\in\F_q$ has at least $8$ roots in $\F_q$, then some nonzero
member of $W_\tau$ has at least $9$ zeros, and $\cC_\tau$ is not near-MDS.
\end{lemma}

\begin{proof}
The member $P=bt+at^2+ct^\tau$, with $A=bt+at^2$ and $B=c$, has degree at most $\tau$, so $m_\infty(P)\ge2$. Its
finite zeros are the roots of $L$.
\end{proof}

\begin{lemma}\label{lem:gcddescent}
\begin{enumerate}[nosep]
\item The kernels of $x^{2^j}+x$ and $x^{2^j}+x^2=(x^{2^{j-1}}+x)^2$ are $\F_{2^{\gcd(j,h)}}$ and
  $\F_{2^{\gcd(j-1,h)}}$.
\item Let $h'\mid h$ and $j'\equiv j\pmod{h'}$. If $x^{2^{j'}}+ax^2+bx$ with $a,b\in\F_{2^{h'}}$ has at least $8$ roots
  in $\F_{2^{h'}}$, then $x^{2^j}+ax^2+bx$ has at least $8$ roots in $\F_q$.
\end{enumerate}
\end{lemma}

\begin{proof}
Part~1 is clear. For part~2, on $\F_{2^{h'}}$ the two polynomials agree, since $x^{2^j}=x^{2^{j'}}$ there.
\end{proof}

Following~\cite{BZ20}, a $3$-set $E$ of residues modulo $h$ is a \emph{Moore exponent set} (for $q=2$ and $n=h$) if
$\det(y_r^{2^\ell})_{r,\ell}\ne0$, with $r=1,2,3$ and $\ell\in E$, whenever $y_1,y_2,y_3\in\F_q$ are
$\F_2$-independent. Equivalently, no nonzero $\sum_{\ell\in E}c_\ell x^{2^\ell}$ has $8$ roots in $\F_q$. Raising such a
polynomial to a power $2^k$ does not change its roots, so $E$ and $E+k$ are Moore exponent sets together.

\begin{lemma}\label{lem:progressions}
Let $2\le j\le h-1$. The set $\{0,1,j\}$ is an arithmetic progression in $\mathbb Z/h$ exactly when $j=2$, $j=h-1$, or
$h$ is odd and $j=(h+1)/2$. In these cases every nonzero $cx^{2^j}+ax^2+bx$ has at most $4$ roots in $\F_q$.
\end{lemma}

\begin{proof}
Consider the middle term of the progression. If it is $0$, the other two terms are $\pm\delta$, so $j=-1$. If it is
$1$, they are $1\pm\delta$, so $\delta=\pm1$ and $j=2$. If it is $j$, they are $j\pm\delta=\{0,1\}$, so $2j\equiv1$.
In each case the difference $\delta$ is a unit modulo $h$. Write $\{0,1,j\}=\{b,b+\delta,b+2\delta\}$
and raise the polynomial to the power $2^{-b}$. The result is $c_0x+c_1x^\theta+c_2x^{\theta^2}$ with
$x^\theta=x^{2^\delta}$, and $\theta$ generates $\mathrm{Gal}(\F_q/\F_2)$. Such a nonzero polynomial has at most $4$
roots~\cite{GQ09}.
\end{proof}

So, apart from the Reed--Solomon case $\tau=q/2$, the twists for which $\langle x,x^2,x^\tau\rangle$ is a Gabidulin
code~\cite{Del78,Gab85,KG05} are $\tau=4$ and $\tau=\sigma$, for the generators $t\mapsto t^2$ and $t\mapsto t^\sigma$
of $\mathrm{Gal}(\F_q/\F_2)$. Lemma~\ref{lem:ninezeros} gives nothing for them. For the others, a trinomial with $8$ roots exists in all cases
we can decide.

\begin{theorem}\label{thm:nmdstwists}
Let $2\le j\le h-2$, and suppose that one of the following holds:
\begin{enumerate}[nosep]
\item $h\le100$;
\item $4j+2<h$ or $4(h-j+1)+2<h$;
\item $\gcd(j,h)\ge2$ or $\gcd(j-1,h)\ge3$;
\item $h$ has a divisor $h'$ with $5\le h'\le100$ such that $j'=j\bmod h'$ satisfies $2\le j'\le h'-2$ and
  $\{0,1,j'\}$ is not an arithmetic progression modulo $h'$.
\end{enumerate}
Then $\cC_{2^j}$ is near-MDS if and only if $h$ is odd and $j\in\{2,(h+1)/2\}$.
\end{theorem}

\begin{proof}
If $h$ is odd and $j\in\{2,(h+1)/2\}$, the code is $\cC_4$ or $\cC_\sigma$. The first is near-MDS by~\cite{TD21},
since its dual is monomially equivalent to their code (Section~\ref{sec:cyclic}), and the second by
Theorem~\ref{thm:main}. Conversely, let $\cC_{2^j}$ be near-MDS. By Lemma~\ref{lem:edgetwists}(3), $\gcd(j,h)=1$; in
particular $h$ is odd if $j=2$. By Lemma~\ref{lem:progressions} it remains to show that $\{0,1,j\}$ is an arithmetic
progression. Suppose it is not. By Lemma~\ref{lem:ninezeros} it suffices to find a trinomial $x^{2^j}+ax^2+bx$ with at
least $8$ roots. In case~1 the computation of Section~\ref{sec:certificate} provides one. In case~2, by~\cite[Theorem~3.2
and Remark~3.3]{BZ20}, a set $\{0,i,J\}$ of integers with $0<i<J$ and $J\ne2i$ is not a Moore exponent set when $h>4J+2$,
for every prime power, including $q=2$. Apply this to $\{0,1,j\}$ or to its translate $\{0,h-j,h-j+1\}$. Here $J\ne2i$
because $j\ne2$ and $j\ne h-1$. Normalise the resulting polynomial as in the proof of Lemma~\ref{lem:ninezeros}. In
case~3 use Lemma~\ref{lem:edgetwists}(3) or Lemma~\ref{lem:gcddescent}(1). In case~4 use
Lemma~\ref{lem:gcddescent}(2) together with case~1 for $h'$.
\end{proof}

\begin{proof}[Proof of Theorem~\ref{thm:unique}]
Part~1 follows from Lemma~\ref{lem:induced} and Proposition~\ref{prop:twists}. For part~2, $\cC_{q/2}$ is not
near-MDS by Lemma~\ref{lem:edgetwists}(2), and for $2\le j\le h-2$ apply Theorem~\ref{thm:nmdstwists}(1).
\end{proof}

The twists are the cyclic codes of~\eqref{eq:Gs}. Let $\tau=2s$ with $2\le s\le q/4$. The computation in
Section~\ref{sec:cyclic}, with $s$ for $\sigma/2$ and without the final substitution $u\mapsto u^\sigma$, shows that
the extension of $\cC_\tau$ to $\F_{q^2}$ is spanned, after rescaling the coordinates by elements of $\F_q^\times$, by the
vectors $(u^j)_{u\in U_{q+1}}$ with $j\in\pm\{s-1,s,s+1\}$. So $\cC_\tau^\perp$ is monomially equivalent to the cyclic
code $\mathcal K_s$ generated by $G_s$. A code is near-MDS if and only if its dual is, so for $h\le100$ the code
$\mathcal K_s$ is near-MDS exactly when $h$ is odd and $s\in\{2,\sigma/2\}$.

\begin{remark}\label{rem:moore}
Every case not covered by Theorem~\ref{thm:nmdstwists} would follow if, for every $h$, the only Moore exponent sets
$\{0,1,j\}$ for $q=2$ were the arithmetic progressions with unit difference (Problem~\ref{prob:classify}). For prime
$h$ this would in turn follow if the Moore property were invariant under $E\mapsto aE$ for units $a$ modulo $h$. Indeed,
by Dirichlet's approximation theorem some $a\le\sqrt h$ moves $\{0,1,j\}$ into a translate of a set $\{0,i,J\}$
with $J<2\sqrt h$, to which~\cite[Theorem~3.2]{BZ20} applies once $h\ge68$. In our computations even the number of
$3$-dimensional kernels in $\langle x^{2^\ell}:\ell\in E\rangle$ is invariant under $E\mapsto aE$. We tested this for
$h\in\{13,17,19,23\}$, including units that are not powers of $2$ modulo $h$. For $q\neq2$ the analogue of
Problem~\ref{prob:classify}(1) fails in general: $\{0,1,3\}$ is a Moore exponent set for $n=7$ and odd $q$, and for
$n=8$ and $q\equiv1\pmod3$~\cite{CMPZ20}.
\end{remark}

\subsection{The computation}\label{sec:certificate}

For $x\in\F_q\setminus\F_2$ put $g_j(x)=(x^{2^j}+x)/(x^2+x)$. Replacing $x$ by $ux$ for a root $u\ne0$, a trinomial
$cx^{2^j}+ax^2+bx$ with $8$ roots may be assumed to have $c=1$ and the root $1$, that is, $b=a+1$. Its kernel is then
$\F_2\cup g_j^{-1}(a)$. So such a trinomial exists if and only if some value of $g_j$ is taken at least $6$ times.

For $5\le h\le27$ and $2\le j\le h-2$ we evaluated $g_j$ on one element of each pair $\{x,x+1\}$ and compared the
values. A repeated value occurs exactly when $\{0,1,j\}$ is not an arithmetic progression. For $28\le h\le100$ a
random search suffices. For random $u\notin\F_2$, Gaussian elimination over $\F_2$ gives the kernel of
$v\mapsto(u^2+u)(v^{2^j}+v)+(u^{2^j}+u)(v^2+v)$, which is $\F_2\cup g_j^{-1}(g_j(u))$. At most $32$ choices of $u$ were
needed for any pair $(h,j)$. For every $5\le h\le100$ and every $j$ with $\{0,1,j\}$ not a progression, the
resulting certificate records one of two things. For $2044$ pairs it records that Lemma~\ref{lem:gcddescent}(1)
applies. For the other $2564$ pairs it records an irreducible polynomial of degree $h$, an element $a$, and three
$\F_2$-independent roots of $x^{2^j}+ax^2+(a+1)x$. A separate program checks every entry: it tests the modulus with
Rabin's irreducibility test, checks the independence of the roots and the vanishing of the trinomial, and checks that
no pair is missing.

We also checked Proposition~\ref{prop:twists} directly. For $q=32$, $64$, $128$ and $256$ we listed every
$\GL(2,q)$-submodule of every $M_d$ of dimension at most $6$, using all lines in the two-dimensional eigenspaces of
the diagonal matrices. Up to field automorphisms, the $6$-dimensional ones are exactly the codes $\cC_{2^j}$ with
$2\le j\le h-1$. Only $\cC_4$ and $\cC_\sigma$ for $h\in\{5,7\}$ are near-MDS, and they hold $4$-designs with
$\lambda=(q-8)/2$. Lemma~\ref{lem:closed} was confirmed by enumerating all such sets $S$ for $5\le h\le24$.

\section{Examples}\label{sec:examples}

This section gives explicit examples. The computations behind it are described in Section~\ref{sec:excomp}.
They are not part of the formalisation, but Proposition~\ref{prop:mobius} is.

\subsection{The twists for small \texorpdfstring{$h$}{h}}\label{sec:twisttable}

Table~\ref{tab:twists} shows, for $5\le h\le13$ and $1\le j\le h-1$, which of the codes $\cC_{2^j}$ of
Section~\ref{sec:unique} are near-MDS. For $j\ge2$ the dual of $\cC_{2^j}$ is monomially equivalent to the cyclic code
$\mathcal K_s$ of~\eqref{eq:Gs} with $s=2^{j-1}$ (Section~\ref{sec:nmdstwists}), so the table also describes these
cyclic codes. For $j\ge2$ these are, by Theorem~\ref{thm:unique}(1), all $[q+1,6]$ codes
with a monomial action of $\PGL(2,q)$ inducing its natural action, up to semilinear equivalence. Near-MDS codes occur
exactly for odd $h$ and $j\in\{2,(h+1)/2\}$, that is, for $j=2$ and $j=2^{-1}$ in $\mathbb Z/h$, and each of them
holds a $4$-$(q+1,6,(q-8)/2)$ design. For even $h$ there is none. Every other code with $2\le j\le h-2$ fails in one of
two ways. Either some five points are dependent, which in the table happens exactly when $\gcd(j,h)\ge2$, as in
Lemma~\ref{lem:edgetwists}(3), or some nonzero member of $W_{2^j}$ has at least seven zeros, as in
Lemma~\ref{lem:ninezeros}.

The entries \textbf{N} and z are determined by the numbers $\lambda(c)$ of dependent $6$-sets containing
$\{0,1,\infty,c\}$. Every $4$-set is equivalent under $\PGL(2,q)$ to $\{0,1,\infty,c\}$ for some $c$, so the dependent
$6$-sets of $\cC_{2^j}$ form a $4$-design exactly when $\lambda(c)$ does not depend on $c$. For $q=128$, for instance,
$\lambda(c)=60$ for every $c$ for $\cC_4$ and $\cC_{16}$, while for $\cC_8$ and $\cC_{32}$ it takes values between $48$
and $86$.

\begin{table}[t]
\caption{The codes $\cC_{2^j}$ for $5\le h\le13$. \textbf{N}: near-MDS, and the dependent $6$-sets form a
$4$-$(q+1,6,\lambda)$ design; M: MDS; d: some five points are dependent; z: any five points are independent, but some
nonzero member of $W_{2^j}$ has at least seven zeros; --: dimension $5$.}
\label{tab:twists}
\setlength{\tabcolsep}{3.6pt}
\begin{tabular}{@{}rr*{12}{c}r@{}}
\toprule
$h$ & $q$ & \multicolumn{12}{c}{$j$} & $\lambda$ \\
\cmidrule(lr){3-14}
 & & 1 & 2 & 3 & 4 & 5 & 6 & 7 & 8 & 9 & 10 & 11 & 12 & \\
\midrule
5 & 32 & -- & \textbf{N} & \textbf{N} & M &  &  &  &  &  &  &  &  & 12 \\
6 & 64 & -- & d & d & d & M &  &  &  &  &  &  &  &  \\
7 & 128 & -- & \textbf{N} & z & \textbf{N} & z & M &  &  &  &  &  &  & 60 \\
8 & 256 & -- & d & z & d & z & d & M &  &  &  &  &  &  \\
9 & 512 & -- & \textbf{N} & d & z & \textbf{N} & d & z & M &  &  &  &  & 252 \\
10 & 1{,}024 & -- & d & z & d & d & d & z & d & M &  &  &  &  \\
11 & 2{,}048 & -- & \textbf{N} & z & z & z & \textbf{N} & z & z & z & M &  &  & 1{,}020 \\
12 & 4{,}096 & -- & d & d & d & z & d & z & d & d & d & M &  &  \\
13 & 8{,}192 & -- & \textbf{N} & z & z & z & z & \textbf{N} & z & z & z & z & M & 4{,}092 \\
\bottomrule
\end{tabular}
\end{table}

\subsection{The two codes for \texorpdfstring{$q=32$}{q=32}}\label{sec:q32}

Let $q=32$, so that $h=5$, $s=4$ and $\sigma=8$, and let $\alpha\in\F_{32}$ be a root of $x^5+x^2+1$. Order the points
of $\PG(1,32)$ as $0,1,\alpha,\dots,\alpha^{30},\infty$. Then $\cC_4$ and $\cC_8$ have the generator matrices
$(I_6\mid P_4)$ and $(I_6\mid P_8)$ of Table~\ref{tab:gen32}. Both are $[33,6,27]$ codes, and by
Corollary~\ref{cor:weights} and the paragraph after it they have the same weight distribution,
\[
\begin{gathered}
A_{27}=1{,}014{,}816,\quad A_{28}=1{,}268{,}520,\quad A_{29}=20{,}296{,}320,\quad A_{30}=64{,}609{,}952,\\
A_{31}=210{,}132{,}384,\quad A_{32}=399{,}584{,}823,\quad A_{33}=376{,}835{,}008 .
\end{gathered}
\]
Their duals are $[33,27,6]$ codes with $A_6=1{,}014{,}816=31\cdot32{,}736$ and $A_7=105{,}033{,}456$.

In the cyclic representation, let $\beta\in\F_{1024}$ have order $33$ and satisfy $\beta+\beta^{-1}=\alpha^7$, so that
$\gamma=\alpha^7$ in~\eqref{eq:Gs}. Then
\[
\begin{aligned}
G_2(x)&=(x^2+\alpha^{14}x+1)(x^4+\alpha^{21}x^3+\alpha^{27}x^2+\alpha^{21}x+1)\\
&=x^6+\alpha^5x^5+\alpha^9x^4+\alpha^{10}x^3+\alpha^9x^2+\alpha^5x+1,\\
G_4(x)&=(x^2+\alpha^{28}x+1)(x^4+\alpha^{4}x^3+\alpha^{2}x^2+\alpha^{4}x+1)\\
&=x^6+\alpha^{19}x^5+\alpha^{11}x^4+\alpha^{30}x^3+\alpha^{11}x^2+\alpha^{19}x+1.
\end{aligned}
\]
The code $\mathcal K_2$ generated by $G_2$, with zeros $\beta^{\pm1},\beta^{\pm2},\beta^{\pm3}$, is the code of Tang
and Ding, and $\mathcal K_4$, with zeros $\beta^{\pm3},\beta^{\pm4},\beta^{\pm5}$, is monomially equivalent to
$\cC_8^\perp$ by Theorem~\ref{thm:main}(3). Both polynomials are instances of~\eqref{eq:Gs}; only $s$ changes.

\begin{table}[t]
\caption{Generator matrices $(I_6\mid P_\tau)$ of $\cC_4$ and $\cC_8$ for $q=32$, where $\alpha^5+\alpha^2+1=0$. The
columns of $I_6$ belong to the points $0,1,\alpha,\dots,\alpha^4$, in this order. Each row of the table gives, for one
of the other points, the corresponding column of $P_4$ and of $P_8$.}
\label{tab:gen32}
\setlength{\tabcolsep}{3.2pt}
\begin{tabular}{@{}c*{6}{c}c*{6}{c}@{}}
\toprule
point & \multicolumn{6}{c}{column of $P_4$} & & \multicolumn{6}{c}{column of $P_8$} \\
\cmidrule(lr){2-7}\cmidrule(l){9-14}
$\alpha^{5}$ & $1$ & $\alpha^{18}$ & $\alpha^{11}$ & $\alpha^{10}$ & $\alpha^{20}$ & $\alpha^{25}$ &  & $\alpha$ & $\alpha^{30}$ & $\alpha^{24}$ & $\alpha^{5}$ & $\alpha^{6}$ & $0$ \\
$\alpha^{6}$ & $\alpha^{21}$ & $\alpha^{12}$ & $\alpha^{19}$ & $\alpha^{26}$ & $\alpha^{20}$ & $\alpha^{6}$ &  & $\alpha$ & $0$ & $\alpha^{30}$ & $\alpha^{24}$ & $\alpha^{5}$ & $\alpha^{6}$ \\
$\alpha^{7}$ & $\alpha^{30}$ & $\alpha^{24}$ & $\alpha^{14}$ & $\alpha^{14}$ & $0$ & $\alpha^{8}$ &  & $\alpha^{28}$ & $\alpha^{5}$ & $\alpha^{30}$ & $\alpha^{22}$ & $\alpha^{4}$ & $\alpha^{5}$ \\
$\alpha^{8}$ & $\alpha^{15}$ & $\alpha^{26}$ & $\alpha^{21}$ & $\alpha^{24}$ & $\alpha^{27}$ & $\alpha^{2}$ &  & $\alpha^{13}$ & $\alpha^{4}$ & $\alpha^{20}$ & $\alpha^{18}$ & $\alpha^{30}$ & $\alpha^{4}$ \\
$\alpha^{9}$ & $\alpha^{16}$ & $\alpha^{20}$ & $\alpha^{27}$ & $\alpha^{28}$ & $\alpha^{27}$ & $0$ &  & $\alpha^{25}$ & $\alpha^{3}$ & $\alpha^{19}$ & $\alpha^{28}$ & $\alpha^{30}$ & $\alpha^{30}$ \\
$\alpha^{10}$ & $\alpha^{16}$ & $0$ & $\alpha^{20}$ & $\alpha^{27}$ & $\alpha^{28}$ & $\alpha^{27}$ &  & $\alpha^{21}$ & $\alpha^{29}$ & $\alpha^{11}$ & $\alpha^{28}$ & $\alpha^{17}$ & $\alpha^{30}$ \\
$\alpha^{11}$ & $\alpha^{4}$ & $\alpha^{14}$ & $\alpha^{7}$ & $\alpha^{19}$ & $\alpha^{4}$ & $\alpha^{12}$ &  & $\alpha^{25}$ & $\alpha^{29}$ & $\alpha^{19}$ & $\alpha^{26}$ & $\alpha^{17}$ & $\alpha^{17}$ \\
$\alpha^{12}$ & $\alpha^{24}$ & $\alpha^{30}$ & $\alpha^{30}$ & $1$ & $\alpha^{2}$ & $\alpha^{9}$ &  & $\alpha^{9}$ & $\alpha^{16}$ & $\alpha^{21}$ & $\alpha^{17}$ & $\alpha^{21}$ & $\alpha^{17}$ \\
$\alpha^{13}$ & $\alpha^{2}$ & $\alpha^{27}$ & $\alpha^{24}$ & $\alpha^{7}$ & $\alpha^{3}$ & $\alpha^{20}$ &  & $\alpha^{25}$ & $\alpha^{16}$ & $\alpha^{6}$ & $\alpha^{8}$ & $\alpha^{13}$ & $\alpha^{21}$ \\
$\alpha^{14}$ & $\alpha^{3}$ & $\alpha^{7}$ & $\alpha^{6}$ & $\alpha^{20}$ & $\alpha^{12}$ & $\alpha^{22}$ &  & $\alpha^{20}$ & $\alpha^{20}$ & $\alpha^{19}$ & $\alpha^{14}$ & $\alpha^{19}$ & $\alpha^{13}$ \\
$\alpha^{15}$ & $\alpha^{14}$ & $\alpha^{9}$ & $\alpha^{4}$ & $\alpha^{3}$ & $\alpha^{27}$ & $\alpha^{22}$ &  & $\alpha^{10}$ & $\alpha^{12}$ & $\alpha^{19}$ & $\alpha^{5}$ & $\alpha^{16}$ & $\alpha^{19}$ \\
$\alpha^{16}$ & $\alpha^{3}$ & $\alpha^{9}$ & $\alpha^{24}$ & $\alpha^{30}$ & $\alpha^{23}$ & $\alpha^{4}$ &  & $\alpha^{14}$ & $\alpha^{18}$ & $0$ & $\alpha^{21}$ & $\alpha^{13}$ & $\alpha^{16}$ \\
$\alpha^{17}$ & $\alpha^{21}$ & $\alpha^{22}$ & $\alpha^{5}$ & $\alpha^{18}$ & $\alpha^{20}$ & $\alpha^{19}$ &  & $\alpha^{12}$ & $\alpha^{15}$ & $\alpha^{25}$ & $\alpha^{21}$ & $\alpha^{8}$ & $\alpha^{13}$ \\
$\alpha^{18}$ & $\alpha^{24}$ & $\alpha^{6}$ & $\alpha^{11}$ & $\alpha^{20}$ & $\alpha^{12}$ & $\alpha^{4}$ &  & $\alpha^{17}$ & $\alpha^{12}$ & $\alpha^{22}$ & $\alpha^{9}$ & $\alpha^{24}$ & $\alpha^{8}$ \\
$\alpha^{19}$ & $\alpha^{12}$ & $\alpha^{22}$ & $\alpha^{22}$ & $\alpha^{9}$ & $\alpha^{30}$ & $\alpha^{11}$ &  & $\alpha^{29}$ & $\alpha^{7}$ & $\alpha^{20}$ & $\alpha^{29}$ & $\alpha^{11}$ & $\alpha^{24}$ \\
$\alpha^{20}$ & $\alpha^{29}$ & $\alpha^{29}$ & $0$ & $\alpha^{8}$ & $\alpha^{16}$ & $\alpha^{26}$ &  & $\alpha^{4}$ & $\alpha^{23}$ & $\alpha^{11}$ & $\alpha^{5}$ & $\alpha^{16}$ & $\alpha^{11}$ \\
$\alpha^{21}$ & $\alpha^{24}$ & $\alpha^{13}$ & $\alpha^{18}$ & $\alpha^{5}$ & $\alpha^{30}$ & $\alpha^{26}$ &  & $\alpha^{24}$ & $\alpha^{10}$ & $\alpha^{15}$ & $\alpha^{13}$ & $\alpha^{28}$ & $\alpha^{16}$ \\
$\alpha^{22}$ & $\alpha^{29}$ & $\alpha^{13}$ & $\alpha^{28}$ & $\alpha^{19}$ & $\alpha^{9}$ & $\alpha^{24}$ &  & $\alpha^{8}$ & $\alpha^{15}$ & $\alpha^{27}$ & $\alpha^{11}$ & $\alpha^{29}$ & $\alpha^{28}$ \\
$\alpha^{23}$ & $\alpha^{26}$ & $\alpha^{11}$ & $\alpha^{20}$ & $\alpha^{24}$ & $\alpha^{9}$ & $\alpha^{25}$ &  & $\alpha^{2}$ & $\alpha^{27}$ & $\alpha^{11}$ & $\alpha^{23}$ & $\alpha^{23}$ & $\alpha^{29}$ \\
$\alpha^{24}$ & $\alpha^{12}$ & $\alpha^{12}$ & $\alpha$ & $\alpha^{13}$ & $\alpha^{18}$ & $\alpha^{13}$ &  & $\alpha^{28}$ & $\alpha^{28}$ & $\alpha^{24}$ & $\alpha^{23}$ & $\alpha^{15}$ & $\alpha^{23}$ \\
$\alpha^{25}$ & $\alpha^{30}$ & $1$ & $\alpha^{4}$ & $\alpha^{8}$ & $\alpha^{21}$ & $\alpha^{26}$ &  & $\alpha^{3}$ & $\alpha^{22}$ & $\alpha^{8}$ & $\alpha^{3}$ & $\alpha^{19}$ & $\alpha^{15}$ \\
$\alpha^{26}$ & $\alpha^{5}$ & $\alpha^{13}$ & $\alpha^{27}$ & $\alpha^{22}$ & $\alpha^{30}$ & $\alpha^{7}$ &  & $\alpha^{17}$ & $\alpha^{14}$ & $\alpha^{21}$ & $1$ & $\alpha^{4}$ & $\alpha^{19}$ \\
$\alpha^{27}$ & $\alpha^{10}$ & $\alpha^{25}$ & $\alpha^{15}$ & $\alpha^{21}$ & $\alpha^{24}$ & $\alpha^{4}$ &  & $\alpha^{15}$ & $\alpha^{18}$ & $\alpha^{17}$ & $\alpha^{19}$ & $\alpha^{21}$ & $\alpha^{4}$ \\
$\alpha^{28}$ & $1$ & $\alpha^{22}$ & $\alpha^{19}$ & $\alpha$ & $\alpha^{19}$ & $\alpha^{26}$ &  & $\alpha^{9}$ & $\alpha^{3}$ & $\alpha^{22}$ & $\alpha^{29}$ & $\alpha^{26}$ & $\alpha^{21}$ \\
$\alpha^{29}$ & $\alpha^{28}$ & $\alpha^{13}$ & $\alpha^{15}$ & $\alpha^{18}$ & $\alpha^{14}$ & $\alpha^{6}$ &  & $\alpha^{23}$ & $\alpha^{20}$ & $\alpha^{22}$ & $\alpha$ & $\alpha$ & $\alpha^{26}$ \\
$\alpha^{30}$ & $\alpha^{13}$ & $\alpha^{24}$ & $\alpha^{23}$ & $\alpha^{2}$ & $\alpha^{7}$ & $\alpha^{13}$ &  & $\alpha^{2}$ & $\alpha^{25}$ & $\alpha^{6}$ & $\alpha^{7}$ & $0$ & $\alpha$ \\
$\infty$ & $\alpha^{19}$ & $\alpha^{27}$ & $\alpha^{7}$ & $\alpha^{2}$ & $\alpha$ & $\alpha^{15}$ &  & $\alpha^{11}$ & $\alpha^{3}$ & $\alpha^{3}$ & $\alpha^{2}$ & $\alpha^{24}$ & $\alpha^{14}$ \\
\bottomrule
\end{tabular}
\end{table}

\subsection{The blocks through a \texorpdfstring{$4$}{4}-set}\label{sec:pictures}

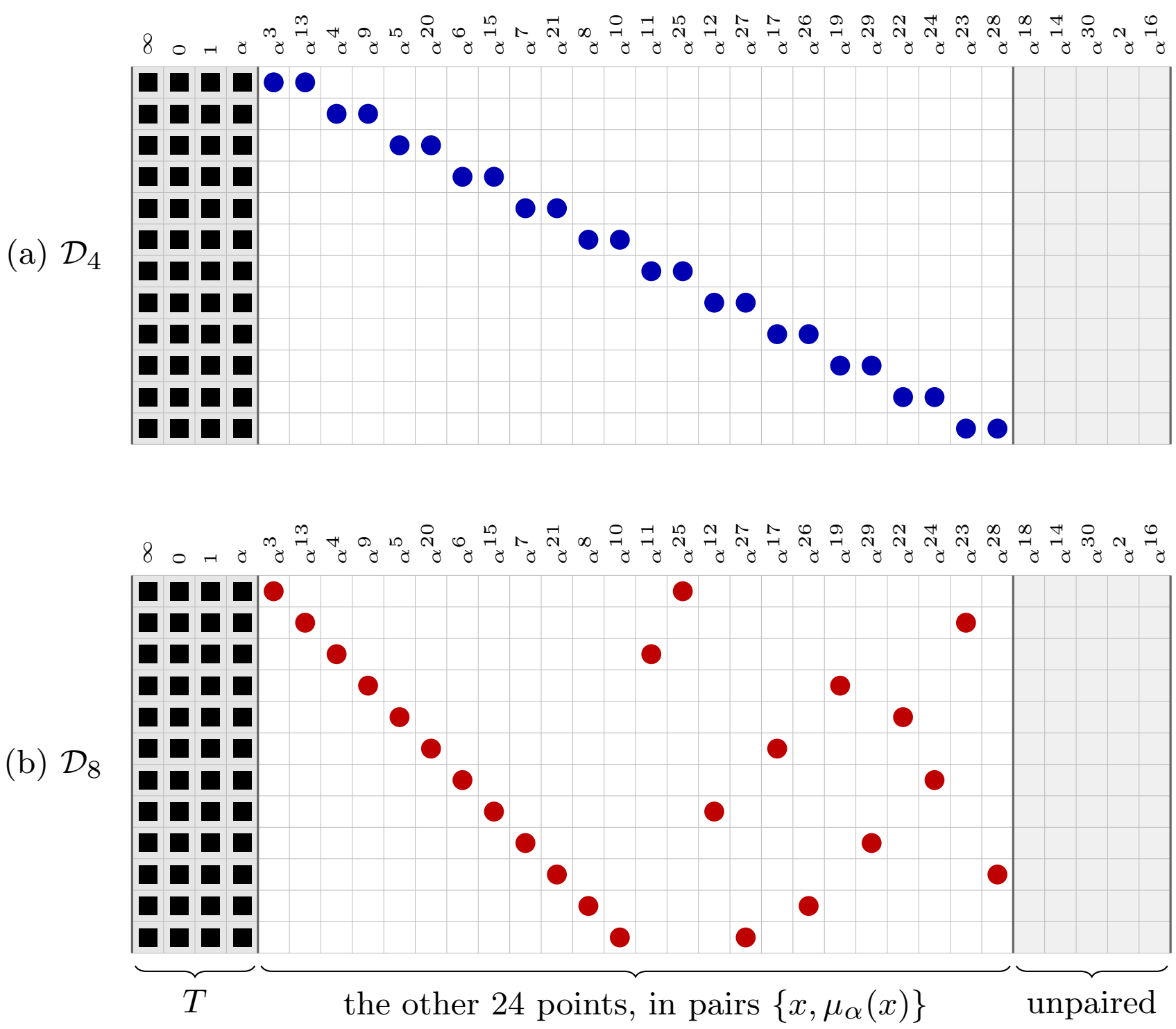
\begin{figure}[t]
\centering
\begin{tikzpicture}[x=1cm,y=1cm,
  tsq/.style={fill=black,rectangle,inner sep=0pt,minimum size=5pt},
  dot/.style={circle,inner sep=0pt,minimum size=5.4pt},
  lab/.style={font=\tiny,inner sep=1pt}]
\begin{scope}[yshift=0cm]
\fill[black!10] (0,0) rectangle (1.200,3.600);
\fill[black!6] (8.400,0) rectangle (9.900,3.600);
\draw[black!25,line width=0.2pt] (0,0.000) -- (9.900,0.000);
\draw[black!25,line width=0.2pt] (0,0.300) -- (9.900,0.300);
\draw[black!25,line width=0.2pt] (0,0.600) -- (9.900,0.600);
\draw[black!25,line width=0.2pt] (0,0.900) -- (9.900,0.900);
\draw[black!25,line width=0.2pt] (0,1.200) -- (9.900,1.200);
\draw[black!25,line width=0.2pt] (0,1.500) -- (9.900,1.500);
\draw[black!25,line width=0.2pt] (0,1.800) -- (9.900,1.800);
\draw[black!25,line width=0.2pt] (0,2.100) -- (9.900,2.100);
\draw[black!25,line width=0.2pt] (0,2.400) -- (9.900,2.400);
\draw[black!25,line width=0.2pt] (0,2.700) -- (9.900,2.700);
\draw[black!25,line width=0.2pt] (0,3.000) -- (9.900,3.000);
\draw[black!25,line width=0.2pt] (0,3.300) -- (9.900,3.300);
\draw[black!25,line width=0.2pt] (0,3.600) -- (9.900,3.600);
\draw[black!60,line width=0.6pt] (0.000,0) -- (0.000,3.600);
\draw[black!25,line width=0.2pt] (0.300,0) -- (0.300,3.600);
\draw[black!25,line width=0.2pt] (0.600,0) -- (0.600,3.600);
\draw[black!25,line width=0.2pt] (0.900,0) -- (0.900,3.600);
\draw[black!60,line width=0.6pt] (1.200,0) -- (1.200,3.600);
\draw[black!25,line width=0.2pt] (1.500,0) -- (1.500,3.600);
\draw[black!25,line width=0.2pt] (1.800,0) -- (1.800,3.600);
\draw[black!25,line width=0.2pt] (2.100,0) -- (2.100,3.600);
\draw[black!25,line width=0.2pt] (2.400,0) -- (2.400,3.600);
\draw[black!25,line width=0.2pt] (2.700,0) -- (2.700,3.600);
\draw[black!25,line width=0.2pt] (3.000,0) -- (3.000,3.600);
\draw[black!25,line width=0.2pt] (3.300,0) -- (3.300,3.600);
\draw[black!25,line width=0.2pt] (3.600,0) -- (3.600,3.600);
\draw[black!25,line width=0.2pt] (3.900,0) -- (3.900,3.600);
\draw[black!25,line width=0.2pt] (4.200,0) -- (4.200,3.600);
\draw[black!25,line width=0.2pt] (4.500,0) -- (4.500,3.600);
\draw[black!25,line width=0.2pt] (4.800,0) -- (4.800,3.600);
\draw[black!25,line width=0.2pt] (5.100,0) -- (5.100,3.600);
\draw[black!25,line width=0.2pt] (5.400,0) -- (5.400,3.600);
\draw[black!25,line width=0.2pt] (5.700,0) -- (5.700,3.600);
\draw[black!25,line width=0.2pt] (6.000,0) -- (6.000,3.600);
\draw[black!25,line width=0.2pt] (6.300,0) -- (6.300,3.600);
\draw[black!25,line width=0.2pt] (6.600,0) -- (6.600,3.600);
\draw[black!25,line width=0.2pt] (6.900,0) -- (6.900,3.600);
\draw[black!25,line width=0.2pt] (7.200,0) -- (7.200,3.600);
\draw[black!25,line width=0.2pt] (7.500,0) -- (7.500,3.600);
\draw[black!25,line width=0.2pt] (7.800,0) -- (7.800,3.600);
\draw[black!25,line width=0.2pt] (8.100,0) -- (8.100,3.600);
\draw[black!60,line width=0.6pt] (8.400,0) -- (8.400,3.600);
\draw[black!25,line width=0.2pt] (8.700,0) -- (8.700,3.600);
\draw[black!25,line width=0.2pt] (9.000,0) -- (9.000,3.600);
\draw[black!25,line width=0.2pt] (9.300,0) -- (9.300,3.600);
\draw[black!25,line width=0.2pt] (9.600,0) -- (9.600,3.600);
\draw[black!60,line width=0.6pt] (9.900,0) -- (9.900,3.600);
\node[tsq] at (0.150,3.450) {};
\node[tsq] at (0.450,3.450) {};
\node[tsq] at (0.750,3.450) {};
\node[tsq] at (1.050,3.450) {};
\node[dot,fill=blue!70!black] at (1.350,3.450) {};
\node[dot,fill=blue!70!black] at (1.650,3.450) {};
\node[tsq] at (0.150,3.150) {};
\node[tsq] at (0.450,3.150) {};
\node[tsq] at (0.750,3.150) {};
\node[tsq] at (1.050,3.150) {};
\node[dot,fill=blue!70!black] at (1.950,3.150) {};
\node[dot,fill=blue!70!black] at (2.250,3.150) {};
\node[tsq] at (0.150,2.850) {};
\node[tsq] at (0.450,2.850) {};
\node[tsq] at (0.750,2.850) {};
\node[tsq] at (1.050,2.850) {};
\node[dot,fill=blue!70!black] at (2.550,2.850) {};
\node[dot,fill=blue!70!black] at (2.850,2.850) {};
\node[tsq] at (0.150,2.550) {};
\node[tsq] at (0.450,2.550) {};
\node[tsq] at (0.750,2.550) {};
\node[tsq] at (1.050,2.550) {};
\node[dot,fill=blue!70!black] at (3.150,2.550) {};
\node[dot,fill=blue!70!black] at (3.450,2.550) {};
\node[tsq] at (0.150,2.250) {};
\node[tsq] at (0.450,2.250) {};
\node[tsq] at (0.750,2.250) {};
\node[tsq] at (1.050,2.250) {};
\node[dot,fill=blue!70!black] at (3.750,2.250) {};
\node[dot,fill=blue!70!black] at (4.050,2.250) {};
\node[tsq] at (0.150,1.950) {};
\node[tsq] at (0.450,1.950) {};
\node[tsq] at (0.750,1.950) {};
\node[tsq] at (1.050,1.950) {};
\node[dot,fill=blue!70!black] at (4.350,1.950) {};
\node[dot,fill=blue!70!black] at (4.650,1.950) {};
\node[tsq] at (0.150,1.650) {};
\node[tsq] at (0.450,1.650) {};
\node[tsq] at (0.750,1.650) {};
\node[tsq] at (1.050,1.650) {};
\node[dot,fill=blue!70!black] at (4.950,1.650) {};
\node[dot,fill=blue!70!black] at (5.250,1.650) {};
\node[tsq] at (0.150,1.350) {};
\node[tsq] at (0.450,1.350) {};
\node[tsq] at (0.750,1.350) {};
\node[tsq] at (1.050,1.350) {};
\node[dot,fill=blue!70!black] at (5.850,1.350) {};
\node[dot,fill=blue!70!black] at (5.550,1.350) {};
\node[tsq] at (0.150,1.050) {};
\node[tsq] at (0.450,1.050) {};
\node[tsq] at (0.750,1.050) {};
\node[tsq] at (1.050,1.050) {};
\node[dot,fill=blue!70!black] at (6.150,1.050) {};
\node[dot,fill=blue!70!black] at (6.450,1.050) {};
\node[tsq] at (0.150,0.750) {};
\node[tsq] at (0.450,0.750) {};
\node[tsq] at (0.750,0.750) {};
\node[tsq] at (1.050,0.750) {};
\node[dot,fill=blue!70!black] at (6.750,0.750) {};
\node[dot,fill=blue!70!black] at (7.050,0.750) {};
\node[tsq] at (0.150,0.450) {};
\node[tsq] at (0.450,0.450) {};
\node[tsq] at (0.750,0.450) {};
\node[tsq] at (1.050,0.450) {};
\node[dot,fill=blue!70!black] at (7.350,0.450) {};
\node[dot,fill=blue!70!black] at (7.650,0.450) {};
\node[tsq] at (0.150,0.150) {};
\node[tsq] at (0.450,0.150) {};
\node[tsq] at (0.750,0.150) {};
\node[tsq] at (1.050,0.150) {};
\node[dot,fill=blue!70!black] at (7.950,0.150) {};
\node[dot,fill=blue!70!black] at (8.250,0.150) {};
\node[lab,rotate=90,anchor=west] at (0.150,3.660) {$\infty$};
\node[lab,rotate=90,anchor=west] at (0.450,3.660) {$0$};
\node[lab,rotate=90,anchor=west] at (0.750,3.660) {$1$};
\node[lab,rotate=90,anchor=west] at (1.050,3.660) {$\alpha$};
\node[lab,rotate=90,anchor=west] at (1.350,3.660) {$\alpha^{3}$};
\node[lab,rotate=90,anchor=west] at (1.650,3.660) {$\alpha^{13}$};
\node[lab,rotate=90,anchor=west] at (1.950,3.660) {$\alpha^{4}$};
\node[lab,rotate=90,anchor=west] at (2.250,3.660) {$\alpha^{9}$};
\node[lab,rotate=90,anchor=west] at (2.550,3.660) {$\alpha^{5}$};
\node[lab,rotate=90,anchor=west] at (2.850,3.660) {$\alpha^{20}$};
\node[lab,rotate=90,anchor=west] at (3.150,3.660) {$\alpha^{6}$};
\node[lab,rotate=90,anchor=west] at (3.450,3.660) {$\alpha^{15}$};
\node[lab,rotate=90,anchor=west] at (3.750,3.660) {$\alpha^{7}$};
\node[lab,rotate=90,anchor=west] at (4.050,3.660) {$\alpha^{21}$};
\node[lab,rotate=90,anchor=west] at (4.350,3.660) {$\alpha^{8}$};
\node[lab,rotate=90,anchor=west] at (4.650,3.660) {$\alpha^{10}$};
\node[lab,rotate=90,anchor=west] at (4.950,3.660) {$\alpha^{11}$};
\node[lab,rotate=90,anchor=west] at (5.250,3.660) {$\alpha^{25}$};
\node[lab,rotate=90,anchor=west] at (5.550,3.660) {$\alpha^{12}$};
\node[lab,rotate=90,anchor=west] at (5.850,3.660) {$\alpha^{27}$};
\node[lab,rotate=90,anchor=west] at (6.150,3.660) {$\alpha^{17}$};
\node[lab,rotate=90,anchor=west] at (6.450,3.660) {$\alpha^{26}$};
\node[lab,rotate=90,anchor=west] at (6.750,3.660) {$\alpha^{19}$};
\node[lab,rotate=90,anchor=west] at (7.050,3.660) {$\alpha^{29}$};
\node[lab,rotate=90,anchor=west] at (7.350,3.660) {$\alpha^{22}$};
\node[lab,rotate=90,anchor=west] at (7.650,3.660) {$\alpha^{24}$};
\node[lab,rotate=90,anchor=west] at (7.950,3.660) {$\alpha^{23}$};
\node[lab,rotate=90,anchor=west] at (8.250,3.660) {$\alpha^{28}$};
\node[lab,rotate=90,anchor=west] at (8.550,3.660) {$\alpha^{18}$};
\node[lab,rotate=90,anchor=west] at (8.850,3.660) {$\alpha^{14}$};
\node[lab,rotate=90,anchor=west] at (9.150,3.660) {$\alpha^{30}$};
\node[lab,rotate=90,anchor=west] at (9.450,3.660) {$\alpha^{2}$};
\node[lab,rotate=90,anchor=west] at (9.750,3.660) {$\alpha^{16}$};
\node[anchor=east,font=\small] at (-0.15,1.800) {(a) $\cD_4$};
\end{scope}
\begin{scope}[yshift=-4.85cm]
\fill[black!10] (0,0) rectangle (1.200,3.600);
\fill[black!6] (8.400,0) rectangle (9.900,3.600);
\draw[black!25,line width=0.2pt] (0,0.000) -- (9.900,0.000);
\draw[black!25,line width=0.2pt] (0,0.300) -- (9.900,0.300);
\draw[black!25,line width=0.2pt] (0,0.600) -- (9.900,0.600);
\draw[black!25,line width=0.2pt] (0,0.900) -- (9.900,0.900);
\draw[black!25,line width=0.2pt] (0,1.200) -- (9.900,1.200);
\draw[black!25,line width=0.2pt] (0,1.500) -- (9.900,1.500);
\draw[black!25,line width=0.2pt] (0,1.800) -- (9.900,1.800);
\draw[black!25,line width=0.2pt] (0,2.100) -- (9.900,2.100);
\draw[black!25,line width=0.2pt] (0,2.400) -- (9.900,2.400);
\draw[black!25,line width=0.2pt] (0,2.700) -- (9.900,2.700);
\draw[black!25,line width=0.2pt] (0,3.000) -- (9.900,3.000);
\draw[black!25,line width=0.2pt] (0,3.300) -- (9.900,3.300);
\draw[black!25,line width=0.2pt] (0,3.600) -- (9.900,3.600);
\draw[black!60,line width=0.6pt] (0.000,0) -- (0.000,3.600);
\draw[black!25,line width=0.2pt] (0.300,0) -- (0.300,3.600);
\draw[black!25,line width=0.2pt] (0.600,0) -- (0.600,3.600);
\draw[black!25,line width=0.2pt] (0.900,0) -- (0.900,3.600);
\draw[black!60,line width=0.6pt] (1.200,0) -- (1.200,3.600);
\draw[black!25,line width=0.2pt] (1.500,0) -- (1.500,3.600);
\draw[black!25,line width=0.2pt] (1.800,0) -- (1.800,3.600);
\draw[black!25,line width=0.2pt] (2.100,0) -- (2.100,3.600);
\draw[black!25,line width=0.2pt] (2.400,0) -- (2.400,3.600);
\draw[black!25,line width=0.2pt] (2.700,0) -- (2.700,3.600);
\draw[black!25,line width=0.2pt] (3.000,0) -- (3.000,3.600);
\draw[black!25,line width=0.2pt] (3.300,0) -- (3.300,3.600);
\draw[black!25,line width=0.2pt] (3.600,0) -- (3.600,3.600);
\draw[black!25,line width=0.2pt] (3.900,0) -- (3.900,3.600);
\draw[black!25,line width=0.2pt] (4.200,0) -- (4.200,3.600);
\draw[black!25,line width=0.2pt] (4.500,0) -- (4.500,3.600);
\draw[black!25,line width=0.2pt] (4.800,0) -- (4.800,3.600);
\draw[black!25,line width=0.2pt] (5.100,0) -- (5.100,3.600);
\draw[black!25,line width=0.2pt] (5.400,0) -- (5.400,3.600);
\draw[black!25,line width=0.2pt] (5.700,0) -- (5.700,3.600);
\draw[black!25,line width=0.2pt] (6.000,0) -- (6.000,3.600);
\draw[black!25,line width=0.2pt] (6.300,0) -- (6.300,3.600);
\draw[black!25,line width=0.2pt] (6.600,0) -- (6.600,3.600);
\draw[black!25,line width=0.2pt] (6.900,0) -- (6.900,3.600);
\draw[black!25,line width=0.2pt] (7.200,0) -- (7.200,3.600);
\draw[black!25,line width=0.2pt] (7.500,0) -- (7.500,3.600);
\draw[black!25,line width=0.2pt] (7.800,0) -- (7.800,3.600);
\draw[black!25,line width=0.2pt] (8.100,0) -- (8.100,3.600);
\draw[black!60,line width=0.6pt] (8.400,0) -- (8.400,3.600);
\draw[black!25,line width=0.2pt] (8.700,0) -- (8.700,3.600);
\draw[black!25,line width=0.2pt] (9.000,0) -- (9.000,3.600);
\draw[black!25,line width=0.2pt] (9.300,0) -- (9.300,3.600);
\draw[black!25,line width=0.2pt] (9.600,0) -- (9.600,3.600);
\draw[black!60,line width=0.6pt] (9.900,0) -- (9.900,3.600);
\node[tsq] at (0.150,3.450) {};
\node[tsq] at (0.450,3.450) {};
\node[tsq] at (0.750,3.450) {};
\node[tsq] at (1.050,3.450) {};
\node[dot,fill=red!75!black] at (1.350,3.450) {};
\node[dot,fill=red!75!black] at (5.250,3.450) {};
\node[tsq] at (0.150,3.150) {};
\node[tsq] at (0.450,3.150) {};
\node[tsq] at (0.750,3.150) {};
\node[tsq] at (1.050,3.150) {};
\node[dot,fill=red!75!black] at (7.950,3.150) {};
\node[dot,fill=red!75!black] at (1.650,3.150) {};
\node[tsq] at (0.150,2.850) {};
\node[tsq] at (0.450,2.850) {};
\node[tsq] at (0.750,2.850) {};
\node[tsq] at (1.050,2.850) {};
\node[dot,fill=red!75!black] at (4.950,2.850) {};
\node[dot,fill=red!75!black] at (1.950,2.850) {};
\node[tsq] at (0.150,2.550) {};
\node[tsq] at (0.450,2.550) {};
\node[tsq] at (0.750,2.550) {};
\node[tsq] at (1.050,2.550) {};
\node[dot,fill=red!75!black] at (6.750,2.550) {};
\node[dot,fill=red!75!black] at (2.250,2.550) {};
\node[tsq] at (0.150,2.250) {};
\node[tsq] at (0.450,2.250) {};
\node[tsq] at (0.750,2.250) {};
\node[tsq] at (1.050,2.250) {};
\node[dot,fill=red!75!black] at (2.550,2.250) {};
\node[dot,fill=red!75!black] at (7.350,2.250) {};
\node[tsq] at (0.150,1.950) {};
\node[tsq] at (0.450,1.950) {};
\node[tsq] at (0.750,1.950) {};
\node[tsq] at (1.050,1.950) {};
\node[dot,fill=red!75!black] at (2.850,1.950) {};
\node[dot,fill=red!75!black] at (6.150,1.950) {};
\node[tsq] at (0.150,1.650) {};
\node[tsq] at (0.450,1.650) {};
\node[tsq] at (0.750,1.650) {};
\node[tsq] at (1.050,1.650) {};
\node[dot,fill=red!75!black] at (3.150,1.650) {};
\node[dot,fill=red!75!black] at (7.650,1.650) {};
\node[tsq] at (0.150,1.350) {};
\node[tsq] at (0.450,1.350) {};
\node[tsq] at (0.750,1.350) {};
\node[tsq] at (1.050,1.350) {};
\node[dot,fill=red!75!black] at (5.550,1.350) {};
\node[dot,fill=red!75!black] at (3.450,1.350) {};
\node[tsq] at (0.150,1.050) {};
\node[tsq] at (0.450,1.050) {};
\node[tsq] at (0.750,1.050) {};
\node[tsq] at (1.050,1.050) {};
\node[dot,fill=red!75!black] at (7.050,1.050) {};
\node[dot,fill=red!75!black] at (3.750,1.050) {};
\node[tsq] at (0.150,0.750) {};
\node[tsq] at (0.450,0.750) {};
\node[tsq] at (0.750,0.750) {};
\node[tsq] at (1.050,0.750) {};
\node[dot,fill=red!75!black] at (8.250,0.750) {};
\node[dot,fill=red!75!black] at (4.050,0.750) {};
\node[tsq] at (0.150,0.450) {};
\node[tsq] at (0.450,0.450) {};
\node[tsq] at (0.750,0.450) {};
\node[tsq] at (1.050,0.450) {};
\node[dot,fill=red!75!black] at (4.350,0.450) {};
\node[dot,fill=red!75!black] at (6.450,0.450) {};
\node[tsq] at (0.150,0.150) {};
\node[tsq] at (0.450,0.150) {};
\node[tsq] at (0.750,0.150) {};
\node[tsq] at (1.050,0.150) {};
\node[dot,fill=red!75!black] at (5.850,0.150) {};
\node[dot,fill=red!75!black] at (4.650,0.150) {};
\node[lab,rotate=90,anchor=west] at (0.150,3.660) {$\infty$};
\node[lab,rotate=90,anchor=west] at (0.450,3.660) {$0$};
\node[lab,rotate=90,anchor=west] at (0.750,3.660) {$1$};
\node[lab,rotate=90,anchor=west] at (1.050,3.660) {$\alpha$};
\node[lab,rotate=90,anchor=west] at (1.350,3.660) {$\alpha^{3}$};
\node[lab,rotate=90,anchor=west] at (1.650,3.660) {$\alpha^{13}$};
\node[lab,rotate=90,anchor=west] at (1.950,3.660) {$\alpha^{4}$};
\node[lab,rotate=90,anchor=west] at (2.250,3.660) {$\alpha^{9}$};
\node[lab,rotate=90,anchor=west] at (2.550,3.660) {$\alpha^{5}$};
\node[lab,rotate=90,anchor=west] at (2.850,3.660) {$\alpha^{20}$};
\node[lab,rotate=90,anchor=west] at (3.150,3.660) {$\alpha^{6}$};
\node[lab,rotate=90,anchor=west] at (3.450,3.660) {$\alpha^{15}$};
\node[lab,rotate=90,anchor=west] at (3.750,3.660) {$\alpha^{7}$};
\node[lab,rotate=90,anchor=west] at (4.050,3.660) {$\alpha^{21}$};
\node[lab,rotate=90,anchor=west] at (4.350,3.660) {$\alpha^{8}$};
\node[lab,rotate=90,anchor=west] at (4.650,3.660) {$\alpha^{10}$};
\node[lab,rotate=90,anchor=west] at (4.950,3.660) {$\alpha^{11}$};
\node[lab,rotate=90,anchor=west] at (5.250,3.660) {$\alpha^{25}$};
\node[lab,rotate=90,anchor=west] at (5.550,3.660) {$\alpha^{12}$};
\node[lab,rotate=90,anchor=west] at (5.850,3.660) {$\alpha^{27}$};
\node[lab,rotate=90,anchor=west] at (6.150,3.660) {$\alpha^{17}$};
\node[lab,rotate=90,anchor=west] at (6.450,3.660) {$\alpha^{26}$};
\node[lab,rotate=90,anchor=west] at (6.750,3.660) {$\alpha^{19}$};
\node[lab,rotate=90,anchor=west] at (7.050,3.660) {$\alpha^{29}$};
\node[lab,rotate=90,anchor=west] at (7.350,3.660) {$\alpha^{22}$};
\node[lab,rotate=90,anchor=west] at (7.650,3.660) {$\alpha^{24}$};
\node[lab,rotate=90,anchor=west] at (7.950,3.660) {$\alpha^{23}$};
\node[lab,rotate=90,anchor=west] at (8.250,3.660) {$\alpha^{28}$};
\node[lab,rotate=90,anchor=west] at (8.550,3.660) {$\alpha^{18}$};
\node[lab,rotate=90,anchor=west] at (8.850,3.660) {$\alpha^{14}$};
\node[lab,rotate=90,anchor=west] at (9.150,3.660) {$\alpha^{30}$};
\node[lab,rotate=90,anchor=west] at (9.450,3.660) {$\alpha^{2}$};
\node[lab,rotate=90,anchor=west] at (9.750,3.660) {$\alpha^{16}$};
\node[anchor=east,font=\small] at (-0.15,1.800) {(b) $\cD_8$};
\end{scope}
\draw[decorate,decoration={brace,mirror,amplitude=3pt}] (0.030,-4.970) -- (1.170,-4.970) node[midway,below=3pt,font=\scriptsize] {$T$};
\draw[decorate,decoration={brace,mirror,amplitude=3pt}] (1.230,-4.970) -- (8.370,-4.970) node[midway,below=3pt,font=\scriptsize] {the other $24$ points, in pairs $\{x,\mu_\alpha(x)\}$};
\draw[decorate,decoration={brace,mirror,amplitude=3pt}] (8.430,-4.970) -- (9.870,-4.970) node[midway,below=3pt,font=\scriptsize] {unpaired};
\end{tikzpicture}
\caption{Incidence matrices of the twelve blocks of $\cD_4$ (a) and of $\cD_8$ (b) that contain
$T=\{\infty,0,1,\alpha\}$, for $q=32$. Rows are blocks and columns are the $33$ points of $\PG(1,32)$. A square marks
a point of $T$, and a dot one of the two further points of a block. After $T$ the columns come in pairs
$\{x,\mu_\alpha(x)\}$ for the involution $\mu_\alpha$ of Proposition~\ref{prop:mobius}. The last five columns are the
unpaired points: the images of $\infty,0,1,\alpha$ under $\mu_\alpha$, and its fixed point $\sqrt\alpha=\alpha^{16}$.}
\label{fig:incidence}
\end{figure}

Figure~\ref{fig:incidence} shows, for $q=32$ and $T=\{\infty,0,1,\alpha\}$, the blocks of $\cD_4$ and of $\cD_8$ that
contain $T$, as incidence matrices: each row is a block, each column a point, and a mark means that the point lies in
the block. Each panel has twelve rows, since $\lambda=(q-8)/2=12$, and the four columns of $T$ are filled in every
row. Every other column has at most one mark, because a $5$-set lies in at most one block, so the twelve blocks
use $24$ of the $29$ points outside $T$. The five remaining columns, the unpaired points $1+\alpha$,
$\alpha/(1+\alpha)$, $1/\alpha$, $\alpha^2$ and $\sqrt\alpha$, are empty in both panels (Theorem~\ref{thm:fivesets}),
and no row occurs in both panels (Proposition~\ref{prop:differ}(1)). For $\cD_4$ the blocks through $T$ are given
explicitly by Tang and Ding's completion formula~\cite{TD21}, written in the coordinate $t$ of $\PG(1,q)$.

\begin{proposition}\label{prop:mobius}
Let $h\ge5$ be odd, $c\in\F_q\setminus\{0,1\}$ and $T=\{0,1,\infty,c\}$, and put
\[
\mu_c(t)=\frac{(1+c)t+c}{t+1+c} .
\]
Then $\mu_c$ is an involution in $\PGL(2,q)$. It maps $\infty,0,1,c$ to the unpaired points $1+c$, $c/(1+c)$, $1/c$,
$c^2$ and fixes the fifth unpaired point $\sqrt c$. The blocks of $\cD_4$ through $T$ are the sets
$T\cup\{u,\mu_c(u)\}$, where $u$ runs over the points outside $T$ that are not unpaired.
\end{proposition}

\begin{proof}
The matrix of $\mu_c$ has equal diagonal entries $1+c$ and determinant $1+c+c^2\neq0$, so its square is a scalar
matrix and $\mu_c$ is an involution. The images of $\infty,0,1,c$ follow by substitution, and $\mu_c(t)=t$ means
$t^2=c$. By the proof of Theorem~\ref{thm:fivesets}, for $u\notin T$ with $u\neq1+c$ the set $T\cup\{u,v\}$ is a
block exactly when $v=(c+u+cu)/(1+c+u)=\mu_c(u)$ and $v\notin T\cup\{u\}$. The point $v$ lies in $T\cup\{u\}$
exactly when $u$ is $\mu_c(0)$, $\mu_c(1)$, $\mu_c(c)$ or $\sqrt c$, and $u=1+c=\mu_c(\infty)$ is unpaired as well.
\end{proof}

In Figure~\ref{fig:incidence} each point after $T$ is followed by its image under $\mu_\alpha$. Each block of
$\cD_4$ then occupies two adjacent columns. For $\cD_8$ this fails, and there is no such involution: no element of $\PGaL(2,32)$ maps each of the $24$ points to its partner. The same holds for
$T=\{0,1,\infty,\alpha\}$ when $q=128$ and $q=512$.

\subsection{An invariant that separates the designs}\label{sec:invariant}

\begin{figure}[t]
\centering
\begin{tikzpicture}[x=1cm,y=1cm]
\begin{scope}[xshift=0.000cm]
\draw[black!12,line width=0.3pt] (0,-1.574) -- (3.100,-1.574);
\node[anchor=east,font=\scriptsize] at (-0.04,-1.574) {$-150$};
\draw[black!12,line width=0.3pt] (0,-1.049) -- (3.100,-1.049);
\node[anchor=east,font=\scriptsize] at (-0.04,-1.049) {$-100$};
\draw[black!12,line width=0.3pt] (0,-0.525) -- (3.100,-0.525);
\node[anchor=east,font=\scriptsize] at (-0.04,-0.525) {$-50$};
\draw[black!12,line width=0.3pt] (0,0.000) -- (3.100,0.000);
\node[anchor=east,font=\scriptsize] at (-0.04,0.000) {$0$};
\draw[black!12,line width=0.3pt] (0,0.525) -- (3.100,0.525);
\node[anchor=east,font=\scriptsize] at (-0.04,0.525) {$50$};
\draw[black!12,line width=0.3pt] (0,1.049) -- (3.100,1.049);
\node[anchor=east,font=\scriptsize] at (-0.04,1.049) {$100$};
\draw[black!12,line width=0.3pt] (0,1.574) -- (3.100,1.574);
\node[anchor=east,font=\scriptsize] at (-0.04,1.574) {$150$};
\draw[black!60,line width=0.4pt] (0,-1.700) -- (0,1.700);
\draw[black!70,line width=0.5pt] (0,0) -- (3.100,0);
\node[font=\scriptsize] at (0.258,-1.920) {$4$};
\node[font=\scriptsize] at (0.775,-1.920) {$5$};
\fill[blue!70!black] (1.112,0) rectangle (1.472,-0.315);
\node[font=\scriptsize] at (1.292,-1.920) {$6$};
\fill[red!75!black] (1.628,0) rectangle (1.988,1.259);
\node[font=\scriptsize] at (1.808,-1.920) {$7$};
\fill[blue!70!black] (2.145,0) rectangle (2.505,-1.574);
\node[font=\scriptsize] at (2.325,-1.920) {$8$};
\fill[red!75!black] (2.662,0) rectangle (3.022,0.630);
\node[font=\scriptsize] at (2.842,-1.920) {$9$};
\node[font=\scriptsize] at (1.550,-2.300) {$k$};
\node[font=\small] at (1.550,2.050) {(a) $q=32$};
\node[font=\scriptsize,rotate=90] at (-1.04,0) {$\Delta_k$};
\end{scope}
\begin{scope}[xshift=4.250cm]
\draw[black!12,line width=0.3pt] (0,-1.562) -- (3.100,-1.562);
\node[anchor=east,font=\scriptsize] at (-0.04,-1.562) {$-15$};
\draw[black!12,line width=0.3pt] (0,-1.041) -- (3.100,-1.041);
\node[anchor=east,font=\scriptsize] at (-0.04,-1.041) {$-10$};
\draw[black!12,line width=0.3pt] (0,-0.521) -- (3.100,-0.521);
\node[anchor=east,font=\scriptsize] at (-0.04,-0.521) {$-5$};
\draw[black!12,line width=0.3pt] (0,0.000) -- (3.100,0.000);
\node[anchor=east,font=\scriptsize] at (-0.04,0.000) {$0$};
\draw[black!12,line width=0.3pt] (0,0.521) -- (3.100,0.521);
\node[anchor=east,font=\scriptsize] at (-0.04,0.521) {$5$};
\draw[black!12,line width=0.3pt] (0,1.041) -- (3.100,1.041);
\node[anchor=east,font=\scriptsize] at (-0.04,1.041) {$10$};
\draw[black!12,line width=0.3pt] (0,1.562) -- (3.100,1.562);
\node[anchor=east,font=\scriptsize] at (-0.04,1.562) {$15$};
\draw[black!60,line width=0.4pt] (0,-1.700) -- (0,1.700);
\draw[black!70,line width=0.5pt] (0,0) -- (3.100,0);
\node[font=\scriptsize] at (0.258,-1.920) {$4$};
\fill[blue!70!black] (0.595,0) rectangle (0.955,-0.066);
\node[font=\scriptsize] at (0.775,-1.920) {$5$};
\fill[blue!70!black] (1.112,0) rectangle (1.472,-0.350);
\node[font=\scriptsize] at (1.292,-1.920) {$6$};
\fill[red!75!black] (1.628,0) rectangle (1.988,1.443);
\node[font=\scriptsize] at (1.808,-1.920) {$7$};
\fill[blue!70!black] (2.145,0) rectangle (2.505,-1.574);
\node[font=\scriptsize] at (2.325,-1.920) {$8$};
\fill[red!75!black] (2.662,0) rectangle (3.022,0.547);
\node[font=\scriptsize] at (2.842,-1.920) {$9$};
\node[font=\scriptsize] at (1.550,-2.300) {$k$};
\node[font=\small] at (1.550,2.050) {(b) $q=128$};
\node[font=\scriptsize,rotate=90] at (-0.88,0) {$\Delta_k$ ($\times10^3$)};
\end{scope}
\begin{scope}[xshift=8.500cm]
\draw[black!12,line width=0.3pt] (0,-1.342) -- (3.100,-1.342);
\node[anchor=east,font=\scriptsize] at (-0.04,-1.342) {$-3$};
\draw[black!12,line width=0.3pt] (0,-0.895) -- (3.100,-0.895);
\node[anchor=east,font=\scriptsize] at (-0.04,-0.895) {$-2$};
\draw[black!12,line width=0.3pt] (0,-0.447) -- (3.100,-0.447);
\node[anchor=east,font=\scriptsize] at (-0.04,-0.447) {$-1$};
\draw[black!12,line width=0.3pt] (0,0.000) -- (3.100,0.000);
\node[anchor=east,font=\scriptsize] at (-0.04,0.000) {$0$};
\draw[black!12,line width=0.3pt] (0,0.447) -- (3.100,0.447);
\node[anchor=east,font=\scriptsize] at (-0.04,0.447) {$1$};
\draw[black!12,line width=0.3pt] (0,0.895) -- (3.100,0.895);
\node[anchor=east,font=\scriptsize] at (-0.04,0.895) {$2$};
\draw[black!12,line width=0.3pt] (0,1.342) -- (3.100,1.342);
\node[anchor=east,font=\scriptsize] at (-0.04,1.342) {$3$};
\draw[black!60,line width=0.4pt] (0,-1.700) -- (0,1.700);
\draw[black!70,line width=0.5pt] (0,0) -- (3.100,0);
\node[font=\scriptsize] at (0.258,-1.920) {$4$};
\fill[blue!70!black] (0.595,0) rectangle (0.955,-0.013);
\node[font=\scriptsize] at (0.775,-1.920) {$5$};
\fill[blue!70!black] (1.112,0) rectangle (1.472,-0.493);
\node[font=\scriptsize] at (1.292,-1.920) {$6$};
\fill[red!75!black] (1.628,0) rectangle (1.988,1.553);
\node[font=\scriptsize] at (1.808,-1.920) {$7$};
\fill[blue!70!black] (2.145,0) rectangle (2.505,-1.574);
\node[font=\scriptsize] at (2.325,-1.920) {$8$};
\fill[red!75!black] (2.662,0) rectangle (3.022,0.527);
\node[font=\scriptsize] at (2.842,-1.920) {$9$};
\node[font=\scriptsize] at (1.550,-2.300) {$k$};
\node[font=\small] at (1.550,2.050) {(c) $q=512$};
\node[font=\scriptsize,rotate=90] at (-0.72,0) {$\Delta_k$ ($\times10^5$)};
\end{scope}
\end{tikzpicture}
\caption{The differences $\Delta_k=n_k(\cD_\sigma)-n_k(\cD_4)$ for $q=32$, $128$ and $512$ ($\sigma=8$, $16$, $32$), where
$n_k$ is the number of unordered pairs of disjoint triples of the derived packing at a $3$-set that are met by exactly
$k$ further triples (Section~\ref{sec:invariant}). Red: more such pairs for $\cD_\sigma$; blue: more for $\cD_4$. The
exact values of $n_k$ are given with the programs of Section~\ref{sec:excomp}.}
\label{fig:delta}
\end{figure}
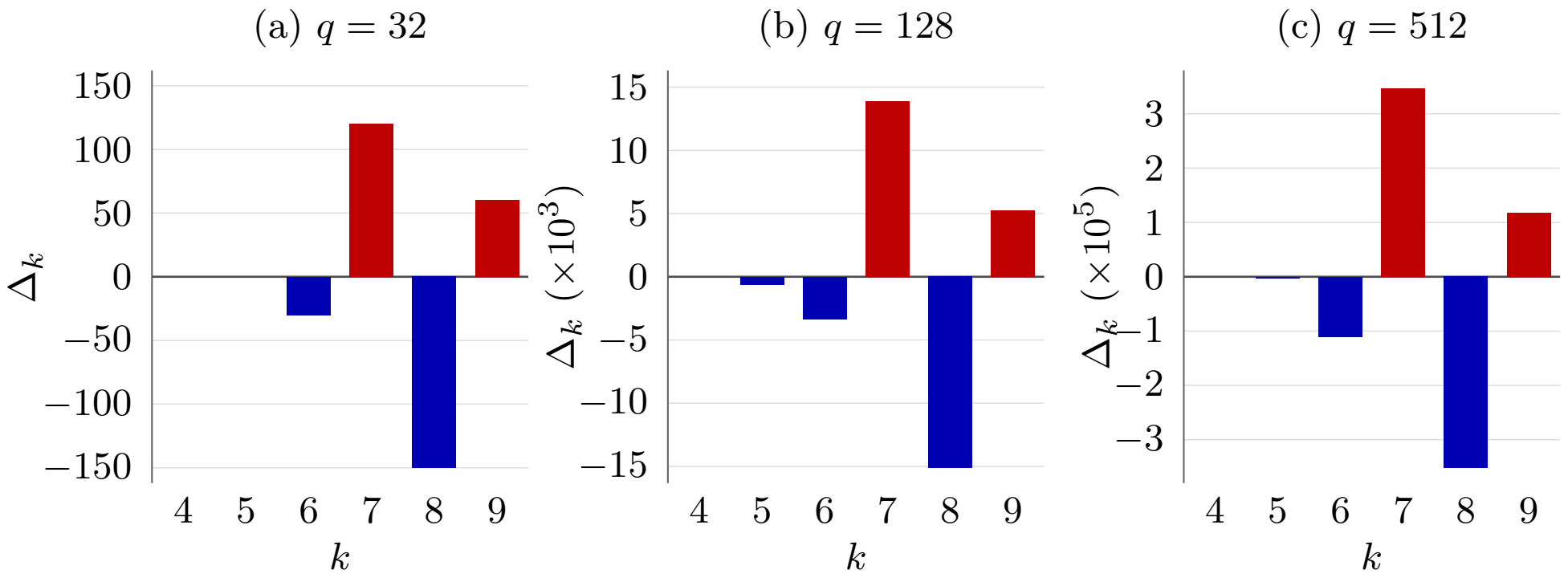

Theorem~\ref{thm:noniso} rests on Lemma~\ref{lem:aut}, and so on the classification of finite simple groups. For
small $q$ an elementary invariant separates the two designs. Let $\cD$ be one of them and let $S$ be a $3$-set. The
\emph{derived packing} $\cD_S=\{B\setminus S: S\subseteq B\in\cD\}$ consists of $(q-8)(q-2)/6$ triples on the $q-2$
points outside $S$, and two points lie in at most one triple, since a $5$-set lies in at most one block. For disjoint
triples $\ell,m\in\cD_S$ let $\kappa(\ell,m)$ be the number of triples of $\cD_S$ that meet both, and let $n_k$ be the
number of unordered pairs of disjoint triples with $\kappa(\ell,m)=k$. Since $\PGL(2,q)$ is $3$-transitive and
preserves $\cD$, all the packings $\cD_S$ are isomorphic, so the numbers $n_k$ do not depend on $S$, and an
isomorphism $\cD_\sigma\to\cD_4$ would preserve them. Figure~\ref{fig:delta} shows the differences
$\Delta_k=n_k(\cD_\sigma)-n_k(\cD_4)$ for $q=32$, $128$ and $512$. They are not all zero. For instance, for $q=32$ the number $n_9$ of pairs met by nine further triples is $570$ for
$\cD_4$ and $630$ for $\cD_8$, and for $q=128$ the number $n_5$ is $1{,}260$ for $\cD_4$ and $630$ for $\cD_{16}$. This
proves $\cD_\sigma\not\cong\cD_4$ for these $q$ without Lemma~\ref{lem:aut}.

Two points $x,y\notin S$ lie in no triple of $\cD_S$ exactly when $y$ is unpaired for $S\cup\{x\}$. By
Theorem~\ref{thm:fivesets} these pairs are the same for both designs, and $\kappa(\ell,m)=9-e(\ell,m)$, where
$e(\ell,m)$ is the number of such pairs with one point in $\ell$ and one in $m$. So the two designs differ in how
their triples lie with respect to the same set of uncovered pairs. This also explains why the differences are small.
The sum $\sum_kn_k$ is the number of pairs of disjoint triples, which depends only on the parameters. The sum
$\sum_kkn_k$ is $9\sum_kn_k$ minus $\sum e(\ell,m)$, and the latter counts, for each uncovered pair $\{x,y\}$, the
disjoint triples $\ell\ni x$ and $m\ni y$; their number is $r^2$, where $r=(q-8)/2$ is the number of triples
through a point, minus the number of points $z$ for which neither $\{x,z\}$ nor $\{y,z\}$ is uncovered. So both
sums are the same for $\cD_4$ and $\cD_\sigma$, that is, $\sum_k\Delta_k=\sum_kk\Delta_k=0$, and the two
distributions differ only in their shape. In all three cases $\Delta_k$ is negative for $k=6,8$ and positive for
$k=7,9$. Coarser invariants agree. For $q=32$ each design
is a single orbit of $\PGaL(2,32)$ whose block stabilisers have order $5$, and for $q=128$ each consists of three
orbits on which $\PGaL(2,128)$ acts regularly.

\subsection{The computations}\label{sec:excomp}

The data of this section were computed with short Python programs that use logarithm tables of $\F_q$ and
$\F_{1024}$. Table~\ref{tab:twists} and the numbers $\lambda(c)$ in Section~\ref{sec:twisttable} use the pencils
through the $4$-sets $\{0,1,\infty,c\}$, as described in Section~\ref{sec:evidence}. The blocks of $\cD_4$ and $\cD_\sigma$ through $\{0,1,\infty\}$ were obtained from the
same pencils, and the orbits and stabilisers by letting $\PGaL(2,q)$ act on these blocks. For the cyclic codes
generated by $G_2$ and $G_4$ we computed the derived packings at three coordinates from the check matrices over
$\F_{1024}$; they have the same numbers $n_k$ as $\cD_4$ and $\cD_8$, and the same holds for $G_2$ and $G_8$ when
$q=128$. For $q=32$, $128$ and $512$ we also checked that~\eqref{eq:Gs} agrees with the product of the minimal
polynomials of $\beta^{\pm(s-1)},\beta^{\pm s},\beta^{\pm(s+1)}$ for $s=2$ and $s=\sigma/2$. The weight distributions above satisfy the
MacWilliams identities.

\section{Concluding remarks and open problems}\label{sec:remarks}

We close with the computational evidence that led to and supports Theorems~\ref{thm:main} and~\ref{thm:noniso},
the formal verification of the results of Sections~\ref{sec:setting} to~\ref{sec:compare}, and the questions that
remain open.

\subsection{Computational evidence}\label{sec:evidence}

The computations in this subsection guided the work. They are evidence, not results, and none of them is used in
the proofs. The computation used in the proof of Theorem~\ref{thm:unique} is described in
Section~\ref{sec:certificate}.

For a power $\tau=2^j$, the forms $A+t^\tau B$ with $\deg A,\deg B\le2$ span $\Sym^2V\otimes V^{(\tau)}$. For each
$h\in\{5,\dots,11\}$ and each $j\in\mathbb Z/h$ we computed, for every $c\in\F_q\setminus\{0,1\}$, the number of
dependent $6$-sets of the evaluation code through $\{0,1,\infty,c\}$, and checked whether some $5$-set is
dependent. Since $\PGL(2,q)$ is $3$-transitive and acts on the code, this decides whether the dependent $6$-sets form a
$4$-design. They do only for odd $h$ and $j\in\{2,(h+1)/2\}$, that is, for the code of Tang and Ding and for
$\cC_\sigma$; for $h\in\{6,8,10\}$ they form a $4$-design for no $j$. The same holds for $5\le h\le13$
(Table~\ref{tab:twists}), and by
Theorem~\ref{thm:unique} the other twists are not even near-MDS when $h\le100$. The count of Section~\ref{sec:curves} suggests
why quadratic $A$ and $B$ are special: the argument needs the degree of the sextic, which is the intersection number
of the two curves, to equal the block size $6=\dim W_\sigma$. For $\deg A,\deg B\le k$ the two curves have classes $(k,1)$ and $(k,2)$ and
meet in $3k$ points, while the code has dimension $2k+2$, and these agree only for $k=2$.

We also sampled a wider class of codes. For $q=2^5,\dots,2^9$ and for $q=25$, $49$, $81$, $121$, $125$, $169$, $243$, $343$
we listed every $\GL_2(q)$-submodule of $\Sym^dV$ spanned by monomials, of dimension $r\in\{6,7,8\}$ with $d\le q-2$,
except those that are Frobenius twists of submodules of smaller degree. The monomial submodules were obtained by
closing sets of exponents under the action of $\GL_2(q)$, using Lucas's theorem for the binomial coefficients of a
substitution $t\mapsto t+a$. For each submodule we drew $60$ random $(r-2)$-sets of points ($40$ for odd $q$) and
counted their completions to dependent $r$-sets. In dimension $6$ the count was constant on the sample only for
$q=2^h$ with $h$ odd, and only for the two families of this paper and their Frobenius twists. For even $q$ in
dimension $6$, Theorem~\ref{thm:unique}(1) replaces this sample by a complete classification. In dimensions $7$ and $8$
every submodule had either no dependent $r$-sets or several values of the count, so none of them holds a $5$- or
$6$-design in this way.

The lemmas of Section~\ref{sec:proof} were checked numerically for $q=32$ and $q=128$, and Theorem~\ref{thm:shells} for
$q=32$. Proposition~\ref{prop:blocks} was checked on $20{,}000$ random $6$-sets for $q=32$ and $q=128$,
Theorem~\ref{thm:fivesets}(1) for every $c$ and $q\le2048$, and Proposition~\ref{prop:differ}(1) for every $4$-set
$\{0,1,\infty,c\}$ and $q\le8192$. For $q=32$, $128$ and $512$,
nauty~\cite{MP14} distinguishes the derived designs of $\cD_\sigma$ and $\cD_4$ at three points, and the numbers
$n_k$ of Section~\ref{sec:invariant} (Figure~\ref{fig:delta}) do so explicitly. The weight
distributions of Corollary~\ref{cor:weights} for $\cC_\sigma$ and $\cC_\sigma^\perp$ satisfy the MacWilliams
identities for $q=32$, $128$ and $512$.

\subsection{Formal verification}\label{sec:lean}

Theorems~\ref{thm:main}, \ref{thm:noniso} and~\ref{thm:unique}, every lemma, proposition, theorem and corollary of
Sections~\ref{sec:setting} to~\ref{sec:examples}, and equations~\eqref{eq:Gs} and~\eqref{eq:count} have been stated
and proved in the Lean~4 proof assistant~\cite{MU21} using the Mathlib library~\cite{Mat20}, for an arbitrary finite
field of order $2^h$, with $h\ge5$ odd in Sections~\ref{sec:setting} to~\ref{sec:compare} and~\ref{sec:examples} and
$h\ge5$ in Section~\ref{sec:unique}. The formal statements follow the wording of the paper. All proofs were replayed
by the kernel of Lean through the Comparator tool~\cite{Cmp}, which checks that the proved statements coincide with
separately written ones and that no axioms other than the permitted ones are used.

\begin{sidewaysfigure}
\centering
\resizebox{\textwidth}{!}{%
\definecolor{SecTwo}{HTML}{5C6F8A}\definecolor{SecThree}{HTML}{2E7D5B}\definecolor{SecFour}{HTML}{B5651D}%
\definecolor{SecFive}{HTML}{6A4C93}%
\begin{tikzpicture}[x=1cm,y=1cm,font=\fontsize{8}{9.4}\selectfont,
  box/.style={rectangle,rounded corners=2pt,align=flush center,text width=2.5cm,minimum height=1.2cm,inner sep=2.5pt,
    line width=.6pt,execute at begin node={\thinmuskip=3mu\medmuskip=4mu\thickmuskip=5mu}},
  arr/.style={-{Stealth[length=1.8mm,width=1.3mm]},line width=.6pt,draw=black!60,rounded corners=2pt},
  astmark/.style={anchor=south west,inner sep=0.6pt,font=\fontsize{9}{9}\selectfont}]
\fill[black!4,rounded corners=4pt] (-1.64,2.05) rectangle (20.34,-1.06);
\node[anchor=north west,inner sep=0pt,font=\fontsize{8.5}{10}\selectfont\itshape,text=black!65] at (-1.44,1.89) {the new family};
\fill[black!4,rounded corners=4pt] (-1.64,-2.15) rectangle (8.84,-6.80);
\node[anchor=north west,inner sep=0pt,font=\fontsize{8.5}{10}\selectfont\itshape,text=black!65] at (-1.44,-2.31) {comparison with $\cD_4$};
\fill[black!4,rounded corners=4pt] (9.86,-2.15) rectangle (20.34,-6.80);
\node[anchor=north east,inner sep=0pt,font=\fontsize{8.5}{10}\selectfont\itshape,text=black!65] at (20.14,-2.31) {uniqueness};
\node[box,draw=SecTwo,line width=.6pt,fill=SecTwo!8] (A) at (0.00,0.00) {\textbf{Lemmas \ref{lem:sigma}, \ref{lem:Wbasic}}\\$\sigma$; action of\\$\PGaL(2,q)$};
\node[box,draw=SecThree,line width=.6pt,fill=SecThree!8] (B) at (3.60,0.00) {\textbf{Lemmas \ref{lem:curves}, \ref{lem:zeros}}\\\textbf{Corollary \ref{cor:never5}}\\zeros of $W_\sigma$};
\node[box,draw=SecThree,line width=.6pt,fill=SecThree!8] (C) at (7.20,0.00) {\textbf{Props. \ref{prop:zerosets}, \ref{prop:N5}}\\\textbf{Equation \eqref{eq:count}}\\blocks through\\a $4$-set};
\node[box,draw=SecThree,line width=1.2pt,fill=SecThree!8] (D) at (11.50,0.00) {\textbf{Theorem \ref{thm:main}}\\\textbf{parts (1), (2)}\\near-MDS codes;\\$4$-design $\cD_\sigma$};
\node[box,draw=SecThree,line width=.6pt,fill=SecThree!8] (E) at (15.10,0.00) {\textbf{Theorem \ref{thm:shells}}\\\textbf{Corollary \ref{cor:weights}}\\weight shells\\and weights};
\node[box,draw=SecThree,line width=1.2pt,fill=SecThree!8] (F) at (18.70,0.00) {\textbf{Theorem \ref{thm:main}(3)}\\\textbf{Proposition \ref{prop:cyclic}}\\\textbf{Equation \eqref{eq:Gs}}\\cyclic codes;\\$\cD_4$ of Tang--Ding};
\node[box,draw=black!45,dashed,fill=black!3,text=black!75] (M) at (0.00,-3.40) {\itshape bialternant\\formula \cite{Mac95}};
\node[box,draw=SecFour,line width=.6pt,fill=SecFour!8] (G) at (3.60,-3.40) {\textbf{Props. \ref{prop:blocks}, \ref{prop:mobius}}\\block conditions;\\blocks of $\cD_4$};
\node[astmark] at ([xshift=-0.4pt,yshift=-1.6pt]G.north east) {$\ast$};
\node[box,draw=SecFour,line width=.6pt,fill=SecFour!8] (H) at (7.20,-3.40) {\textbf{Theorem \ref{thm:fivesets}}\\\textbf{Proposition \ref{prop:differ}}\\$\cD_\sigma\cap\cD_4=\emptyset$};
\node[box,draw=SecFive,line width=1.2pt,fill=SecFive!8] (T) at (11.50,-3.40) {\textbf{Theorem \ref{thm:nmdstwists}}\\\textbf{Theorem \ref{thm:unique}}\\the two families\\are the only ones};
\node[astmark] at ([xshift=-0.4pt,yshift=-1.6pt]T.north east) {$\ast$};
\node[box,draw=SecFive,line width=.6pt,fill=SecFive!8] (S) at (15.10,-3.40) {\textbf{Lemmas \ref{lem:edgetwists}--\ref{lem:progressions}}\\twists that are\\not near-MDS};
\node[astmark] at ([xshift=-0.4pt,yshift=-1.6pt]S.north east) {$\ast$};
\node[box,draw=black!45,dashed,fill=black!3,text=black!75] (Y) at (18.70,-3.40) {\itshape linearised\\polynomials\\\cite{GQ09,BZ20}};
\node[box,draw=black!45,dashed,fill=black!3,text=black!75] (N) at (0.00,-5.90) {\itshape overgroups of\\$\PGL(2,q)$\\\cite{DM96,Cam81}};
\node[box,draw=SecFour,line width=.6pt,fill=SecFour!8] (I) at (3.60,-5.90) {\textbf{Lemma \ref{lem:aut}}\\$\Aut(\cD)=\PGaL(2,q)$};
\node[astmark] at ([xshift=-0.4pt,yshift=-1.6pt]I.north east) {$\ast$};
\node[box,draw=SecFour,line width=1.2pt,fill=SecFour!8] (J) at (7.20,-5.90) {\textbf{Theorem \ref{thm:noniso}}\\\textbf{Corollary \ref{cor:codes}}\\$\cD_\sigma\not\cong\cD_4$};
\node[astmark] at ([xshift=-0.4pt,yshift=-1.6pt]J.north east) {$\ast$};
\node[box,draw=black!45,dashed,fill=black!3,text=black!75] (K) at (11.50,-5.90) {\itshape certificate of\\Section \ref{sec:certificate}};
\node[box,draw=SecFive,line width=.6pt,fill=SecFive!8] (U) at (15.10,-5.90) {\textbf{Lemmas \ref{lem:induced}--\ref{lem:closed}}\\\textbf{Proposition \ref{prop:twists}}\\invariant codes\\are twists};
\node[astmark] at ([xshift=-0.4pt,yshift=-1.6pt]U.north east) {$\ast$};
\node[box,draw=black!45,dashed,fill=black!3,text=black!75] (X) at (18.70,-5.90) {\itshape invariant codes\\\cite{DTT21}};
\draw[arr] (A) -- (B);
\draw[arr] (B) -- (C);
\draw[arr] (C) -- (D);
\draw[arr] (D) -- (E);
\draw[arr] ([xshift=6mm]C.north) |- (18.70,1.65) -- (F.north);
\draw[arr] (C) -- (H);
\draw[arr] (M) -- (G);
\draw[arr] (G) -- (H);
\draw[arr] (H) -- (J);
\draw[arr] (N) -- (I);
\draw[arr] (I) -- (J);
\draw[arr] (D) -- (T);
\draw[arr] (H) -- (T);
\draw[arr] ([yshift=3mm]S.west) -- ([yshift=3mm]T.east);
\draw[arr] (Y) -- (S);
\draw[arr] (K) -- (T);
\draw[arr] (U.west) -- (13.30,-5.90) -- (13.30,-3.70) -- ([yshift=-3mm]T.east);
\draw[arr] (X) -- (U);
\end{tikzpicture}}
\caption{Dependencies among the formal proofs of the numbered results of this paper (legend in
Section~\ref{sec:lean}). All 59 Lean statements are certified by the Lean Comparator, run 20261128-section5b.}
\label{fig:deps}
\end{sidewaysfigure}
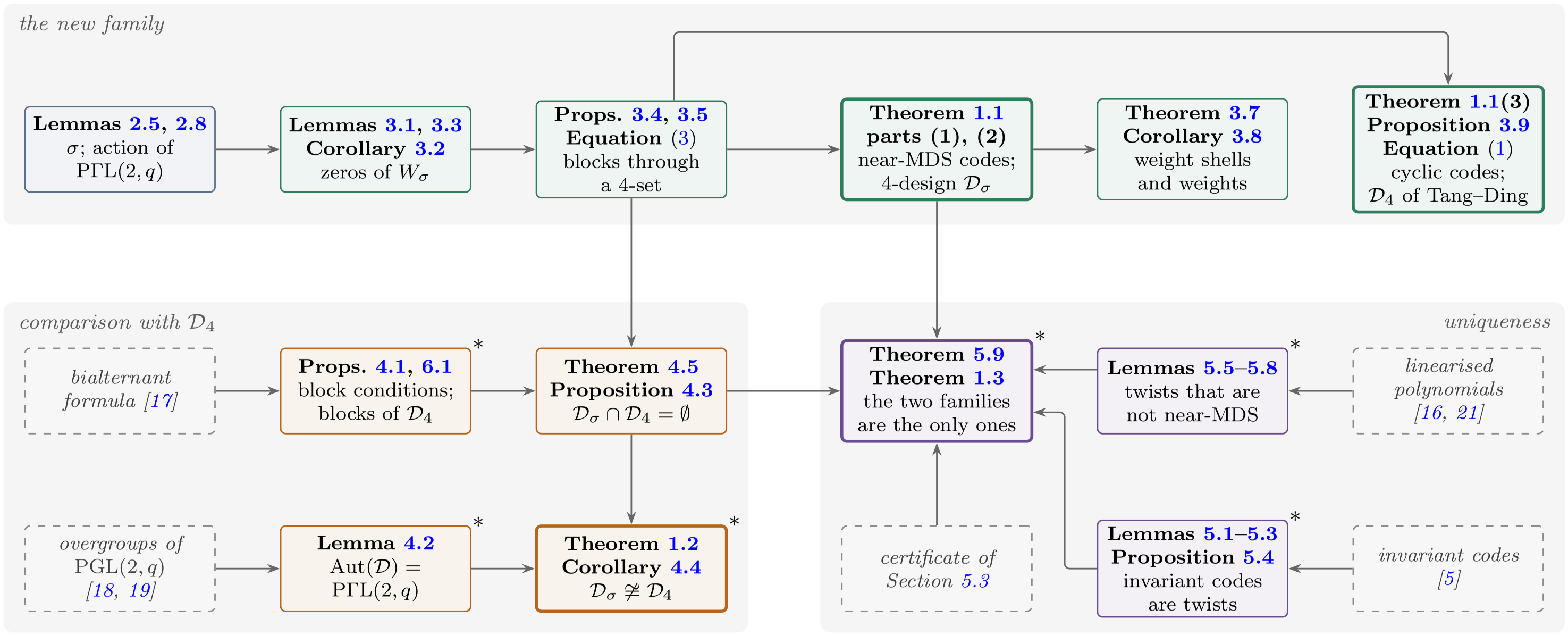

Figure~\ref{fig:deps} shows how the results depend on each other in these proofs. The boxes are grouped by theme:
the new family (Sections~\ref{sec:setting} and~\ref{sec:proof}), the comparison with $\cD_4$
(Section~\ref{sec:compare} and Proposition~\ref{prop:mobius}) and uniqueness (Section~\ref{sec:unique}). Each box
collects results that are proved one after another. An arrow points from a box to a box whose formal proofs use one of
its Lean statements, as read off the proof terms, and arrows implied by other arrows are omitted. The dashed boxes are
the cited results and the computation that enter the formalisation as axioms, described below. An asterisk marks a box
with a result whose formal proof uses one of them. Thick frames mark the main theorems.

Five cited results that are not available in Mathlib, and the computation of Section~\ref{sec:certificate}, enter
as axioms, one for each. The first is the bialternant formula~\cite{Mac95}. It is used only to express the
conditions of Proposition~\ref{prop:blocks} by Schur functions, and the determinant form of these conditions is
verified without it. The second is the statement that every subgroup of the symmetric group on $\PG(1,q)$ containing
$\PGL(2,q)$ either lies in $\PGaL(2,q)$ or contains the alternating group, which follows from the results cited in
the proof of Lemma~\ref{lem:aut}. It is used for Lemma~\ref{lem:aut}, Theorem~\ref{thm:noniso} and
Corollary~\ref{cor:codes}, and these three results are verified also with this statement as an explicit hypothesis,
without the axiom. The third is~\cite[Theorem~11]{DTT21} for the permutation action of $\PGL(2,q)$, used in
Proposition~\ref{prop:twists} for $d\equiv0$. The fourth is the bound of~\cite{GQ09} for polynomials
$c_0x+c_1x^\theta+c_2x^{\theta^2}$, used in Lemma~\ref{lem:progressions}. The fifth is~\cite[Theorem~3.2 and
Remark~3.3]{BZ20} for $q=2$, used in case~2 of Theorem~\ref{thm:nmdstwists}. The computation is stated as one axiom:
for $5\le h\le100$ and $2\le j\le h-2$ with $\{0,1,j\}$ not an arithmetic progression modulo $h$,
$\gcd(j,h)=1$ and $\gcd(j-1,h)\le2$, some $x^{2^j}+ax^2+bx$ has at least $8$ roots in $\F_q$. It is the content of
the certificate of Section~\ref{sec:certificate}, which the program described there checks, and it is used in cases~1
and~4 of Theorem~\ref{thm:nmdstwists} and in Theorem~\ref{thm:unique}(2). All other results use only the standard
axioms of Lean.

The formal statements are written with the definitions of Sections~\ref{sec:setting} and~\ref{sec:twists}. In the
formalisation Schur polynomials are defined by the Jacobi--Trudi determinant, and the bialternant formula is used only
for the Schur-function form of Proposition~\ref{prop:blocks}. The design $\cD_4$ enters the formal statements as the
set of dependent $6$-sets of $\cC_4$, as in Section~\ref{sec:compare}, and Proposition~\ref{prop:cyclic}(2)
identifies it formally with the design of Tang and Ding. In Theorem~\ref{thm:unique}(1) an action of $\PGL(2,q)$ is a
homomorphism $\rho$ into the group of invertible linear maps of $\F_q^{q+1}$ with $(\rho(g)x)_P=s_P\,x_{g^{-1}P}$ for
nonzero scalars $s_P$ depending on $g$ and $P$. In the
formal proof of Theorem~\ref{thm:nmdstwists} the near-MDS property of $\cC_4$ is derived from
Proposition~\ref{prop:zerosets}(1) and Theorem~\ref{thm:fivesets}(1), instead of being taken from~\cite{TD21}.
In some places the formal proofs take routes different from the arguments given above. The example of
Section~\ref{sec:group}, Remark~\ref{rem:moore}, the computations of Section~\ref{sec:examples} and the remarks in the
text between the numbered results, such as the expanded weight distributions after Corollary~\ref{cor:weights}, are
not part of the formalisation.

\subsection{Open problems}

For $h\le100$, Theorem~\ref{thm:unique} shows that the two families are the only ones with this symmetry in
characteristic $2$. Two questions remain.

\begin{problem}\label{prob:classify}
\begin{enumerate}[nosep]
\item Show that for every $h\ge5$ and $2\le j\le h-2$ the set $\{0,1,j\}$ is a Moore exponent set for $q=2$ and $n=h$
  only if it is an arithmetic progression modulo $h$. By Theorem~\ref{thm:nmdstwists} and Remark~\ref{rem:moore}, this
  would remove the restriction $h\le100$ from Theorem~\ref{thm:unique}.
\item Determine the near-MDS codes of length $q+1$ and dimension $6$ over $\F_q$, $q$ odd, on which $\PGL(2,q)$ acts
  by monomial maps inducing its natural action on $\PG(1,q)$, and decide whether any of them holds a $4$-design.
\end{enumerate}
\end{problem}

The two families hold their designs for different reasons. For $\cD_4$ the completion of a $4$-set is a M\"obius
involution, which in characteristic $2$ has exactly one fixed point (Proposition~\ref{prop:mobius}). For
$\cD_\sigma$ the completion map is not a M\"obius map (Figure~\ref{fig:incidence}),
and the uniformity comes from the involution $\Phi$ of Lemma~\ref{lem:curves}(3).

\begin{problem}\label{prob:unify}
Find a single argument that proves the design property for both $\cD_4$ and $\cD_\sigma$, for instance one that
uses only that $x$, $x^2$ and $x^\tau$ span a Gabidulin code.
\end{problem}

The two twists are related. Write the automorphisms of $\F_q$ as powers $\phi^j$, $j\in\mathbb Z/h$, of the Frobenius
map $\phi(t)=t^2$. Then $t\mapsto t^4$ is $\phi^2$, and $t\mapsto t^\sigma$ is $\phi^{(h+1)/2}$, the square root of
$\phi$, since $(t^\sigma)^\sigma=t^2$ by Lemma~\ref{lem:sigma}. So the two families correspond to $j=2$ and
$j=2^{-1}$ in $\mathbb Z/h$. By Lemma~\ref{lem:progressions}, these and $j=-1$ are exactly the $j$ for which
$\{0,1,j\}$ is an arithmetic progression, that is, for which $x$, $x^2$ and $x^{2^j}$ span a Gabidulin code. The
relation $(t^\sigma)^\sigma=t^2$ is also what closes the system of equations in the proof of
Proposition~\ref{prop:differ}(1). For even $h$ the element $2$ is not invertible in $\mathbb Z/h$, and by
Theorem~\ref{thm:unique} there is no such near-MDS code for even $h\le100$. An argument in which the
exponents $2$ and $2^{-1}$ play symmetric roles might be a starting point for Problem~\ref{prob:unify}.

Finally, the weight-$6$ designs here are not $5$-designs, since a $5$-set lies in at most one block and some $5$-sets
lie in none. The computations above suggest that this construction is specific to dimension $6$.

\begin{problem}[Tang and Ding~\cite{TD21}]\label{prob:strength}
Find an infinite family of linear codes holding $t$-designs with $t\ge5$, and a linear code holding a nontrivial
$6$-design.
\end{problem}

\backmatter

\section*{Statements and Declarations}

\bmhead{Funding} This research (J.-L. Kim) was
supported in part by the Brain Korea 21 FOUR (Fostering Outstanding Universities for Research) funded by the Ministry
of Education (MOE, Korea) and National Research Foundation of Korea (NRF) under Grant No.~4120240415042, and by the
Basic Science Research Program through the National Research Foundation of Korea (NRF) funded by the Ministry of
Science and ICT under Grant No.~RS-2025-24534992.

\bmhead{Competing interests} The authors have no competing interests to declare that are relevant to the content of
this article.

\bmhead{Data availability} The Lean formalisation of the results of Sections~\ref{sec:setting}
to~\ref{sec:compare} (Section~\ref{sec:lean}), with the versions of Lean and Mathlib and the axioms used by each
theorem, is available at \url{https://github.com/LeGenAI/nmds-four-designs}. The same repository contains the
certificate of Section~\ref{sec:certificate} and the programs that produce and check it, and the programs that
compute the data of Section~\ref{sec:examples}. The scripts used for the
numerical checks described in Section~\ref{sec:evidence} are available from the corresponding author on reasonable
request.

\bmhead{Use of generative AI} The construction, the proofs, the numerical checks and the
Lean formalisation were developed with substantial assistance from the AI system Claude (Anthropic), under the direction of the authors.
The authors checked every argument and computation and take full responsibility for the content.

\bibliography{references}

\end{document}